\documentclass[twocolumn]{aastex62}
\usepackage[encapsulated]{CJK}
\usepackage{mathrsfs}
\usepackage{subfigure}
\usepackage{epstopdf}
\usepackage{amsmath}
\usepackage{longtable}
\usepackage{booktabs}
\usepackage{hyperref} 
\usepackage{overpic}
\def\etal {et al.~}
\def\eg   {e.g.,~}
\newbox\grsign \setbox\grsign=\hbox{$>$} \newdimen\grdimen \grdimen=\ht\grsign
\newbox\laxbox \newbox\gaxbox
\setbox\gaxbox=\hbox{\raise.5ex\hbox{$>$}\llap
     {\lower.5ex\hbox{$\sim$}}}\ht1=\grdimen\dp1=0pt
\setbox\laxbox=\hbox{\raise.5ex\hbox{$<$}\llap
     {\lower.5ex\hbox{$\sim$}}}\ht2=\grdimen\dp2=0pt

\shorttitle{CMa OB1 survey}
\shortauthors{Lin \etal}

\def\co   {$^{12}$CO }                             
\def\xco  {$^{13}$CO }                             
\def\xxco {C$^{18}$O }                             

\newcommand{\kms}{km\,s$^{\rm -1}$}

\definecolor{malachite}{rgb}{0.34, 0.7, 0.22}

\begin{document}
\begin{CJK*}{UTF8}{gbsn}
\title{Three isotopologues CO survey of Molecular Clouds in the CMa OB1 complex}

\correspondingauthor{Yan Sun}
\email{yansun@pmo.ac.cn, linzh@pmo.ac.cn}

\author{Zehao lin}
\affiliation{Purple Mountain Observatory, Chinese Academy of Sciences, Nanjing 210008, People's Republic of China}
\affiliation{University of Science and Technology of China, 96 Jinzhai Road, Hefei 230026, People's Republic of China}
\affiliation{Shanghai Normal University, Shanghai 200234, People's Republic of China}

\author{Yan sun}
\affiliation{Purple Mountain Observatory, Chinese Academy of Sciences, Nanjing 210008, People's Republic of China}

\author{Ye Xu}
\affiliation{Purple Mountain Observatory, Chinese Academy of Sciences, Nanjing 210008, People's Republic of China}

\author{Ji Yang}
\affiliation{Purple Mountain Observatory, Chinese Academy of Sciences, Nanjing 210008, People's Republic of China}

\author{Yingjie li}
\affiliation{Purple Mountain Observatory, Chinese Academy of Sciences, Nanjing 210008, People's Republic of China}

\begin{abstract}
Using the Purple Mountain Observatory 13.7-m millimeter telescope at Delingha in China, we have conducted a large-scale simultaneous
survey of $^{12}$CO, $^{13}$CO, and \xxco ($J$ = 1--0) toward the CMa OB1 complex with a sky coverage of 16.5
deg$^{\rm 2}$ ($221\fdg5\leq l\leq227\degr$, $-2\fdg5\leq b\leq0\fdg5$).
Emission from the CMa OB1 complex is found in the range 7 km\,s$^{-1}\ \leq$ V$_\mathrm{LSR}\ \leq$ 25 km\,s$^{-1}$.
The large-scale structure, physical properties and chemical abundances of the molecular clouds are presented.
A total of 83 \xxco molecular clumps are identified with GaussClumps algorithm within the mapped region.
We find that 94\% of these \xxco molecular clumps are gravitationally bound. The relationship between their size and mass
indicates that none of the \xxco clump has the potential to form high-mass stars.
Using semi-automatic IDL algorithm, we newly discover 85 CO
outflow candidates in the mapped area, including 23 bipolar outflows candidates.
Additionally, a comparative study reveals evidence for significant variety of physical properties,
evolutionary stages, and levels of star formation activity in different sub-regions of the CMa OB1 complex.
\end{abstract}


\section{Introduction}\label{Section1}
The CMa OB1 complex is a frequently studied and relatively nearby star-forming region in the third Galactic quadrant.
A rich collection of objects lies in this region, including more than 30 nebulae \citep{1978ApJ...223..471H},
two H{\sc ii} regions \citep{1959ApJS....4..257S}, three water masers \citep{1994A&AS..103..541B},
as well as a large number of young stellar objects \citep[YSOs;][]{2013ApJ...772...45E,2019ApJS..240...26S}.

Many achievements have been made in previous CO surveys toward the CMa OB1 complex.
\cite{1980gmcg.work..211B} first observed and studied CO molecules in this region, although their resolution was only 8.8$\arcmin$. \cite{1980gmcg.work..211B} revealed molecular emission arising from two distinct clouds, which appeared to be associated with the H{\sc ii} region $S$292 and $S$296.
Later, \cite{2004PASJ...56..313K} made large-scale
observation of \xco ($J$ = 1--0) line in this region using the NANTEN 4-m telescope at 2.6$\arcmin$ resolution.
They detected three massive star-forming \xco clouds and six intermediate- or low-mass \xco clouds within the CMa OB1 complex.
The same telescope was used to map the CMa OB1 complex using the \co ($J$ = 1--0) line \citep{2013ApJ...772...45E}.
\cite{2013ApJ...772...45E} revealed that 93\% ``\co clumps'' are sub-virial, and they obtained a power-law fit to high-mass portion of the ``\co clumps'' distribution, $N(\log M) \propto M^\alpha$, which had a power-law index of $-0.7\pm0.3$.
More recent large-scale \co and \xco observations of the CMa OB1 complex were made with the ARO 12-m antenna \citep{2020A&A...633A.147B}. Although conducted with improved spatial resolution,
the root mean square (RMS) noise levels of the ARO survey were 0.79--1.31~K (for $^{12}$CO), and 0.33--0.58~K (for $^{13}$CO), with a velocity resolution of 0.65~\kms.
To date, \xxco observation has been conducted toward the $l$=224$\degr$ field \citep[with sky coverage of about 200 arcmin${^2}$;][]{2016A&A...594A..58O}, using the Mopra radio telescope.

However, limited by their relatively low spatial resolution, and/or sensitivity, to date no unbiased molecular outflow survey has been
carried out towards the CMa OB1 complex. In addition, there is a lack of large-scale \xxco observations toward the CMa OB1 complex,
which is generally optically thin and therefore allows us to probe the interior of molecular clouds (MCs) therein.

In this paper, we present a new $J$ = 1--0 $^{12}$CO/$^{13}$CO/\xxco survey towards the CMa OB1 complex with a sky coverage of 16.5
deg$^2$ ($221\fdg5\leq l\leq227\degr$, $-2\fdg5\leq b\leq0\fdg5$), which is part of the
ongoing project Milky Way Imaging Scroll Painting \citep[MWISP; refer to][ etc., for project details and initial results]
{2014AJ....147...46Z,2015ApJ...798L..27S,2017ApJS..229...24D,2017ApJS..230...17S,2019ApJS...240..9}.
This set of \co and \xco data, which has a high spatial resolution, high velocity resolution, and high sensitivity
have allowed us to carry out the first unbiased search for molecular outflows in this region.
Besides, this is the first time the CMa OB1 complex has been observed at a large-scale with \xxco line,
which allowed us to carry out the first search for dense molecular clumps that are related to star formation.

The remainder of this paper is organized as follows. In Section~\ref{Section2}, we describe the data quality of the CMa OB1 complex.
The large-scale structure, physical properties and chemical abundances of the MCs are presented in Section~\ref{Section3}. Next, in Section~\ref{Section4},
we identify the \xxco clumps and analyze their physical properties. In Section~\ref{Section5},
we search for CO molecular outflow candidates and investigate their physical properties.
In Section~\ref{Section6}, we discuss the analysis of the
physical association between the compact \xxco clumps and the outflow candidates.
Finally, in Section~\ref{Section7} a summary of our results is given.

\section{Data}\label{Section2}
The observations used in this study were conducted during September 2017 to April 2018 using the 13.7-m
millimeter-wavelength telescope of the Purple Mountain Observatory~(PMO) in Delingha, China.
Using the nine-beam superconducting spectroscopic array receiver (SSAR), working in sideband
separation mode and employing the fast Fourier transform spectrometer (FFTS) \citep{2012ITTST...2..593S},
the lines of three CO isotopologues ($^{12}$CO, $^{13}$CO, and \xxco ($J$ = 1--0)) were observed simultaneously. The \co line was observed at the upper sideband (USB) with a main-beam width of $\sim$ 48$\arcsec$, and the \xco and \xxco lines were observed at the lower sideband (LSB) with a main-beam width of $\sim$ 50$\arcsec$\footnote{For more information, see the details presented in the current status report of the
telescope \url{http://www.radioast.csdb.cn/yhhjindex.php}.}.
Both bandwidths were 1000 MHz with 16,384 channels, resulting in velocity separations of about
0.159~\kms for \co and 0.166~\kms for \xco and C$^{18}$O.
The typical noise temperatures including the atmosphere at 110~GHz and 115~GHz are 140~K and 250~K, respectively. A detailed description of the instrument was given by \cite{2012ITTST...2..593S}. All observations were conducted by using position-switch On-The-Fly (OTF) mode.

\begin{figure}[t]
\centering
\includegraphics[height=0.18\textheight]{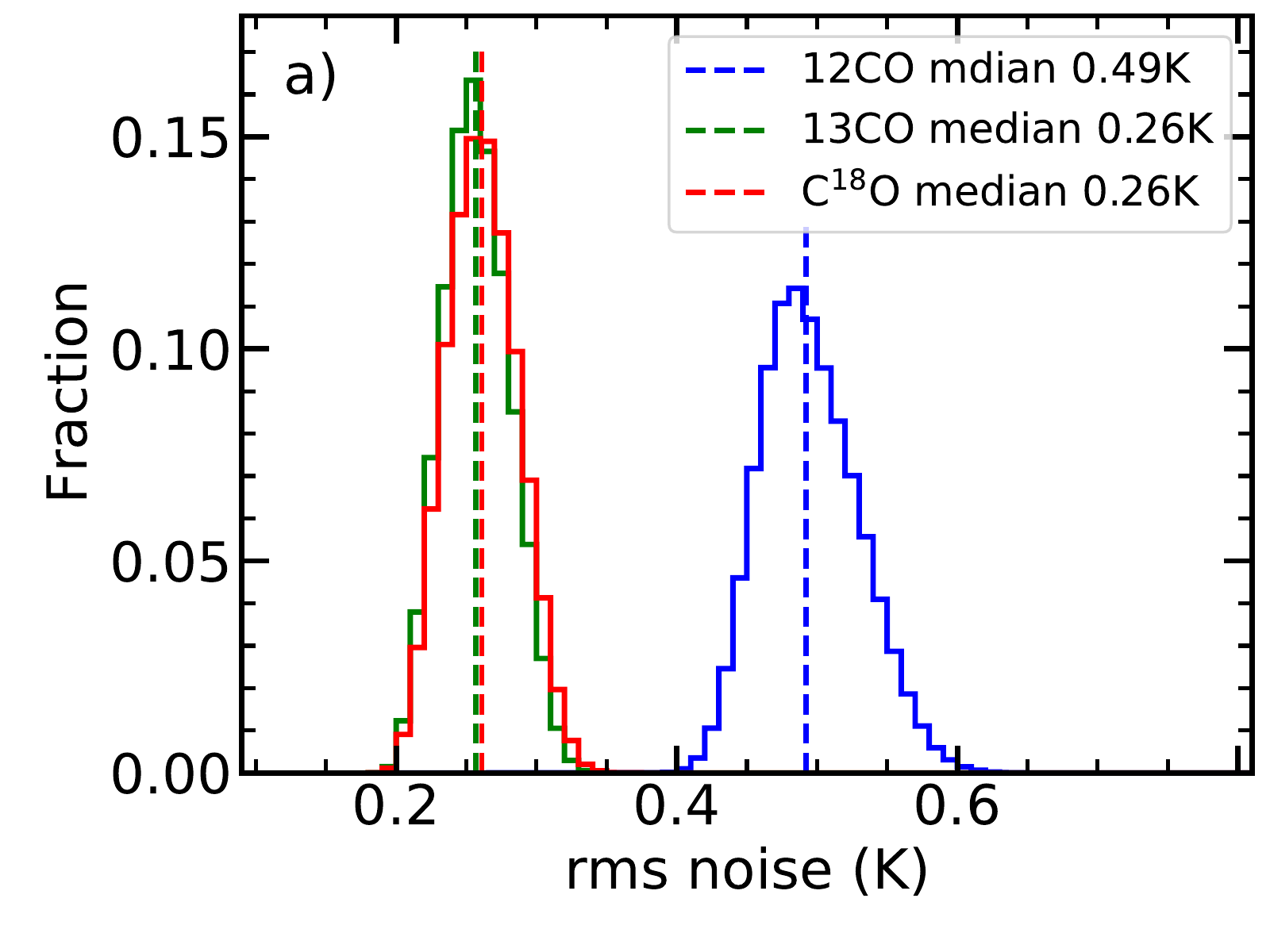}
\includegraphics[height=0.18\textheight]{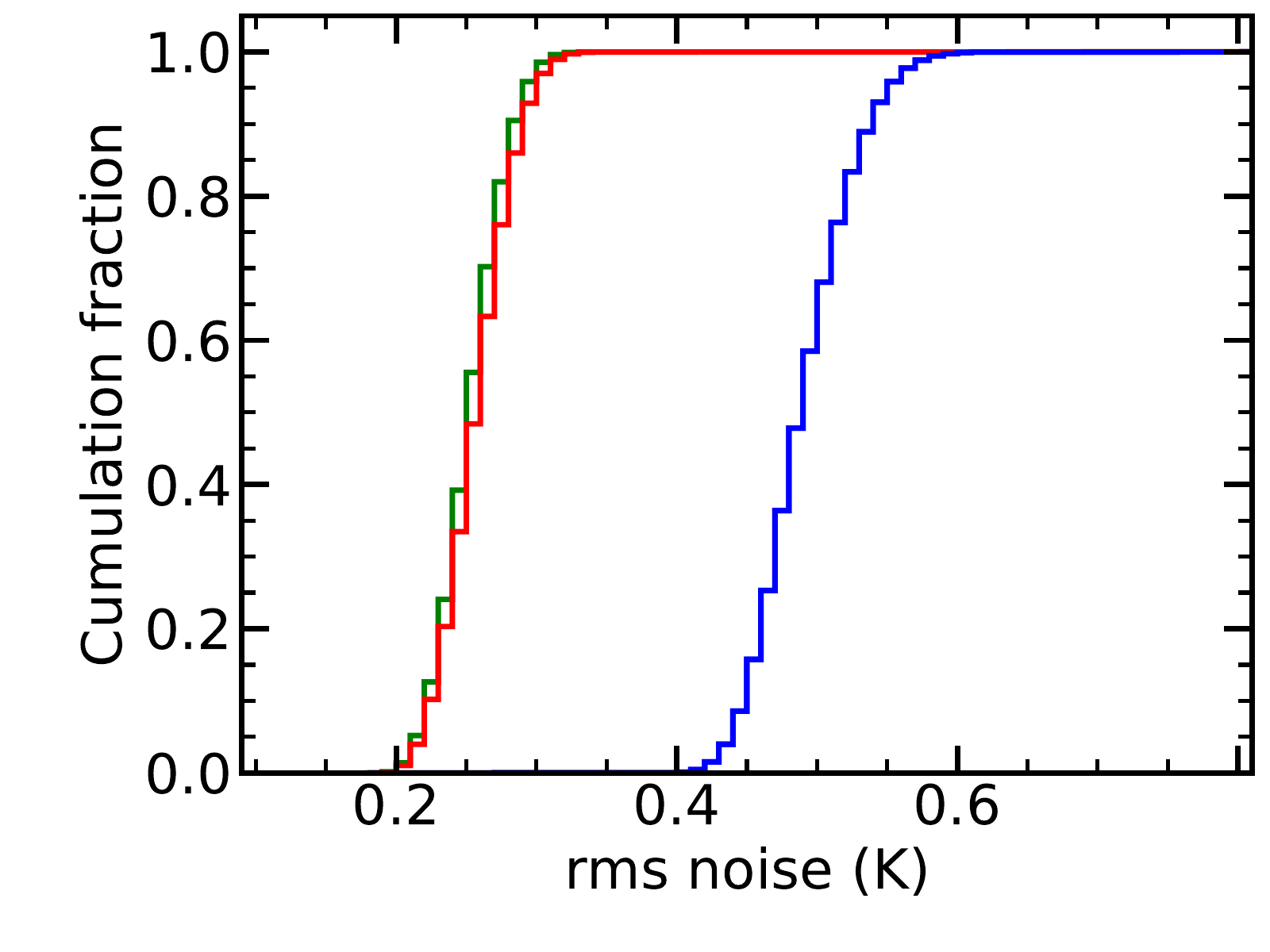}
\caption{($a$) Histograms of the rms noise levels ($\sigma$) for \co (blue), \xco (green), and \xxco (red).
($b$) Cumulative distributions of $\sigma$.}
\label{Fig:RMSdistribution}
\end{figure}

A first-order baseline was applied to the spectra in this study.
All the data were sampled every 30$\arcsec$.
The RMS distributions are presented in Figure~\ref{Fig:RMSdistribution}.
The typical RMS noise levels of the spectra are 0.49~K for \co ($J$ = 1--0) at a
velocity resolution of 0.159 km\,s$^{-1}$, and 0.26~K for \xco ($J$ = 1--0) and \xxco ($J$ = 1--0) at 0.166 km\,s$^{-1}$.
Please also refer to \cite{2019ApJS...240..9} for the detailed observations and data-reduction strategies.
The velocity channels were re-gridded into 0.168~\kms in this study, which facilitated comparisons between the
channel-by-channel emission from the three CO lines.

\section{Large-scale Molecular Clouds}\label{Section3}
\subsection{Molecular Gas Distributions}\label{Section3.1}
\begin{figure}[!ht]
\centering
\includegraphics[height=0.24\textheight]{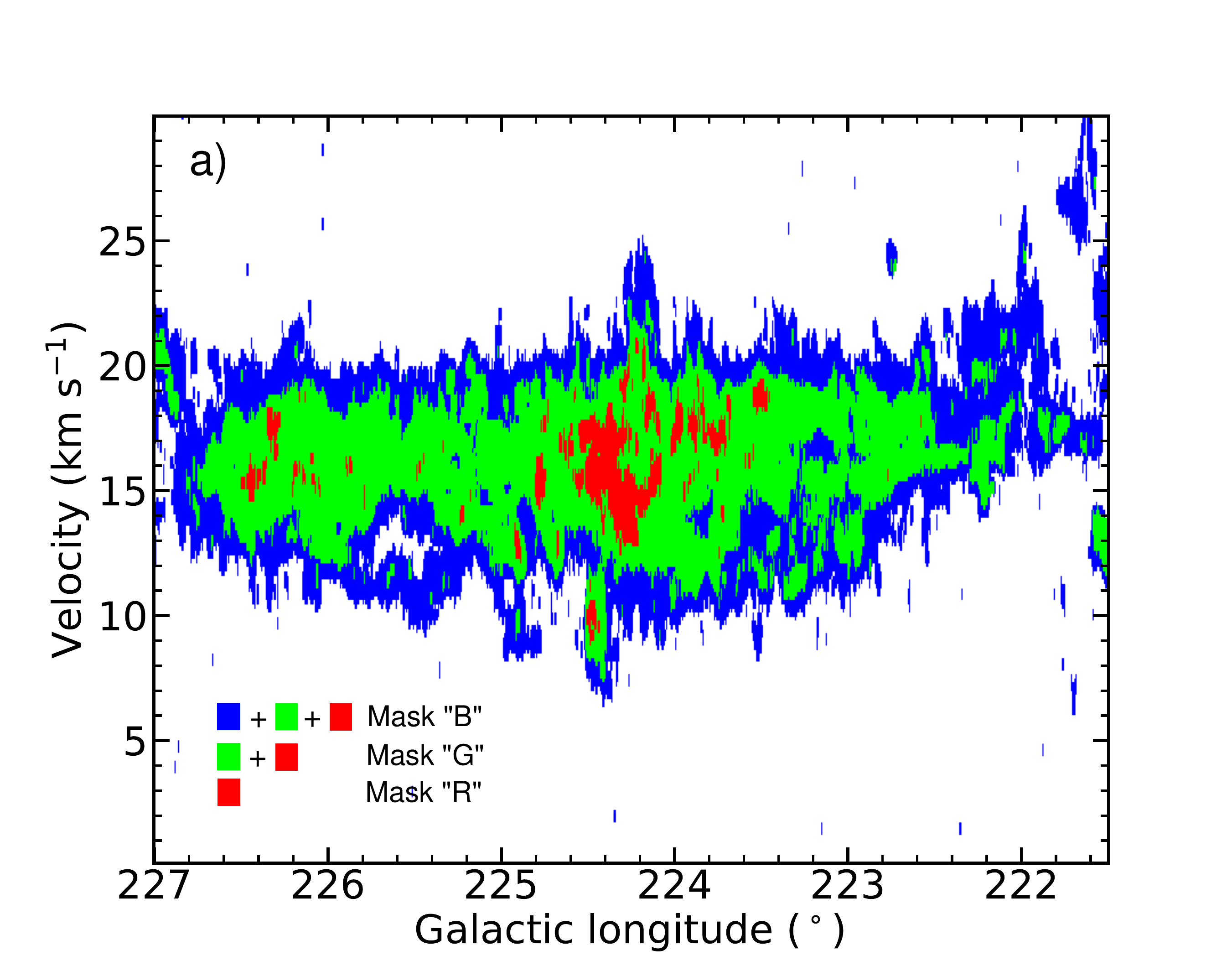}
\includegraphics[height=0.24\textheight]{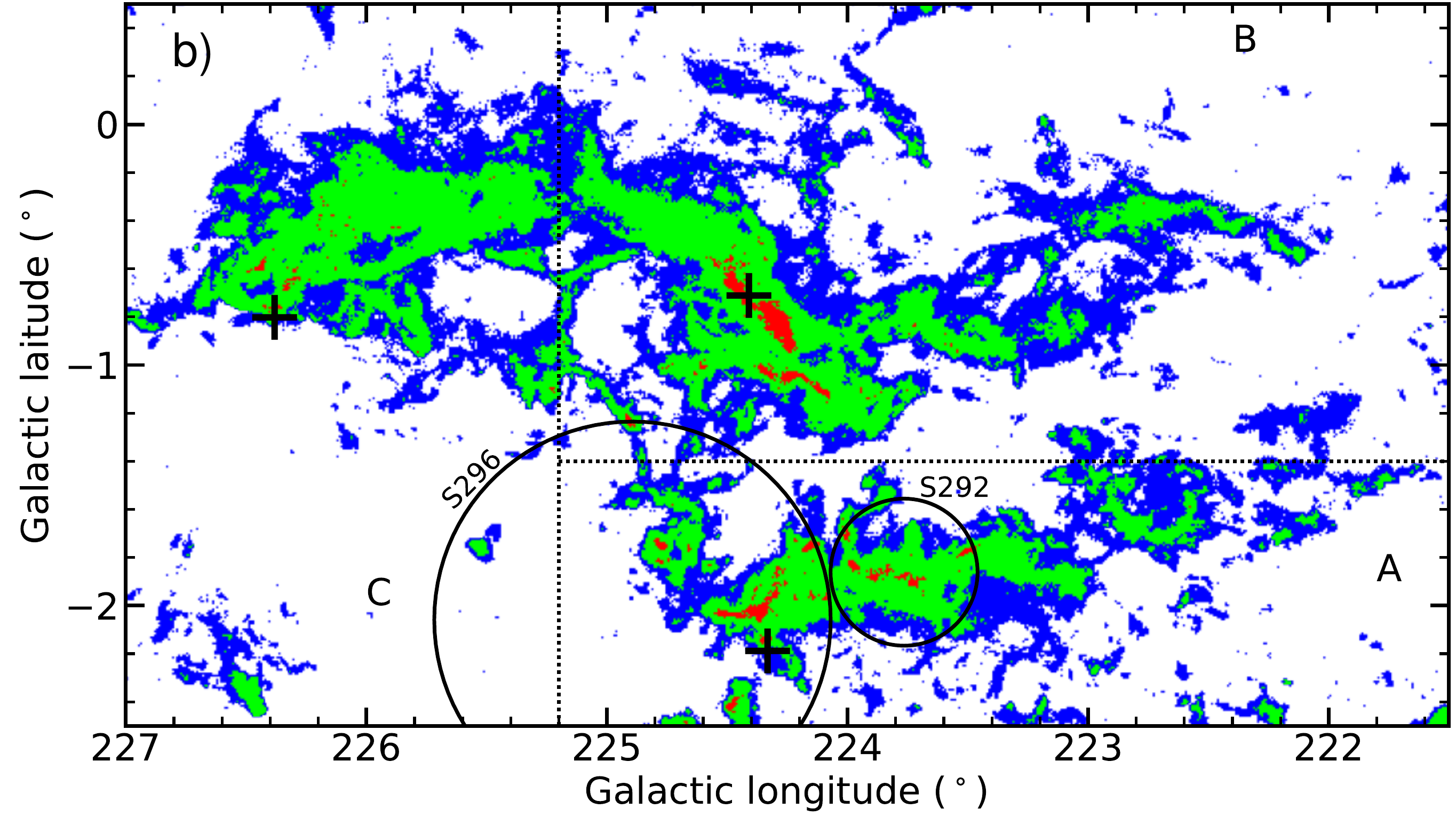}
\caption{($a$) Longitude--velocity map, integrated over Galactic latitudes $b$ = $-2\fdg5$ to $0\fdg5$.
($b$) Longitude-latitude map, integrated over velocities $V_{\rm LSR}$ = 0 to 30~\kms.
The crosses represent of positions of the three known water masers \citep{1994A&AS..103..541B}.
The two known H{\sc ii} regions from \cite{1959ApJS....4..257S} are also shown as circles.
The blue, green and red colors in both maps indicate the mask regions with detectable $^{12}$CO, $^{13}$CO, and \xxco emission, respectively, as described in \S\ref{Section3.1}.}
\label{Fig:tracer}
\end{figure}

Figure~\ref{Fig:tracer} presents longitude-velocity~($l$-$v$) and longitude-latitude~($l$-$b$) maps of all
three CO isotopologue emission lines in the CMa OB1 complex, which were obtained by integrating the emission over Galactic latitude $b$ = $-2\fdg5$ to $0\fdg5$, and local standard of rest (LSR) velocity $V_{\rm LSR}$ = 0 to 30~\kms, respectively. To reduce the influence of noise, the $^{12}$CO, $^{13}$CO, and \xxco data were integrated over
all channels which had at least three contiguous channels above 3$\sigma$ (i.e., $\gtrsim$ 1.5 K for $^{12}$CO, and $\gtrsim$ 0.75 K for \xco and C$^{18}$O).
Similar to \cite{2020ApJS..246....7S}, we define mask ``B'' as the region where \co emission existed,
mask ``G'' as the region where \co and \xco emission were detected, and mask ``R'' as the region where
$^{12}$CO, $^{13}$CO, and \xxco emission were simultaneously present.

The CMa OB1 complex is concentrated within the $V_{\rm LSR}$ range of 7 to 25~\kms,
and is located in the Local arm. The main part of the CMa OB1 complex appears as a single, merged structure, as detected via $^{12}$CO emission, while it shows many fragmented and discontinuous structures via $^{13}$CO emission
Base on the morphology of \xco emission, we roughly divided the CMa OB1 complex into three sub-regions,
as shown in Figure~\ref{Fig:tracer} (i.e., sub-regions A, B, and C). Each sub-region contains an extended structure with an angular area larger than 1500 arcmin$^{\rm 2}$, which corresponds to the three most massive \xco clouds
3 (sub-region A), 4 (sub-region B), and 12 (sub-region C) of \cite{2004PASJ...56..313K}, respectively.
Coincidently, each sub-region resides one known water maser~\citep{1994A&AS..103..541B},
which are indicated as crosses in Figure~\ref{Fig:tracer}. The two known connected H{\sc ii} regions
$S$292 and $S$296 are also shown in the Figure~\ref{Fig:tracer}, which are marked as circles~\citep{1959ApJS....4..257S}.

\begin{figure}[!t]
\centering
\includegraphics[width=0.28\textwidth]{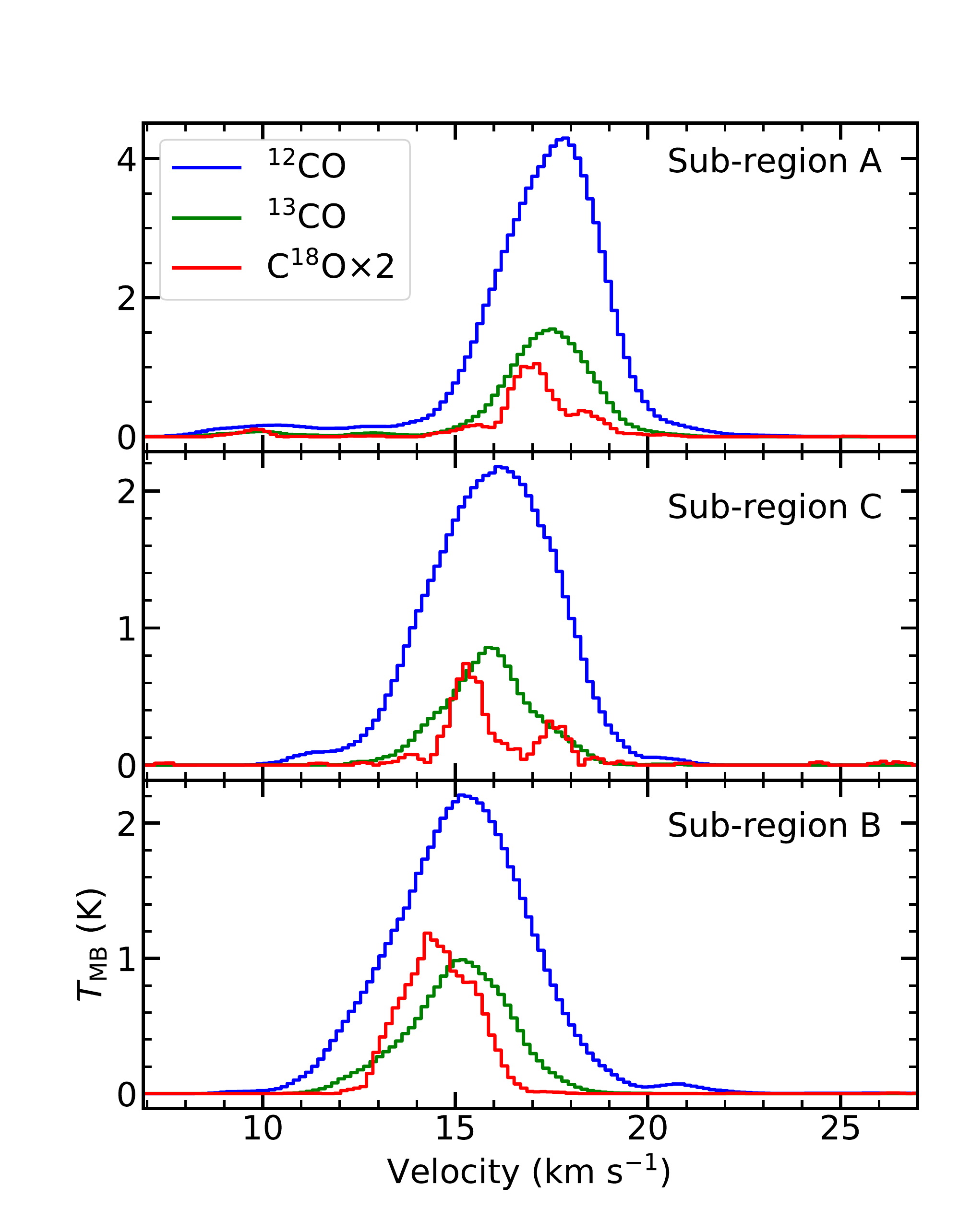}
\caption{Averaged spectra of \co (blue), \xco (green), and \xxco (red) for each region.
Note that only those spectra with at least three contiguous channels $\gtrsim$ 3$\sigma$ are averaged.}\label{Fig:specturm}
\end{figure}

The averaged spectra of \co (blue), \xco (green), and \xxco (red) for each sub-region are shown in Figure~\ref{Fig:specturm}.
Obviously, the total integrated intensity, LSR velocity, and line width of each sub-region for each isotopologue is different; these values
are summarized in Table~\ref{table:cloud1}.
The LSR velocity of sub-region A is apparently different from the other two sub-regions,
which may be due to the feedback from H{\sc ii} region.
The \xxco line profile of sub-region C seems to exhibit two velocity components~(one at $\sim$15.5~\kms, and another at $\sim$17.5~\kms).
The full-width half-maximum (FWHM) line width is the largest in sub-region B, while it is the smallest in sub-region A. Generally, the results revealed from our new \xco data are consistent with those of \cite{2004PASJ...56..313K}.

\begin{figure}[!ht]
\centering
\includegraphics[height=0.2\textheight]{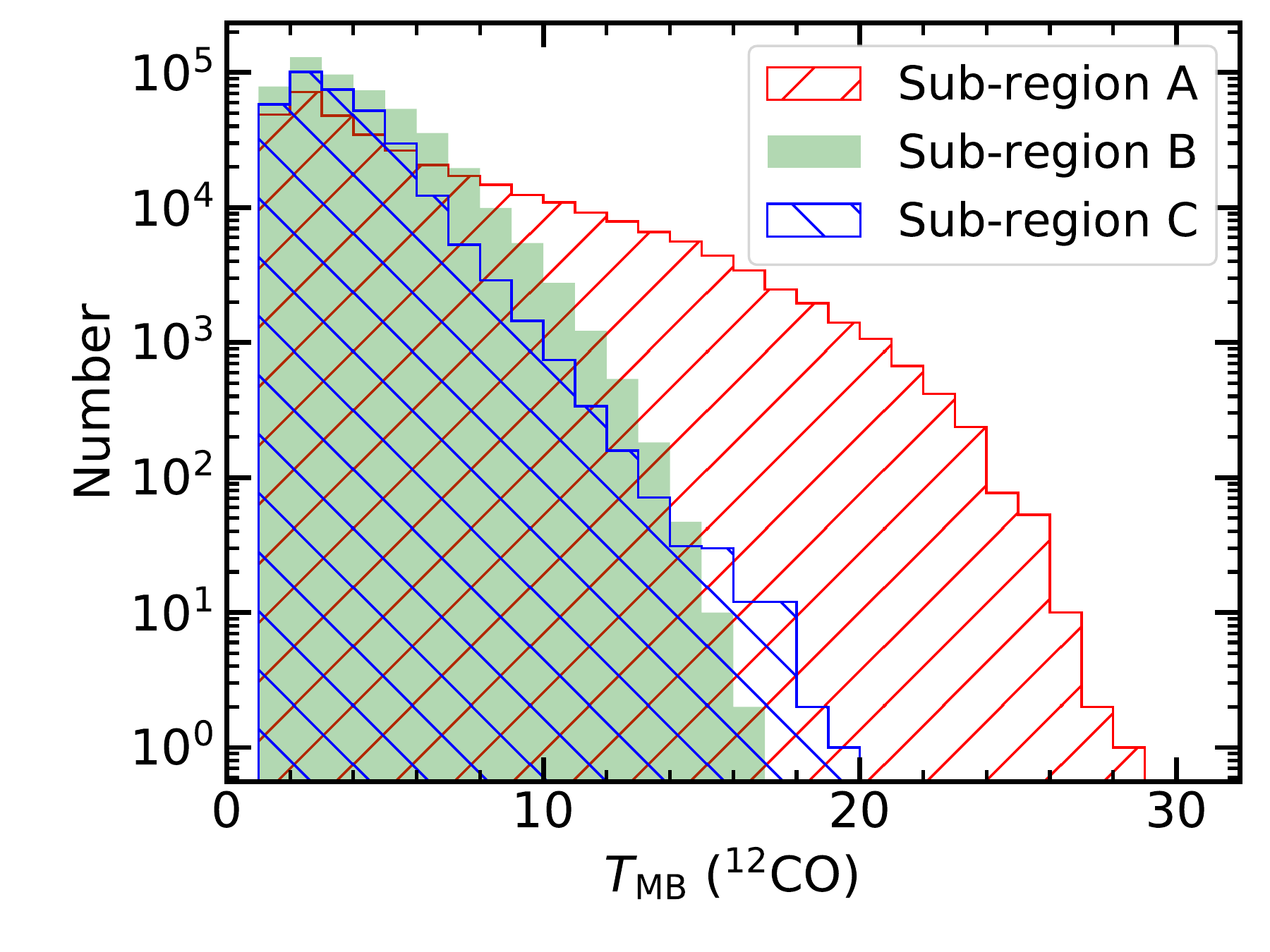}
\includegraphics[height=0.2\textheight]{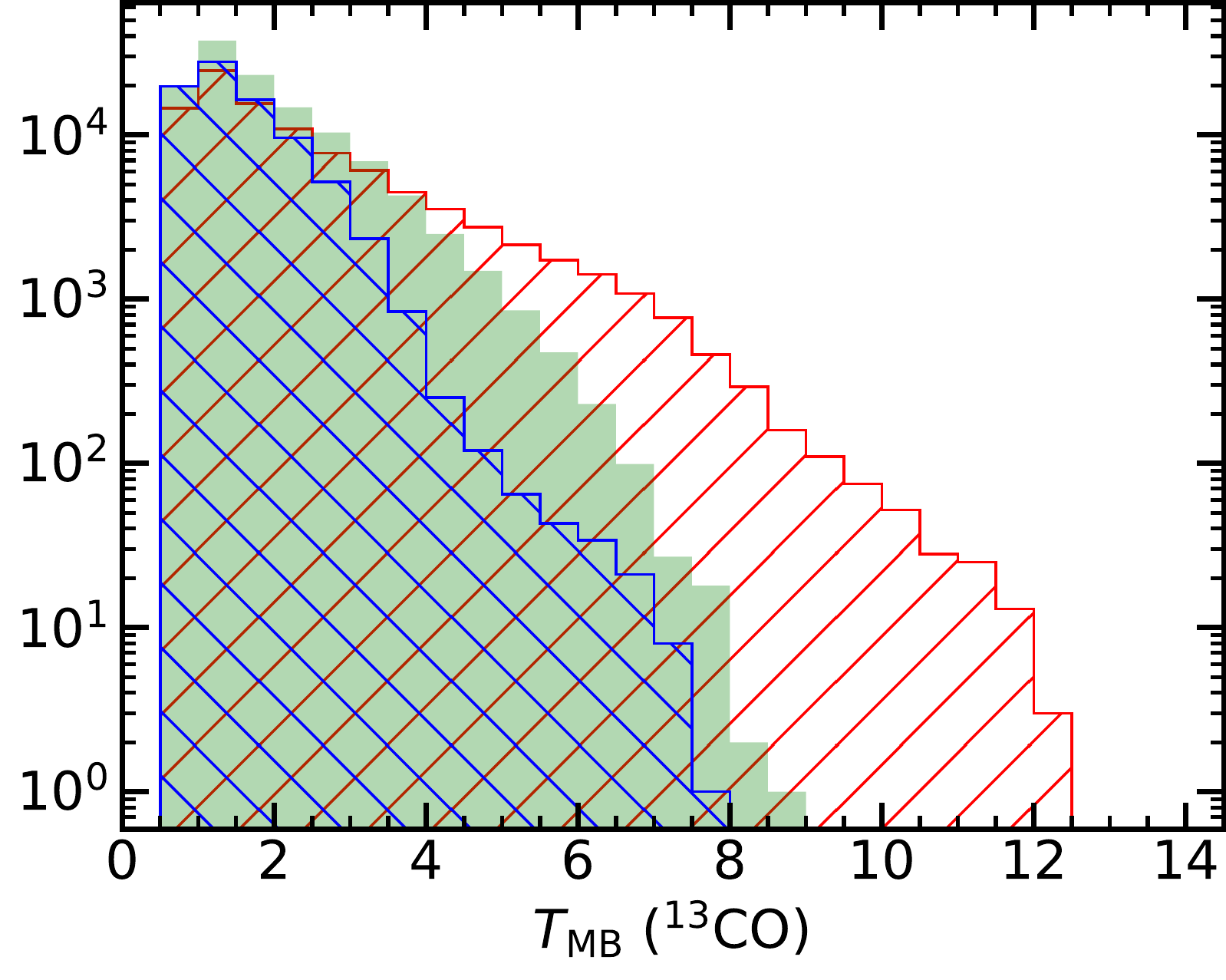}
\includegraphics[height=0.2\textheight]{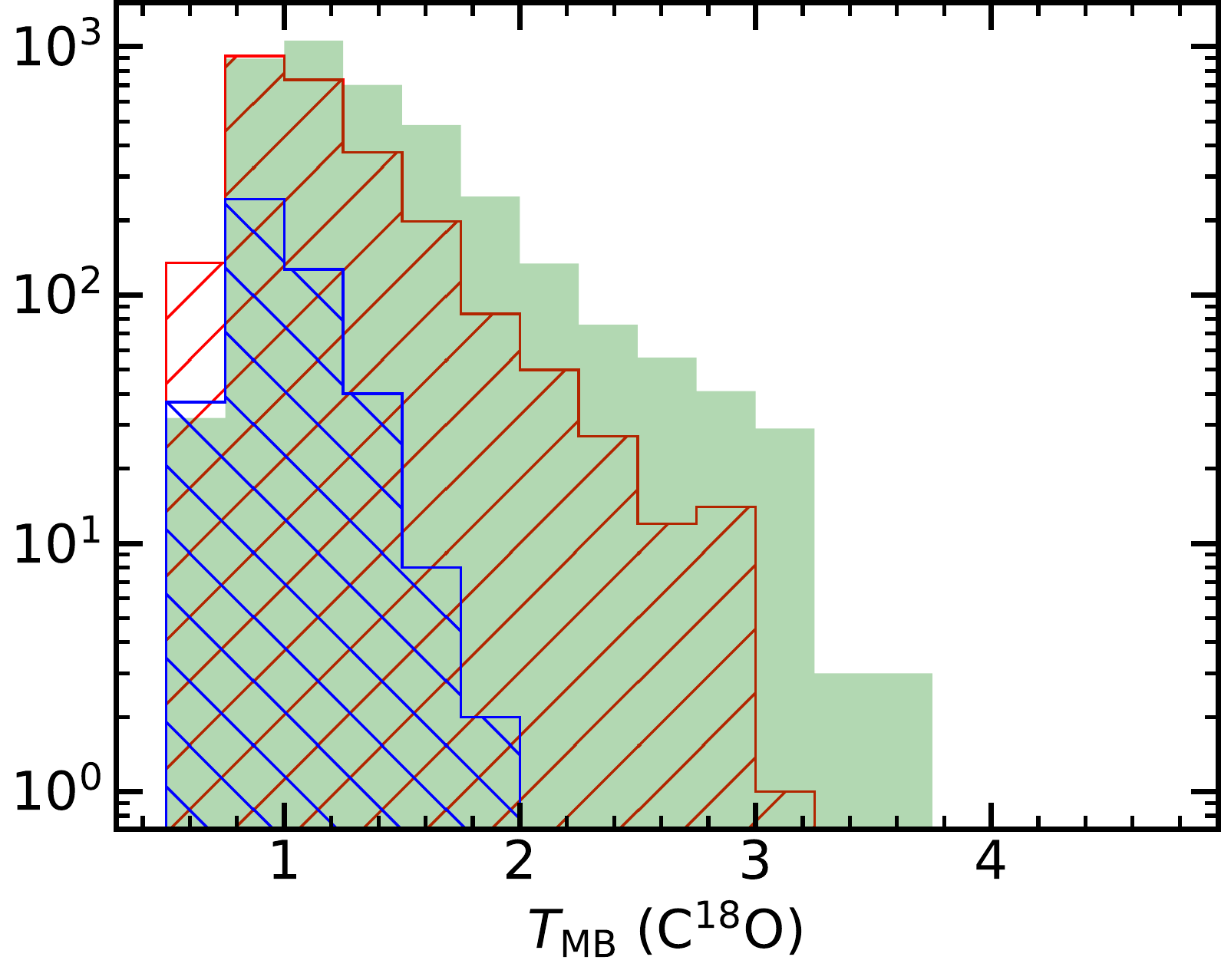}
\caption{Histograms of the main-beam temperatures of the $^{12}$CO, $^{13}$CO, and C$^{18}$O emission for each region.
Note that only voxels with at least three continuous channels above 3$\sigma$ are plotted.}
\label{Fig:Tmbdistribution}
\end{figure}

Figure~\ref{Fig:Tmbdistribution} presents the distributions of $T_\mathrm{MB}$ of the three sub-regions for each isotopologue.
Note that only those voxels with at least three continuous channels above 3$\sigma$ are plotted.
We can see that sub-region A dominates the high value wing of the $T_{\rm MB}$($^{12}$CO) and $T_{\rm MB}$($^{13}$CO),
yet shows weaker \xxco emission than sub-region B.
Instead, sub-region B exhibits the strongest \xxco emission, and the weakest \co emission.
All these results could be interpreted as the different evolutionary stages of star formation for sub-regions A, B, and C,
which will be discussed in detail in \S\ref{Section3.3}.

\begin{figure}[!ht]
\centering
\includegraphics[width=0.88\textwidth]{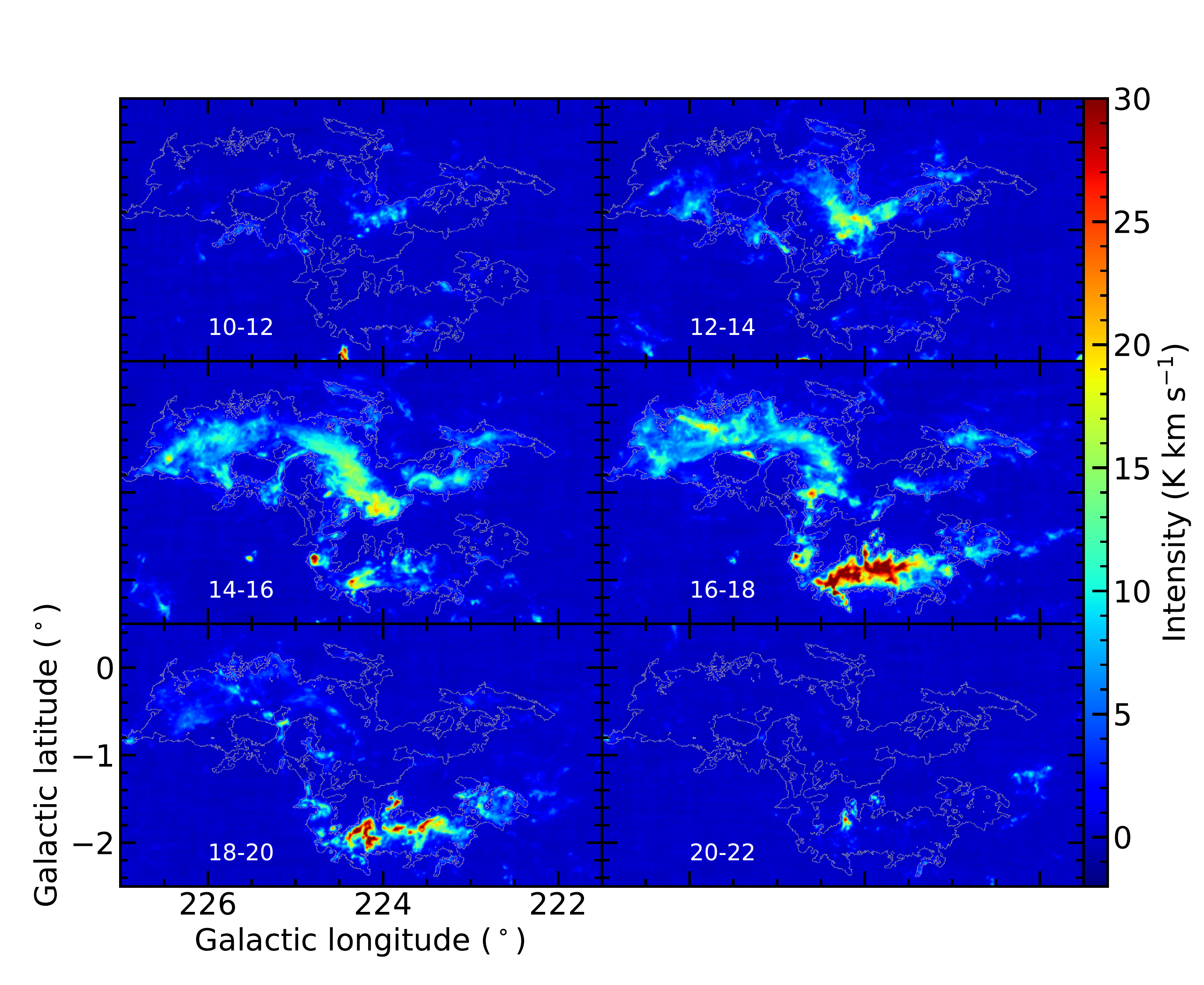}
\caption{Velocity channel maps of \co emission in the CMa OB1 complex. The gray contour indicates the boundary of
the largest \co Mc, which was defined as an independent consecutive structure in $l$--$b$ space in Figure~\ref{Fig:tracer}(b).
Each map was integrated over a velocity interval of 2~\kms, and the velocity ranges are marked in each panel.}
\label{Fig:cloudchannalmap12}
\end{figure}

\begin{figure}[!ht]
\centering
\includegraphics[width=0.87\textwidth]{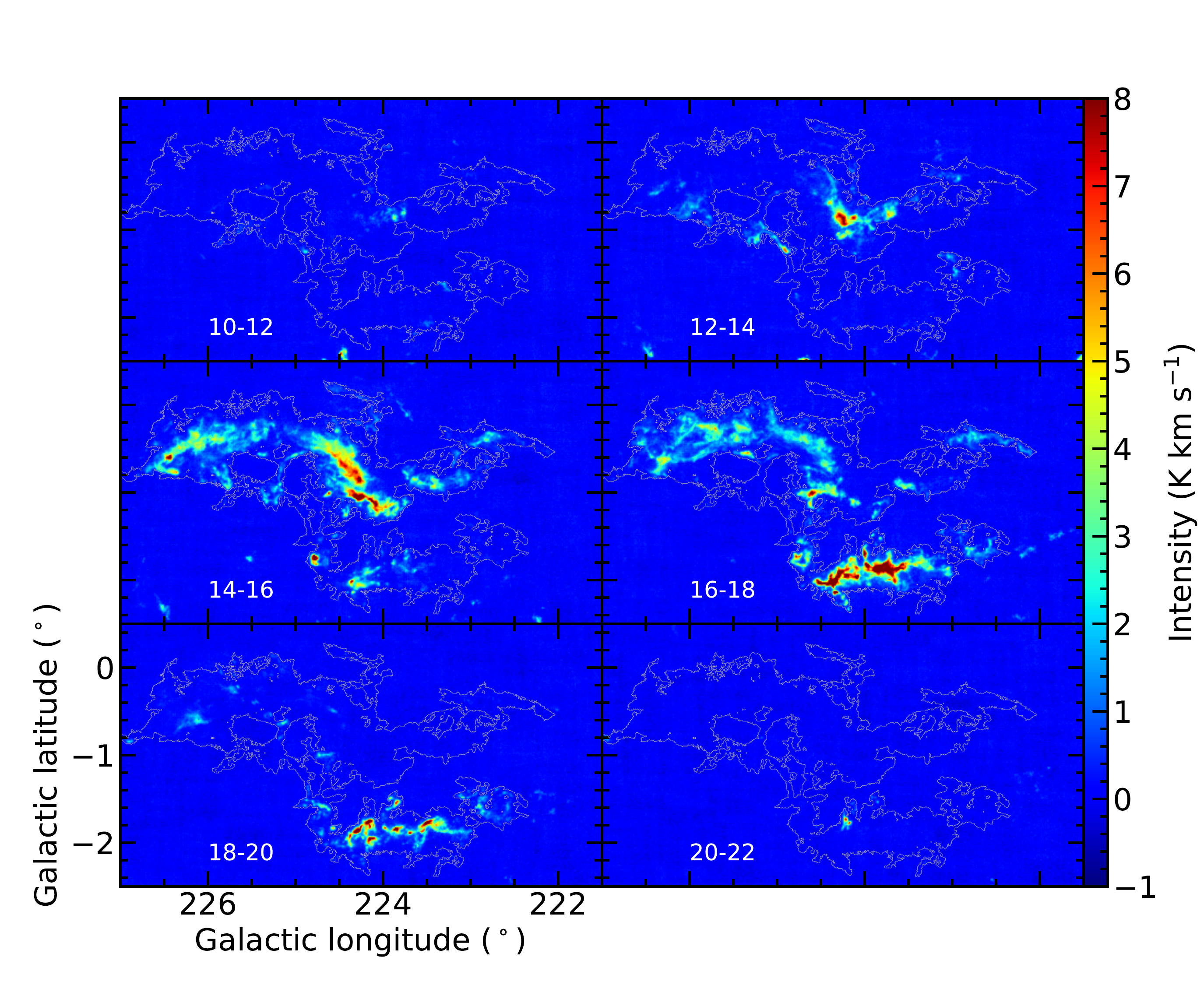}
\caption{
Same as Fig.~\ref{Fig:cloudchannalmap12}, but for \xco emission.}
\label{Fig:cloudchannalmap13}
\end{figure}

\begin{figure}[!ht]
\centering
\includegraphics[width=0.88\textwidth]{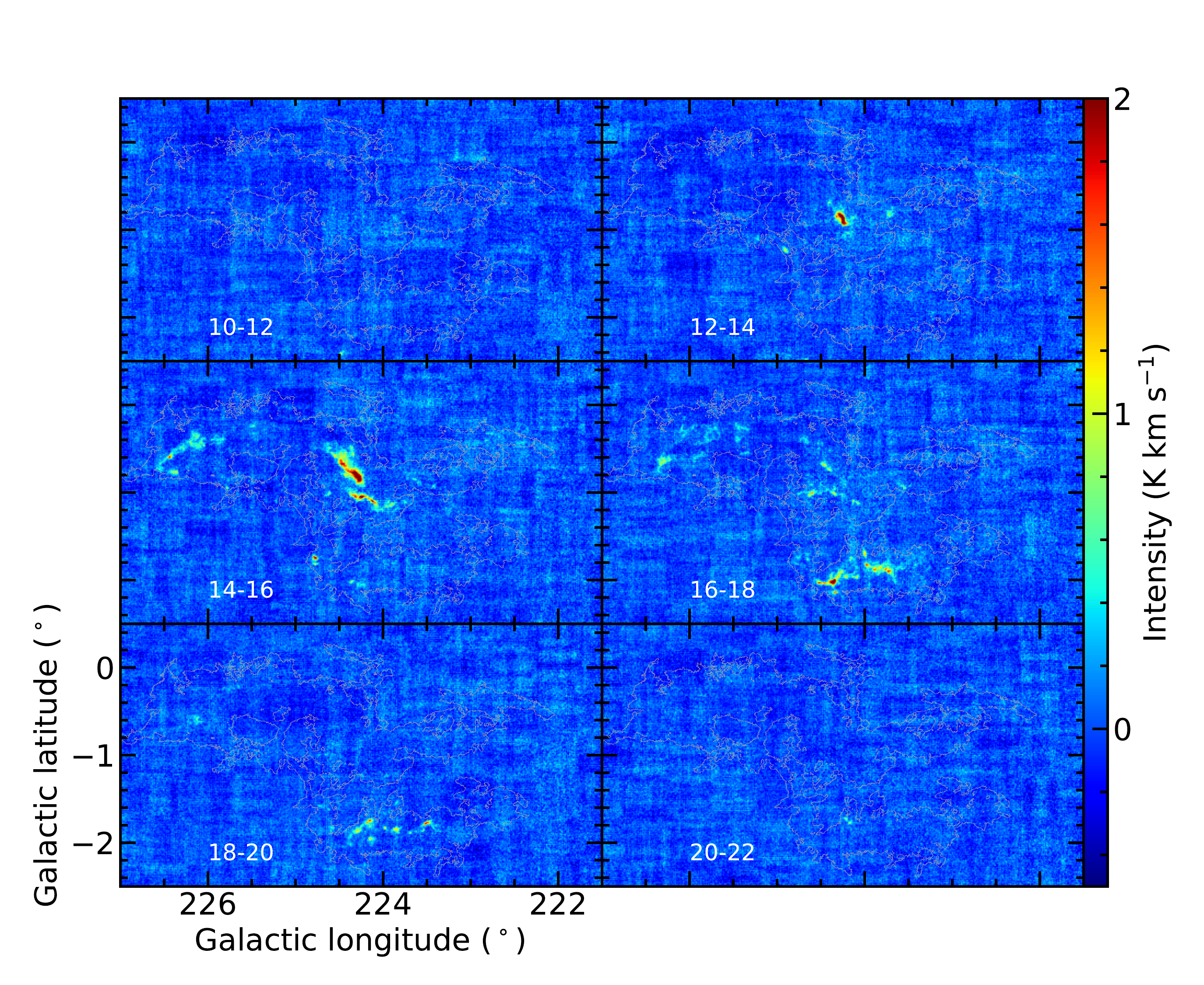}
\caption{
Same as Fig.~\ref{Fig:cloudchannalmap12}, but for \xxco emission.}
\label{Fig:cloudchannalmap18}
\end{figure}

Figures~\ref{Fig:cloudchannalmap12}--\ref{Fig:cloudchannalmap18} present the channel maps
of $^{12}$CO, $^{13}$CO, and \xxco lines, respectively. Interestingly, the molecular gas emission is characterized by filamentary
structures, especially in the denser gas traced by \xco and \xxco emission. The filamentary structures in the major parts of the mapped region have been studied by using $Herschel$ data~\citep{2019ApJS..240...26S}.
These molecular line data with additional velocity information, might be helpful for identifying the velocity-coherent filaments, but this theme seems beyond the scope of this study.

\subsection{Physical Properties of Molecular Cloud\label{Section3.2}}
The distance to the CMa OB1 complex has been measured by various methods.
Using photometric observations of OB stars in the CMa OB1 complex, \cite{1974A&A....37..229C}
determined a distance of 1150$\pm$140 pc for the association. Later, a physical connection between
the OB association and the MCs in the CMa OB1 was confirmed~\citep[e.g.,][]{1980ApJ...242..121M, 1980gmcg.work..211B}.
NANTEN CO(1-0) observations were used to derive the kinematic distances of ``\co clumps'' in the CMa OB1 complex,
which were constrained to a range of 800 to 1000 pc~\citep{2013ApJ...772...45E}.
Considering the much higher accuracy of the photometric method, we adopt a value of 1150~pc as the distance to the CMa OB1 complex.

Using the equations $I(\mathrm{CO})_{i,j}\ =\ \int{T_\mathrm{MB}dv}$ and
$I(\mathrm{CO})\ =\ a^2 \sum\limits_{i,j}{I(\mathrm{CO})_{i,j}}$, where $a$ (equal to 30$\arcsec$)
is the angular size of each pixel, we estimate the total integrated intensities and areas of the emission of all three CO isotopologues (see Table \ref{table:cloud1}).
The total integrated intensities of $^{12}$CO, $^{13}$CO, and \xxco emission were found to be $2.1\times10^5$, $2.5\times10^4$,
and $3.5\times10^2$ K\,km\,s$^{-1}$\,arcmin$^2$, within the three mask areas (i.e., mask ``B'', ``G'', ``R'') in Figure~\ref{Fig:tracer}(b).

In this study, the properties of the molecular gas were estimated under the assumption
that all the molecular gas is in local thermodynamic equilibrium (LTE).
We assume that \co emission is optically thick. Based on these, the excitation temperature ($T_\mathrm{ex}$)
can be estimated by \citep{1998AJ....116..336N},
\begin{equation}\label{Equ:tex}
 T_\mathrm{ex}=\frac{5.53}{\ln \left(1 + \frac{5.53}{T_\mathrm{MB,12}}\right)},
\end{equation}
where $T_\mathrm{MB,12}$ is the peak main beam temperature of $^{12}$CO.
Assuming that different isotopologues have the same $T_\mathrm{ex}$, the optical depths of $^{13}$CO ($\tau_{13}$),
and C$^{18}$O ($\tau_{18}$) were estimated as \citep{1998ApJS..117..387K,2010ApJ...721..686P}:

\begin{gather}
 \tau_{13} = -\ln\left\{1-\frac{T_\mathrm{MB,13}}{5.29}[(e^{5.29/T_\mathrm{ex}}-1)^{-1}-0.164]^{-1}\right\} \label{Equ:tau1},\\
 \tau_{18} = -\ln\left\{1-\frac{T_\mathrm{MB,18}}{5.27}[(e ^{5.27/T_\mathrm{ex}}-1)^{-1}-0.166]^{-1}\right\} \label{Equ:tau2},
\end{gather}

where $T_\mathrm{MB,13}$ and $T_\mathrm{MB,18}$ are the peak main beam temperatures of \xco and C$^{18}$O, respectively.
Considering the correction factor for an optical depth of $f=\frac{\tau}{1-e^{-\tau}}$ \citep{2015ApJS..219...20L},
the total integrated intensities of \xco and \xxco after optical depth correction can be estimated as:

\begin{gather}
  I_{13}=\frac{\tau_{13}}{1-e^{-\tau_{13}}}\times\frac{1+0.88/T_\mathrm{ex}}{1-e^{-5.29/T_\mathrm{ex}}}\int{T_\mathrm{MB,13}dv} \label{Equ:odc1},\\
  I_{18}=\frac{\tau_{18}}{1-e^{-\tau_{18}}}\times\frac{1+0.88/T_\mathrm{ex}}{1-e^{-5.27/T_\mathrm{ex}}}\int{T_\mathrm{MB,18}dv} \label{Equ:odc2}.
\end{gather}

The column densities of \xco and \xxco were estimated as $N_{13}=2.42\times10^{14} I_{13}$, and $N_{18}=2.54\times10^{14}I_{18}$~\citep{1991ApJ...374..540G,1997ApJ...476..781B}.
The isotopic ratios of [$^{12}$C/$^{13}$C] = 77,
[$^{16}$O/$^{18}$O] = 560~\citep{1994ARA&A..32..191W}, and [H$_2$/\co] abundance ratio of 1.1$\times10^{5}$~\citep{1982ApJ...262..590F} were used to derive H$_2$ column density.

The H$_2$ column densities traced by \co can be estimated by adopting a CO-to-H$_2$ conversion
factor ($X_\mathrm{CO}$ $\equiv$ $N_{\rm H_2}$/$I_\mathrm{CO}$).
Using \co and \xco data, \cite{2015ApJ...812....6B,2018ApJ...866...19B} first examined the relation between the $I_\mathrm{CO}$ and $X_\mathrm{CO}$ = $N_{\rm H_2}$/$I_\mathrm{CO}$ (where $N_{\rm H_2}$ is derived from $^{13}$CO). They calculated the $X_\mathrm{CO}$ factor in both 3D data cubes and 2D images from all channels and pixels where both \co and \xco are detected. Using MWISP \co and \xco data, \cite{2020ApJS..246....7S} derived $X_\mathrm{CO}$ for a $\sim$110 deg$^{\rm 2}$ region in the Galactic coordinate range of $l$=[129\fdg75, 140\fdg25] and $b$=[$-$5\fdg25, +5\fdg25] (hereafter the G130 region). All of these studies confirmed that $X_\mathrm{CO}$ varies from region to region. Therefore, following their analytical approach, we derived the $X_\mathrm{CO}$ for all locations that had both \co and \xco detections (mask `G' region), then applied $X_\mathrm{CO}$ to locations that had \co detections (mask `B' region) to the calculated column densities.

\begin{figure}[!ht]
  \centering
  \includegraphics[height=0.28\textheight]{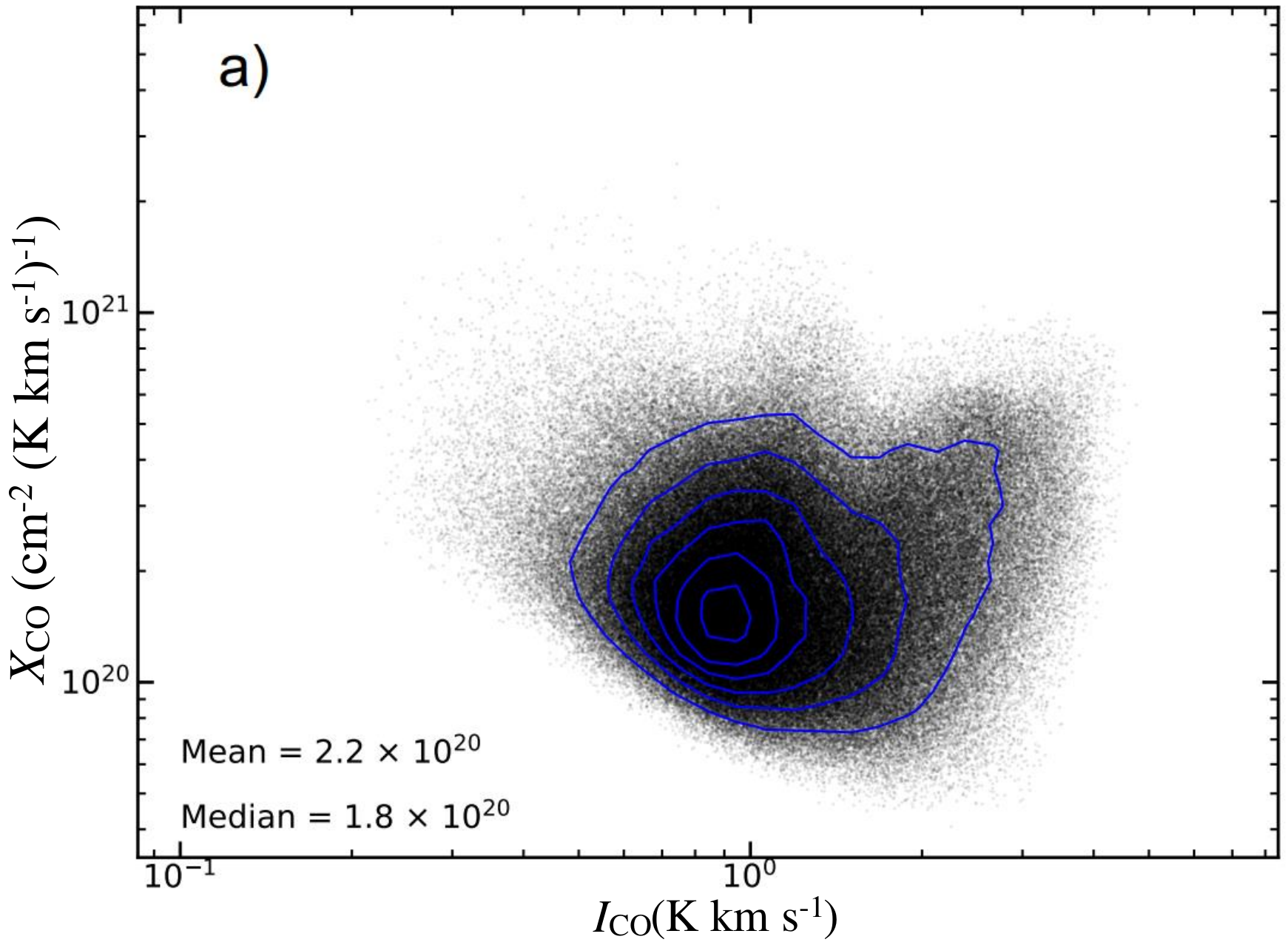}
  \includegraphics[height=0.28\textheight]{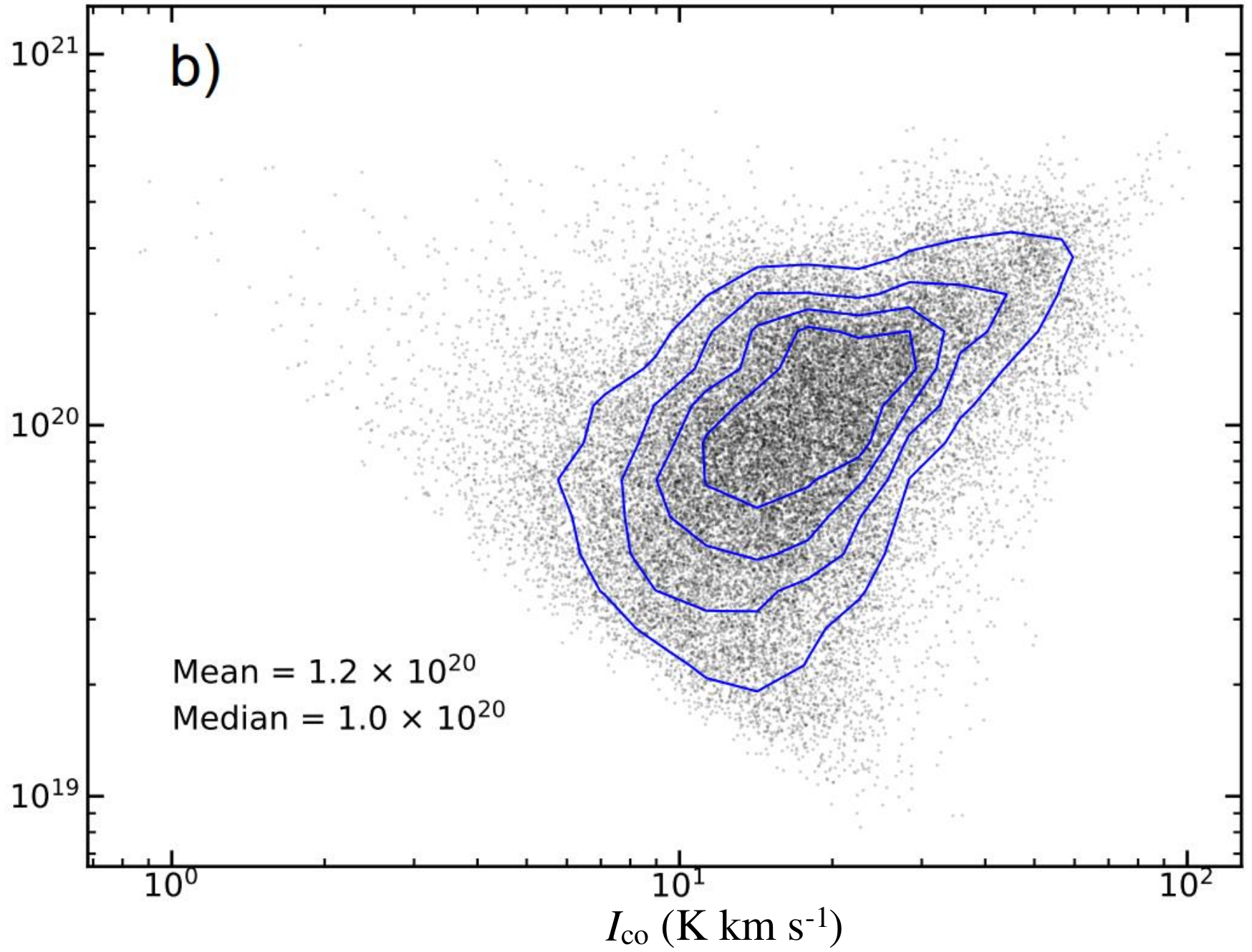}
  \caption{Relations between $N_\mathrm{H_2}$/$I_{CO}$ and $I_\mathrm{CO}$. ($a$) Each black point represents
  a voxel with a channel width of 0.168~\kms which showed both \xco and \co detections.
  The contours show the surface density of points. ($b$) Each black point is integrated over all
  velocity channels with detectable \co emission, regardless of whether \xco was detected.
}\label{Fig:X}
\end{figure}

Similar to figure 5 of \cite{2018ApJ...866...19B}, in Figure~\ref{Fig:X}(a), we have presented the results of our analysis of $X_\mathrm{CO}$ (i, j, v) for each channel (or voxel) and pixel (i.e., this calculation was conducted in 3D). Each black point represents a channel that showed both \xco and \co detections. $N_{\rm H_2}$ and $I_\mathrm{CO}$ (i.e., $I_\mathrm{CO}$ = $T_\mathrm{MB}\times$0.168~\kms) were both calculated here per channel. Similar to figure 9 of \cite{2018ApJ...866...19B}, in Figure~\ref{Fig:X}(b), have presented our results of our analysis of $N_{\rm H_2}$(i, j) for each pixel. Each quantity was integrated across all velocity channels with detectable \co emission, regardless of whether \xco emission is detected.

The $X_\mathrm{CO}$ and $I_\mathrm{CO}$ relationships (shown in both panels of Figure~\ref{Fig:X}) confirm that $X_\mathrm{CO}$ decreases as $I_\mathrm{CO}$ increases for faint $I_\mathrm{CO}$ (i.e. it has a negative slope), but $X_\mathrm{CO}$ increases as $I_\mathrm{CO}$ increases for brighter $I_\mathrm{CO}$ (i.e. a positive slope is seen). A
caveat of this result is that the power-law fit to our data is unreliable due to the relatively small range that the $I_\mathrm{CO}$ data covers (i.e. $\sim$ an order of magnitude) and the large scatter of the
data. Consequently, only the mean and median values of $X_\mathrm{CO}$ are indicated in each panel.

As discussed by \cite{2018ApJ...866...19B}, the relation in Figure~\ref{Fig:X}(b) represents the practical conversion laws, since what is used is the total unmasked $I_\mathrm{CO}$ data across all velocity channels with detectable \co emission. Therefore, the derived $X_\mathrm{CO}$ factors derived from the voxels containing both \co and \xco emission in Figure~\ref{Fig:X}(a) represent upper limits. However, because a large velocity coverage integrated, any non-relevant components will enhance $I_\mathrm{CO}$ and lower $X_\mathrm{CO}$; thus, the derived $X_\mathrm{CO}$ conversion factors in the right panel represents lower limits.

As suggested by \cite{2020ApJS..246....7S}, the practical X$_\mathrm{CO}$ conversion factor
might fall in the range between the values in Figure~\ref{Fig:X}(a--b).
We have adopted a mean value of $X_\mathrm{CO}$ = 1.7 $\times$ 10$^{20}$ cm$^{\rm -2}$\,(K\,km\,s$^{-1}$)$^{-1}$ in this work. It should be noted that the derived value of $X_\mathrm{CO}$ largely depends on the uncertainties
of the isotopic ratio of [$^{12}$C/$^{13}$C] and abundance ratio of [H$_2$]/[$^{12}$CO]. Nevertheless, our value
is still very similar to the value of 1.8 $\times$ 10$^{20}$ cm$^{\rm -2}$\,(K\,km\,s$^{-1}$)$^{-1}$
estimated by \cite{2001ApJ...547..792D} for the local gas.

Knowing the $N_{\rm H_2}$ value in each pixel, we then calculated the mass traced by each isotopologue
by integrating the column density over the area of each mask region as:

\begin{gather}
    M\ =\ 2\mu_\mathrm{H} m_\mathrm{H} a^2 d^2\sum{N_\mathrm{H_2}} \label{Equ:M18}.
\end{gather}

where $\mu_\mathrm{H}$ \citep[=1.36,][]{1983QJRAS..24..267H} is the mean atomic weight per proton in the interstellar medium (ISM) and $m_\mathrm{H}$ is the mass of proton. The molecular gas surface density, $\Sigma$, is
simply the mass divided by the projected area, $\Sigma = M/A$.

The area ratios traced by $^{12}$CO, $^{13}$CO, and \xxco were also calculated as $\mathbb{A}_{\rm 12}^{\rm 13}$=$A_{\rm ^{13}CO}$/$A_{\rm ^{12}CO}$, $\mathbb{A}_{\rm 12}^{\rm 18}$=$A_{\rm C^{18}O}$/$A_{\rm ^{12}CO}$,
where $A_{\rm ^{12}CO}$, $A_\mathrm{^{13}CO}$, and $A_{\rm C^{18}O}$ are areas
with detectable \co~(mask `B'), \xco~(mask `G'), and \xxco~(mask `R') emission, respectively.
Similarly, we also calculate the mass ratios,
$\mathbb{M}_{\rm 12}^{\rm 13}$=$M_{\rm ^{13}CO}$/$M_{\rm ^{12}CO}$, $\mathbb{M}_{\rm 12}^{\rm 18}$=$M_{\rm C^{18}O}$/$M_{\rm ^{12}CO}$,
where $M_{\rm ^{12}CO}$, $M_{\rm ^{13}CO}$, and $M_{\rm C^{18}O}$ are the total masses contained in each area, respectively.

The derived properties of the molecular gas are all summarized in Tables~\ref{table:cloud2}--\ref{tab:ratio}.
As expected, the excitation temperatures, column densities, and surface gas densities increase as
we move from region `B', i.e., the outskirts of MCs, to the denser regions `G' and `R'. These statistical results are consistent with those revealed by \cite{2020ApJS..246....7S} in G130 region.
In addition, the physical properties seem to vary considerably in different sub-regions.
The H$_2$ column density, surface gas density, $\mathbb{A}_{\rm 12}^{\rm 18}$, and $\mathbb{M}_{\rm 12}^{\rm 18}$ in sub-region C
are much smaller than those in the other two sub-regions. Such significant variations likely reflect the different levels of star-formation activities and/or the different evolutionary stages of the MCs.

\begin{figure}[!ht]
  \centering
  \includegraphics[height=0.18\textheight]{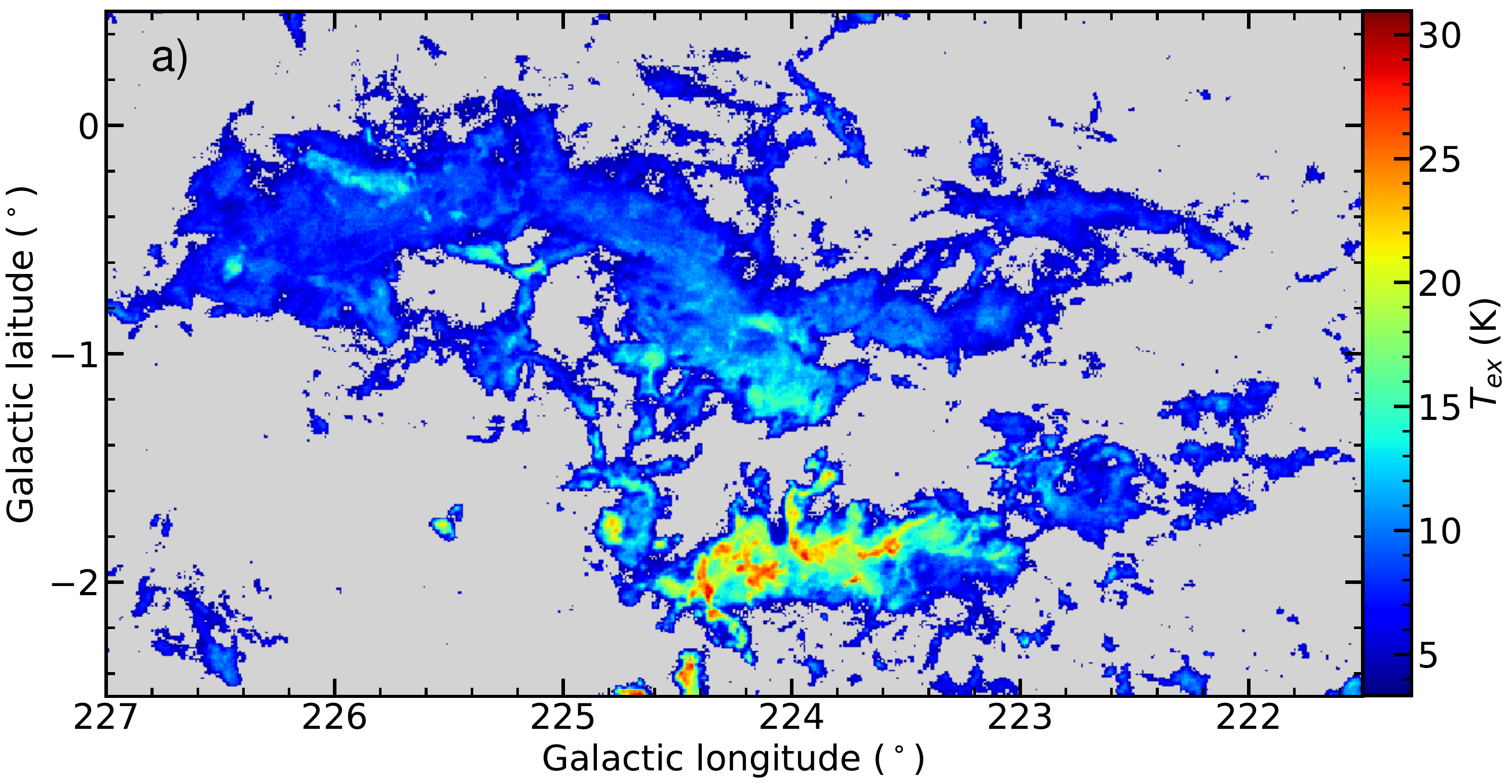}
  \includegraphics[height=0.18\textheight]{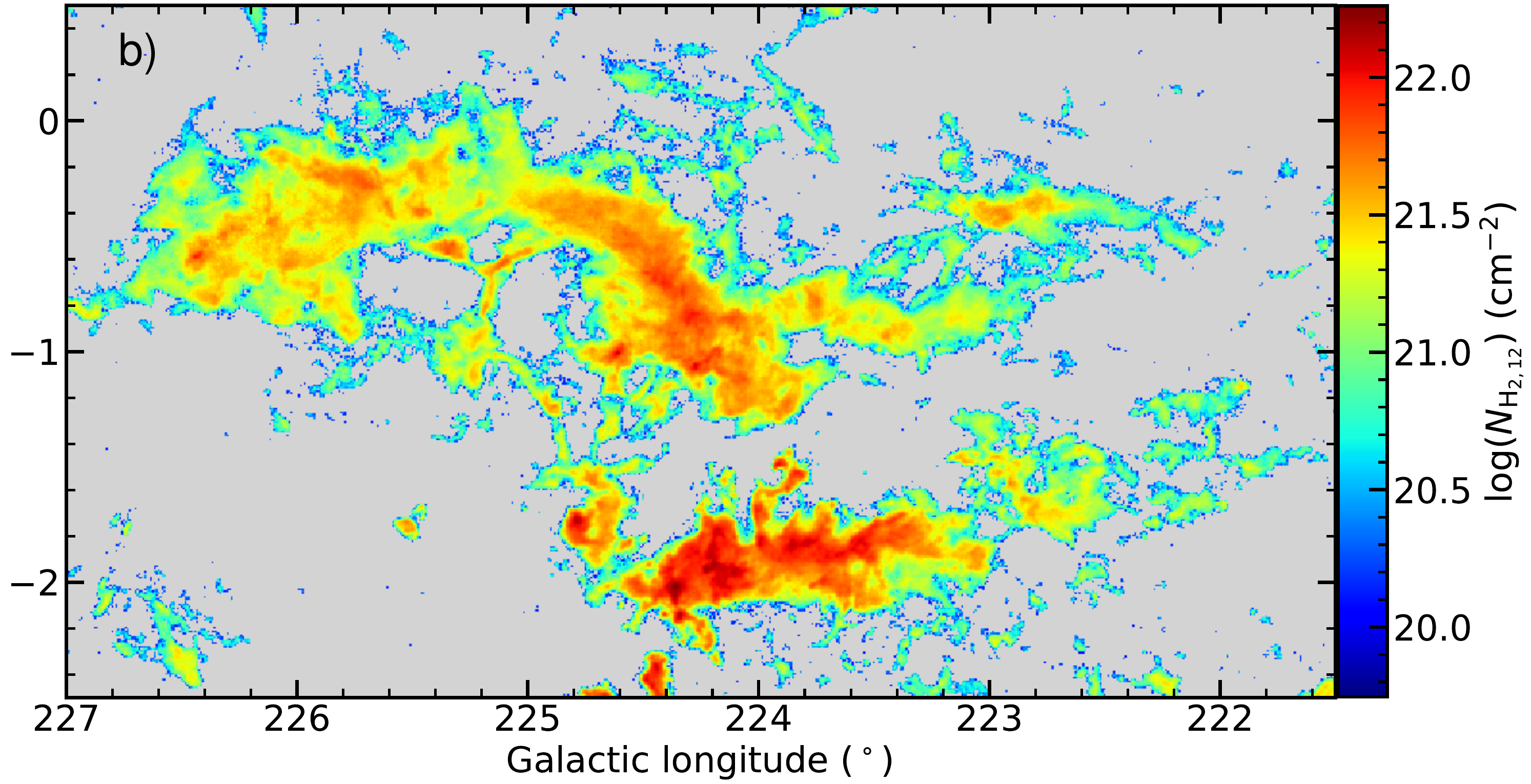}
  \includegraphics[height=0.18\textheight]{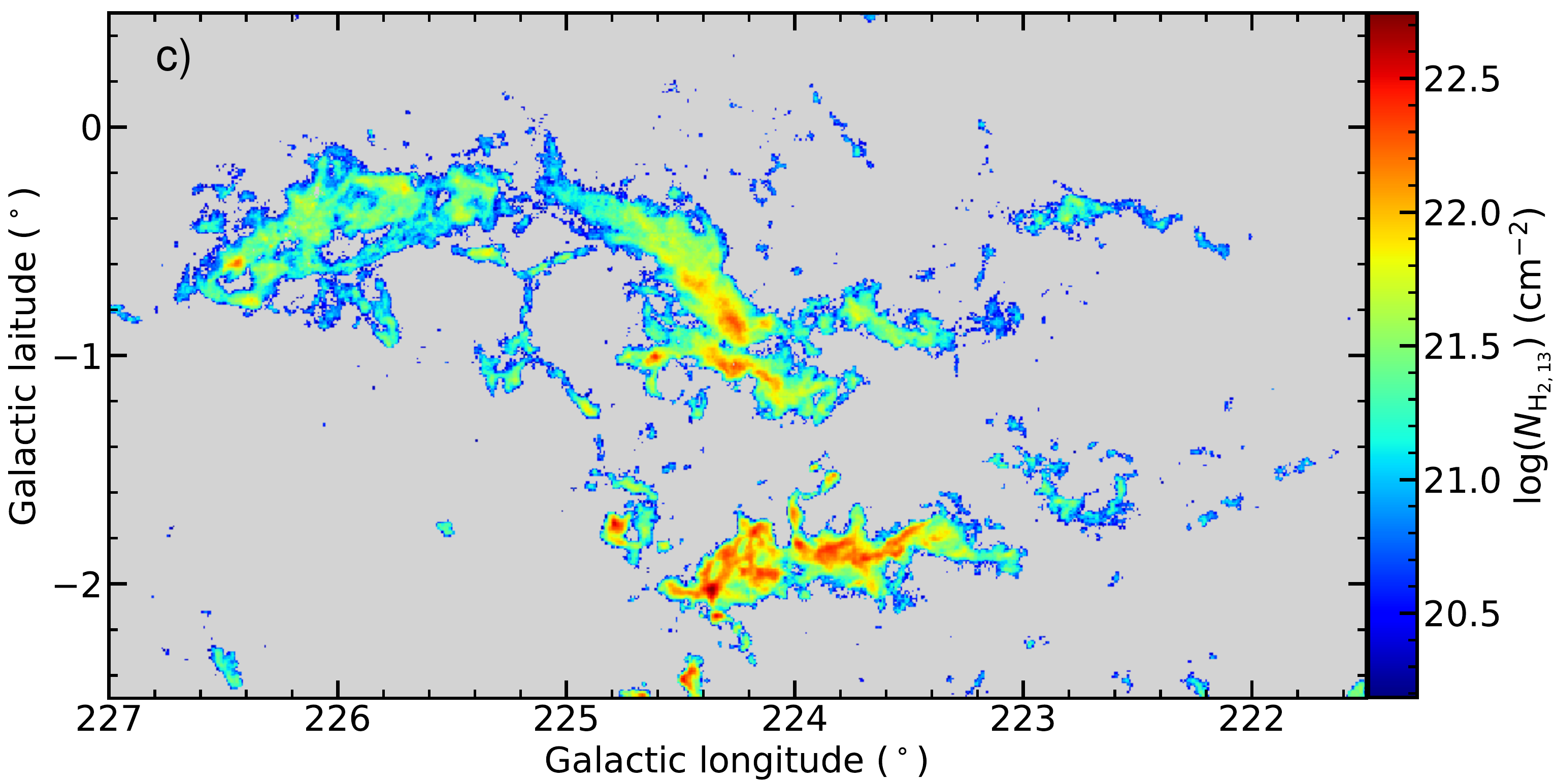}
  \includegraphics[height=0.18\textheight]{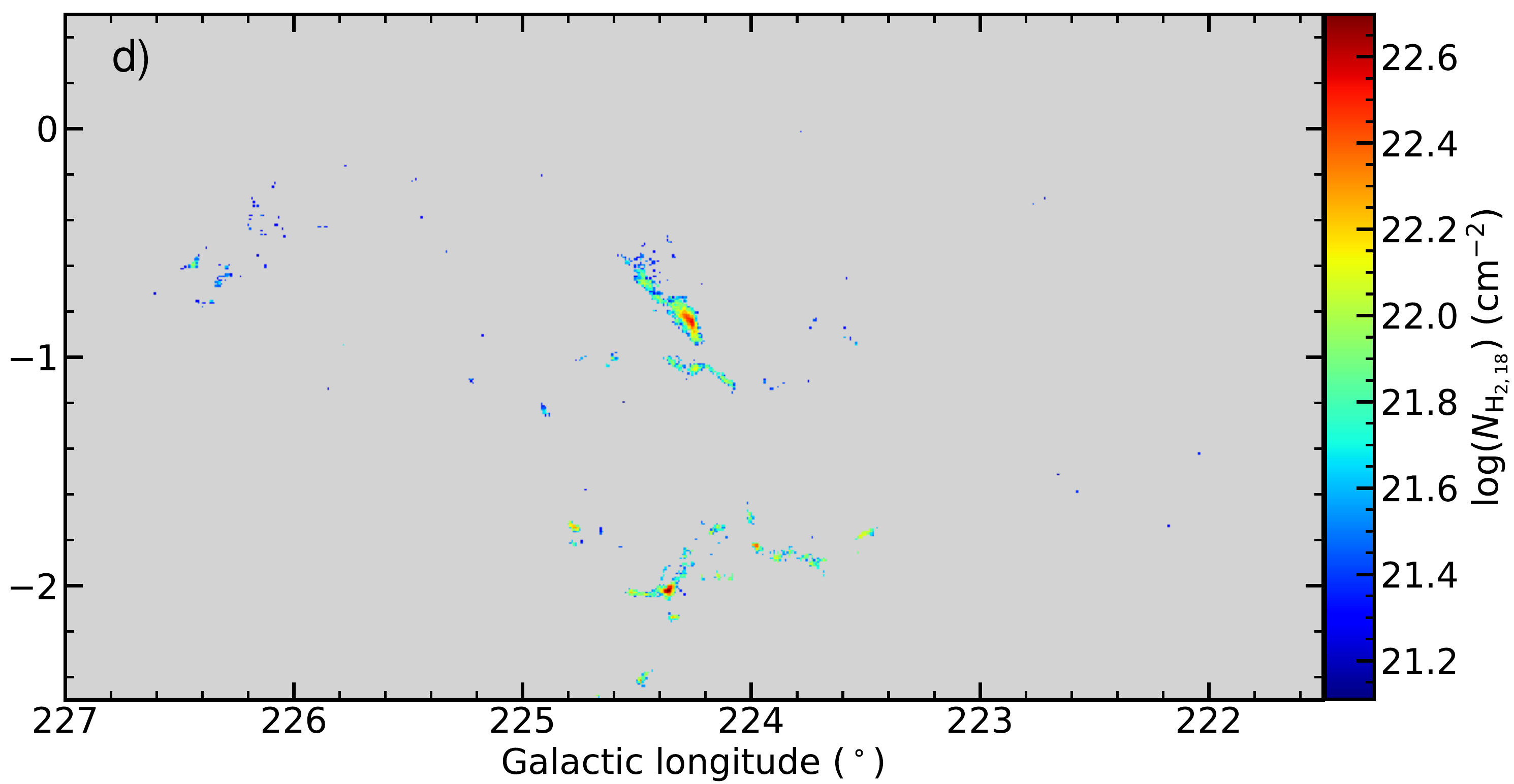}
  \caption{Maps of ($a$) excitation temperature distributions, H$_2$ column density distributions traced by ($b$) $^{12}$CO, ($c$) $^{13}$CO, and ($d$) C$^{18}$O.}\label{Fig:TaN}
\end{figure}

Maps of excitation temperature and H$_2$ column density are also presented in Figure~\ref{Fig:TaN}.
It is noticeable that a large fraction of the area has the $T_{\rm ex}<$10~K. In addition, as the
hottest region, sub-region A still has a mean value of $T_{\rm ex}$ below 10~K.
This suggests that a considerable amount of the observed molecular gas is excited in regions with low excitation conditions, which was also suggested by many other studies~\citep[e.g.,][]{2008ApJ...680..428G,2008ApJ...679..481P}.

\subsection{Abundance Ratio}
\label{Section3.3}
\begin{figure}[!t]
\centering
\includegraphics[height=0.2\textheight]{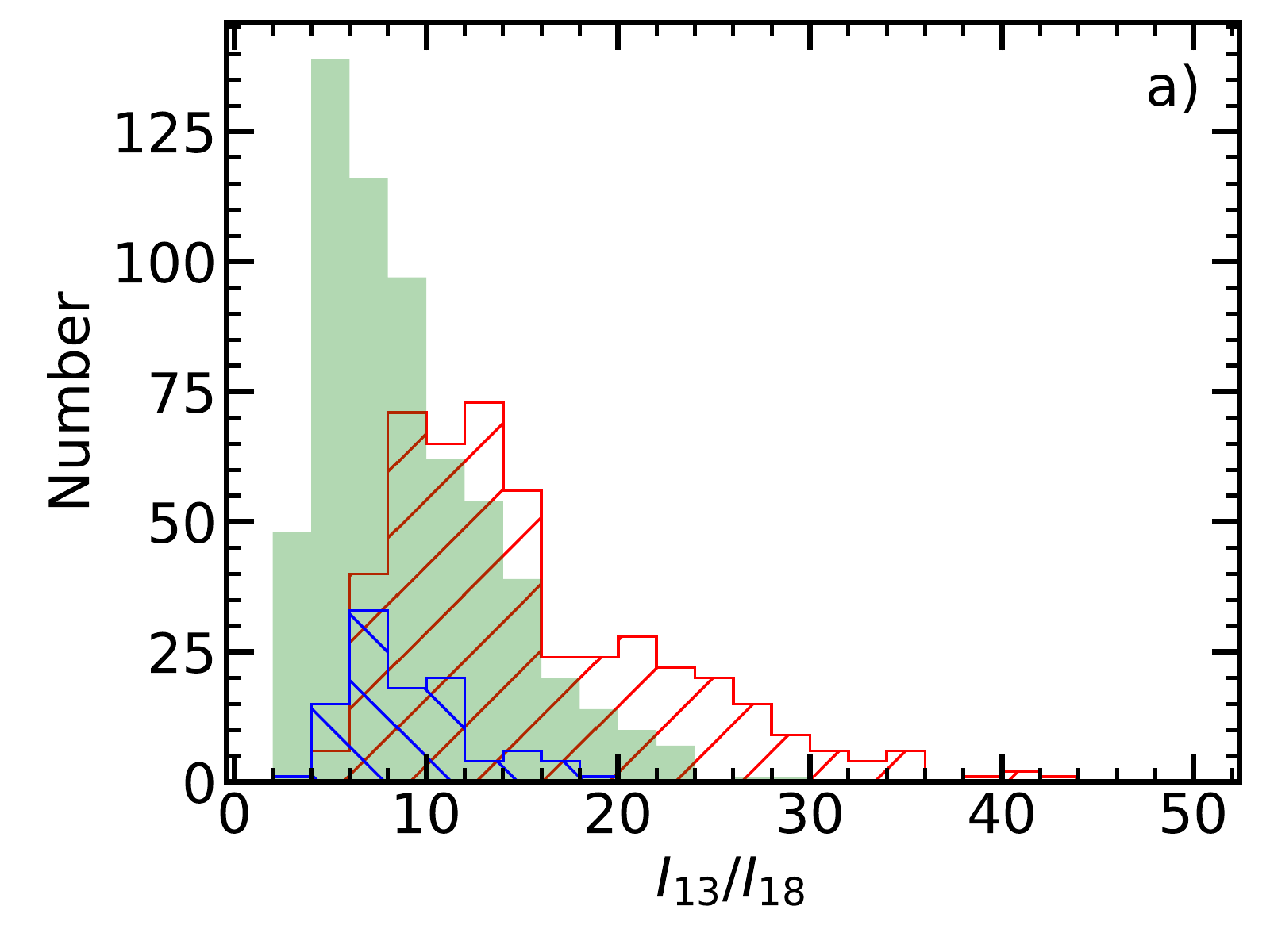}
\includegraphics[height=0.2\textheight]{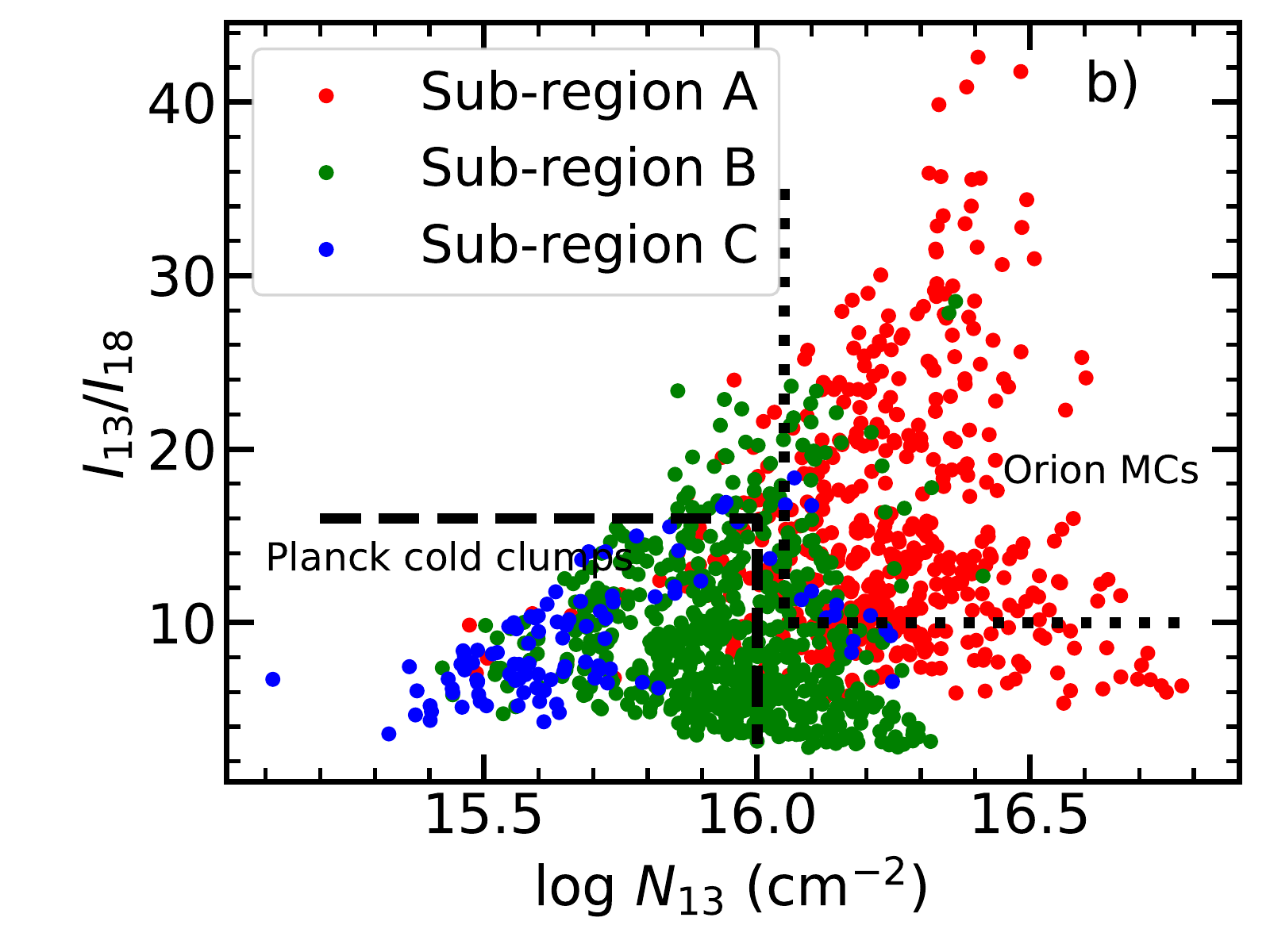}
\hspace{-0.21cm}
\includegraphics[height=0.2\textheight]{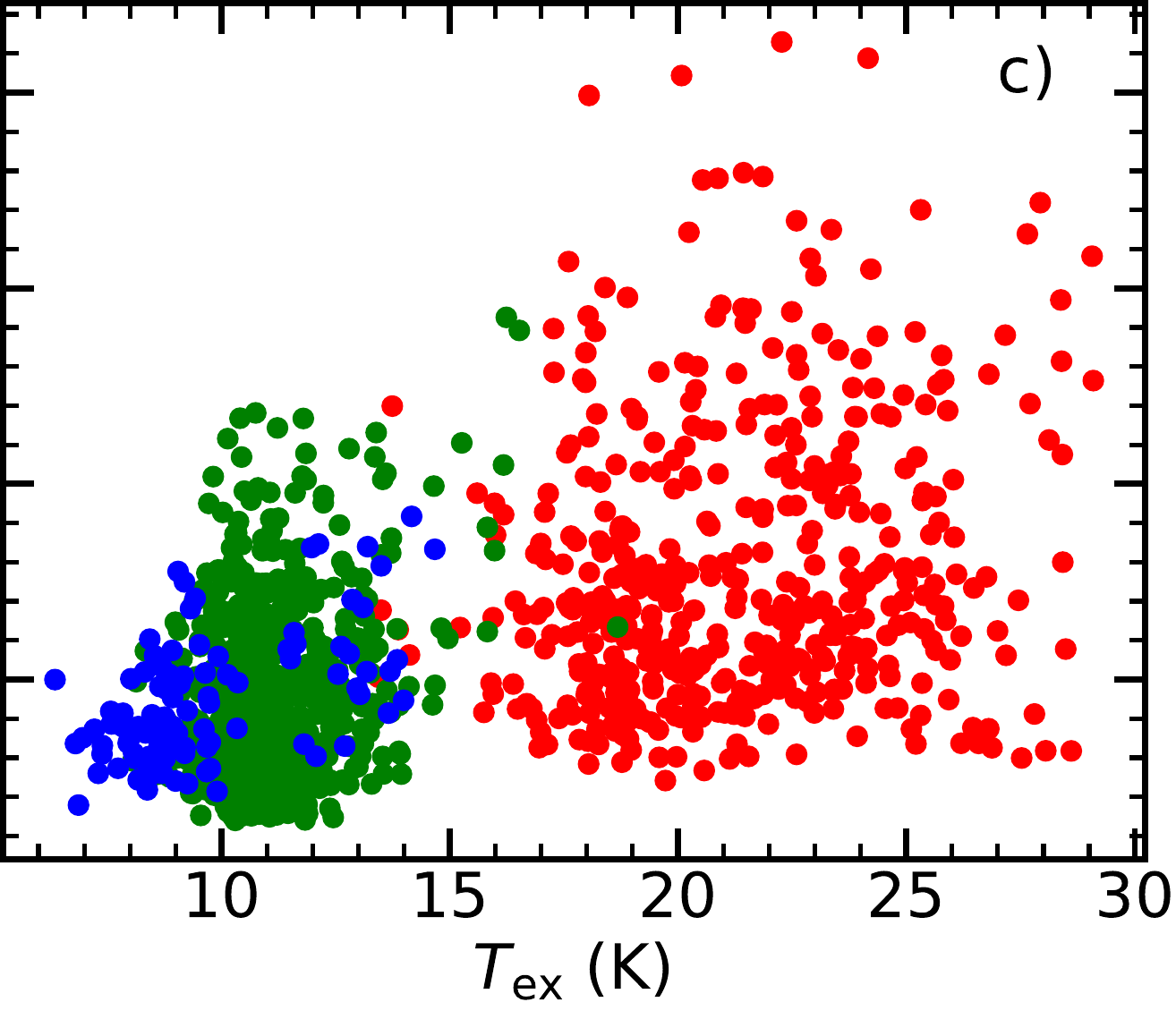}
\caption{($a$) Histograms of abundance ratios $X_{\rm ^{13}CO}$/$X_{\rm C^{18}O}$ of all three sub-regions.
($b$) Relation between $X_{\rm ^{13}CO}$/$X_{\rm C^{18}O}$ and $N_{\rm ^{13}CO}$.
The dashed and dotted lines indicate the approximate boundaries
of type {\sc i} and type {\sc ii} MCs~\citep{2019ApJS..243...25W}, respectively, into which the Planck cold clumps \citep{2012ApJ...756...76W},
and the Orion MCs fall into~\citep{2014A&A...564A..68S}.
($c$) Relation between $X_{\rm ^{13}CO}$/$X_{\rm C^{18}O}$ and $T_{\rm ex}$.
Note that only those pixels with both detectable \xco and \xxco emission (mask ‘R’ region) have been plotted.}
\label{Fig:Icodistribution}
\end{figure}

Following the methodology adopted in previous studies \citep[\eg][]{2014A&A...564A..68S,2019ApJS..243...25W}, in this section we systematically examined in this section the abundance ratio of [$\rm ^{13}CO/C^{18}O$] assuming that the abundance ratio is proportional to the total integrated intensity ratio (with optical depth correction), $X_{\rm ^{13}CO}$/$X_{\rm C^{18}O}\ \propto\ I_{13}/I_{18}$.
The derived abundance ratios $X_{\rm ^{13}CO}$/$X_{\rm C^{18}O}$ of all three sub-regions are plotted as histograms in Figure~\ref{Fig:Icodistribution}(a).
The relationships between $X_{\rm ^{13}CO}$/$X_{\rm C^{18}O}$ and $N_{\rm ^{13}CO}$ are shown in Figure~\ref{Fig:Icodistribution}(b).
In addition, Figure~\ref{Fig:Icodistribution}(c) presents the relationship between $X_{\rm ^{13}CO}$/$X_{\rm C^{18}O}$ and
$T_\mathrm{ex}$.

We find that the abundance ratios of $X_{\rm ^{13}CO}$/$X_{\rm C^{18}O}$ have mean values of 15.4, 9.1, and 9.0, and median values of 13.6, 8.0, and 8.2, for sub-regions A, B, and C, respectively.
These values are much larger than the solar system's value (5.5), suggesting that UV photons from the nearby OB stars and/or H{\sc ii} regions have a strong influence on the chemistry of clouds in the CMa OB1 complex.
Since \xxco is selectively dissociated by UV emission more effectively than $^{13}$CO,
the abundance variations of these isotopologues across different environments have been interpreted as
an indicator of the evolutionary stages of MCs~\citep[e.g., ][]{1988ApJ...334..771V, 2014A&A...564A..68S}.
According to the relation $X_{\rm ^{13}CO}$/$X_{\rm C^{18}O}$-$N_{\rm ^{13}CO}$,
\cite{2019ApJS..243...25W} defined two types of MCs.
The boundaries of type {\sc i} and type {\sc ii} MCs are approximately indicated by the dashed and
dotted lines in Figure~\ref{Fig:Icodistribution}(b).
Type {\sc i} MCs are characterized by a lower abundance ratio $X_{\rm ^{13}CO}$/$X_{\rm C^{18}O}$,
and lower \xco column density with relatively lower excitation temperature than type {\sc ii} MCs.
Therefore, type {\sc i} MCs are probably quiescent clouds or in the early stage of star formation.
On the contrary, type {\sc ii} MCs contain active star-forming MCs and are strongly influenced by newly formed stars. Clearly, we can classify sub-region C as a type {\sc i} cloud, and sub-region A as a type {\sc ii} cloud. Meanwhile, sub-region B appears to fall into a region between
type {\sc i} and type {\sc ii}, where star-formation has occurred but the feedback from the young
generation seems not significant. This view is supported by our results of the ongoing star-formation census in \S\ref{Section5}.

\section{\xxco clumps\label{Section4}}
\subsection{Clump Identification\label{Section4.1}}
Since clumps/cores are the basic unit of star formation, they are helpful to understand
the fragmentation and star formation of MCs~\citep[\eg][]{2007ARA&A..45..339B}.
The physical properties of the clumps/cores in the CMa OB1 complex were studied using the NANTEN CO and $Herschel/Hi-GAL$ data~\citep[e.g.,][]{2013ApJ...772...45E}.
Considering much denser gas it can probe than $^{12}$CO, \xxco is more suitable for tracing compact molecular
clumps/cores, which are more likely to form stars.

\begin{figure}[!ht]
\centering
\includegraphics[width=0.88\textwidth]{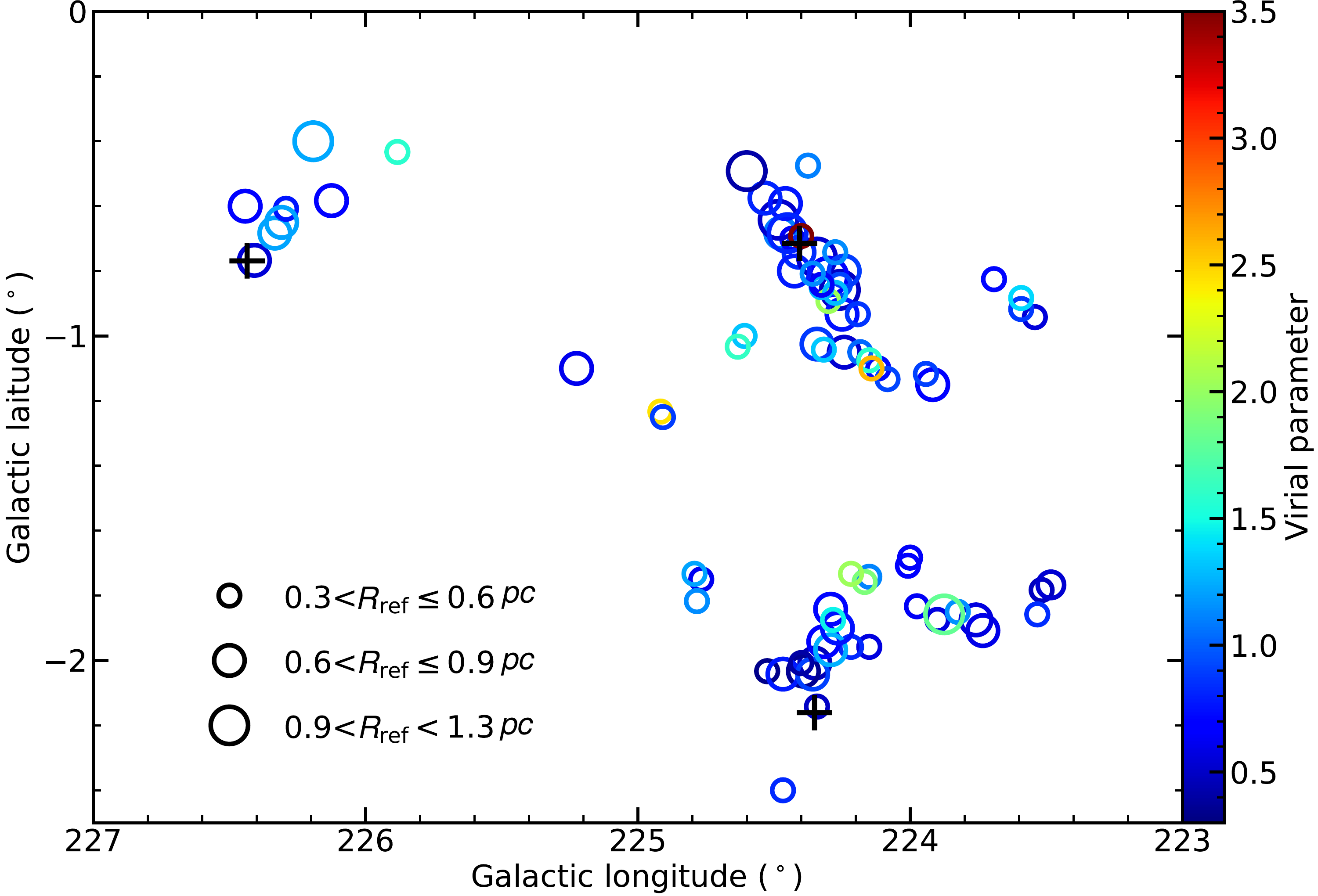}
\caption{Distributions of \xxco clumps. The circles and crosses represent the positions of the C$^{18}$O clumps and the three known water masers, respectively.}
\label{Fig:clumpdistribution}
\end{figure}

We use the CUPID (part of the STARLINK project software\footnote{\url{http://starlink.eao.hawaii.edu/starlink}}) clump-finding algorithm GaussClumps \citep{1990ApJ...356..513S} to identify compact clumps in the 3D FITS cube of C$^{18}$O.
We set the parameters \texttt{MAXSKIP} = 100, \texttt{MINPIX} = 8, and \texttt{NPEAK} = 5.
Here, \texttt{MAXSKIP} determines the maximum amount of iterations that the Gaussian fits are performed. If the number of failed attempts to fit consecutive clumps reaches \texttt{MAXSKIP}, Gaussclumps stops searching for further clumps. A large value of \texttt{MAXSKIP} may allow the user to identify more clumps; however, it becomes more time consuming to do so. We tested three cases, \texttt{MAXSKIP} = 50, 100, and 400, which revealed that the number of clumps was the same for the latter two cases, and were both larger than \texttt{MAXSKIP} = 50. Therefore, we set \texttt{MAXSKIP} = 100 in this study.
Next, \texttt{MINPIX} determines the minimum number of pixels each clump contains. Considering that a clump must contain at least one beam size (nearly four pixels) in the $l$-$b$ space, and at least two channels in velocity space, we finally use \texttt{MINPIX} = 8.
\texttt{NPEAK} determines the minimum peak emission of each clump. We assumed a peak main beam temperature of each clump of $\geqslant$ 5$\sigma$. Thus, we take \texttt{NPEAK} = 5 as our final selection.
Finally, 83 clumps in total were identified in the whole mapped area.
The measured parameters of each clump including the position, central velocity ($V_{\rm LSR}$), angular size ($A$), velocity dispersion ($dv$), peak main beam temperatures of \xxco ($T_{\mathrm{C}^{18}\mathrm{O}}$) and \co ($T_{^{12}\mathrm{CO}}$) are listed in Table~\ref{table:clumps}. The spatial distributions of these clumps are also shown in Figure~\ref{Fig:clumpdistribution}, of which 32, 42, and nine \xxco clumps
are located at sub-regions A, B, and C, respectively.

\subsection{Physical Properties of \xxco Clumps\label{Section4.2}}
As stated in \S\ref{Section3.2}, here we also assume that $T_\mathrm{ex}$ is the same of all CO isotopologues.
For simplicity, we also assume that $T_\mathrm{ex}$ and $\tau$ are uniform within each clump.
Then following the Equations~\ref{Equ:tex} and~\ref{Equ:tau2}, we adopt the peak main beam temperatures of \co and \xxco
(columns 4 and 5 of Table~\ref{table:clumps}) to estimate $T_\mathrm{ex}$ and $\tau$ for each clump.

The effective radius obtained after beam deconvolution ($R_\mathrm{eff}$) is given by $R_\mathrm{eff} = d \times \sqrt{\frac{A}{\pi}-\frac{\theta_\mathrm{MB}^2}{4}}$, where $d$ is the distance (which was assumed to be 1.15 kpc), $\theta_\mathrm{MB}$ is the FWHM
beam width, and $A$ is the angle area of each clump.
Following \cite{1988ApJ...333..821M}, virial mass $M_\mathrm{vir}$ can be estimated by
$M_\mathrm{vir}=210 R_\mathrm{eff} (\Delta v)^2$, where $\Delta v=\sqrt{8\ln2} dv$.
Virial parameter can be calculated by $\alpha_\mathrm{vir}$=$M_\mathrm{vir}$/$M_\mathrm{LTE}$.
The derived physical properties are summarized in Table~\ref{table:clumps}.

\begin{figure}[!ht]
\centering
\includegraphics[height=0.19\textheight]{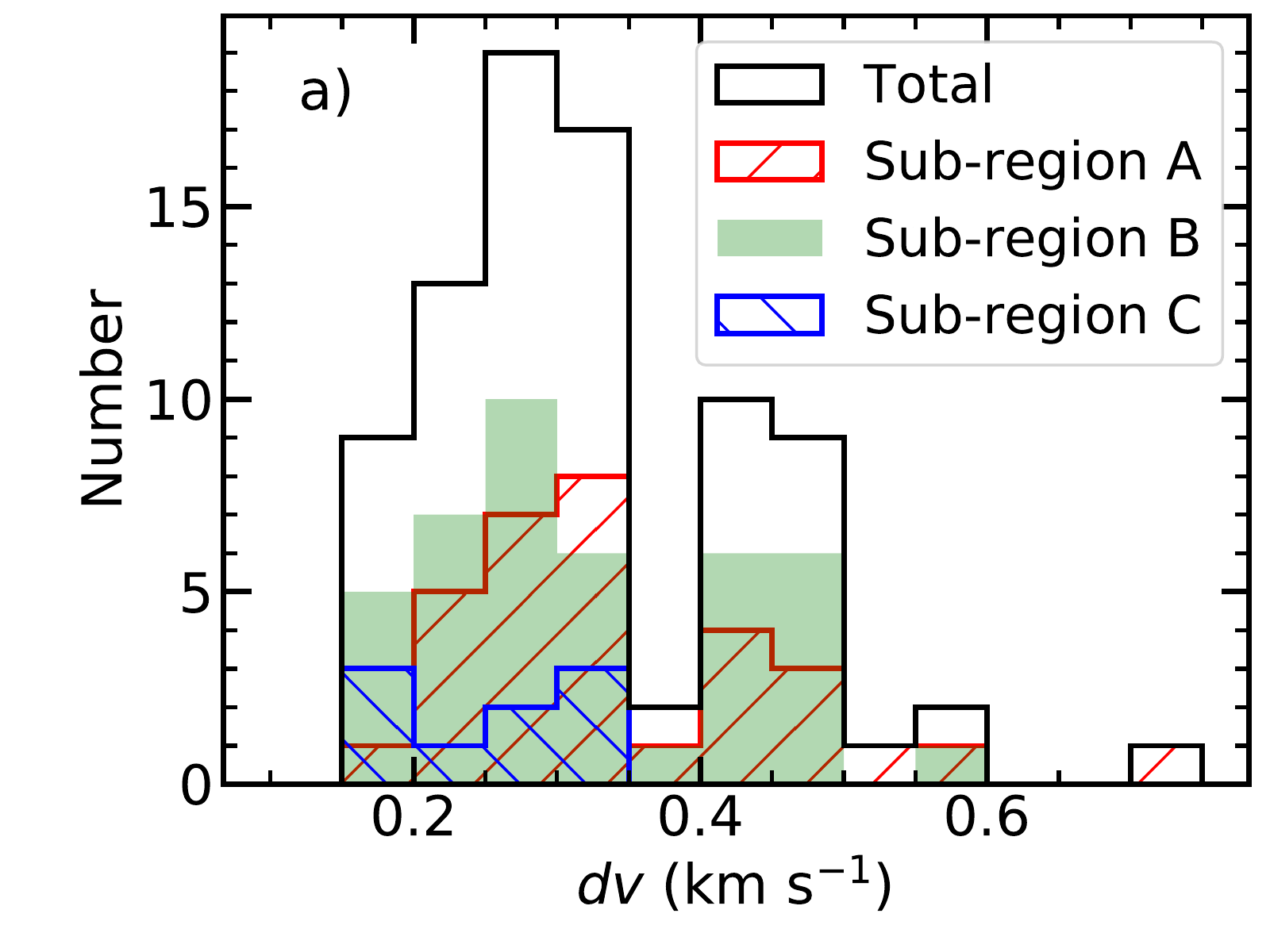}
\includegraphics[height=0.19\textheight]{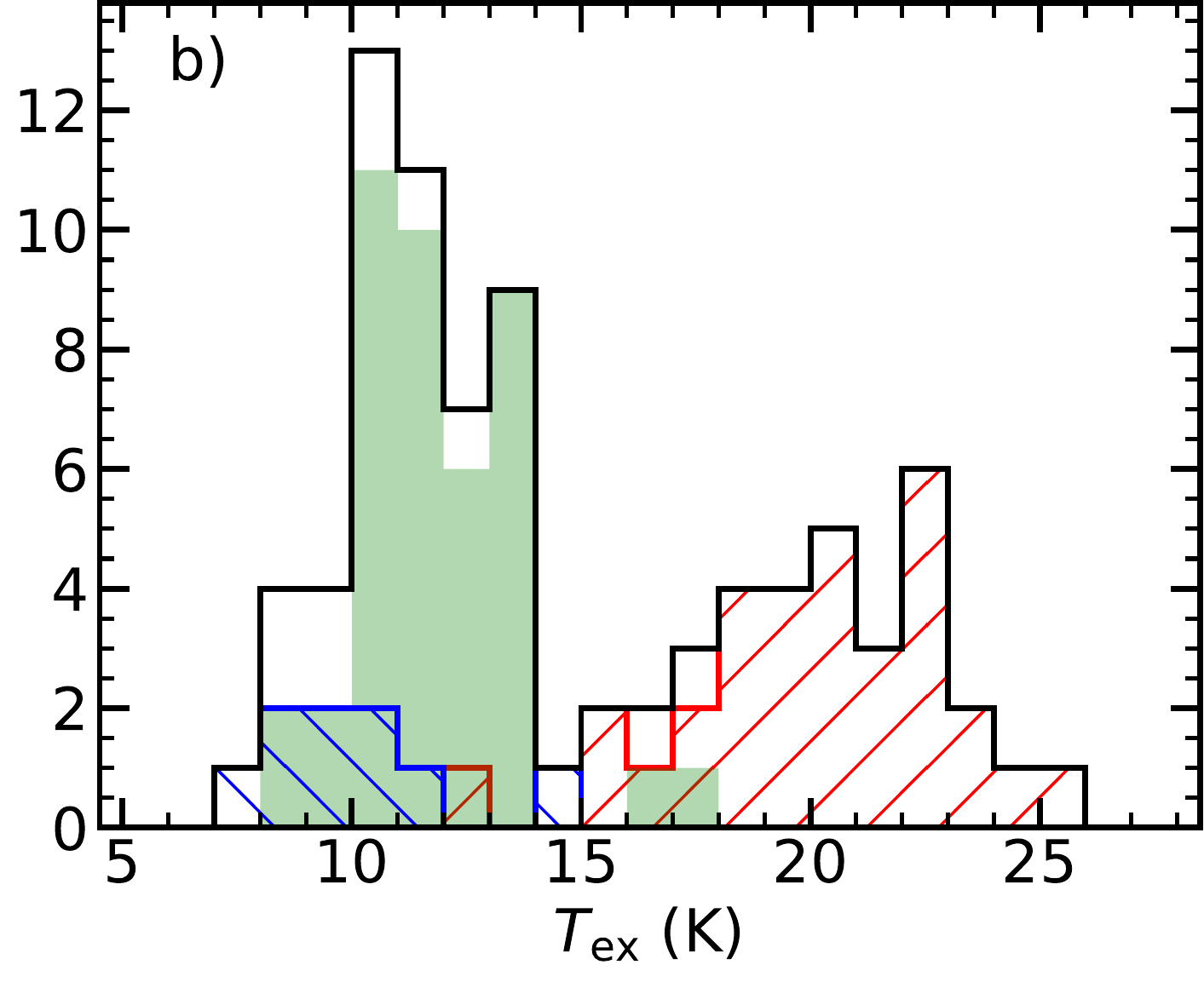}
\includegraphics[height=0.19\textheight]{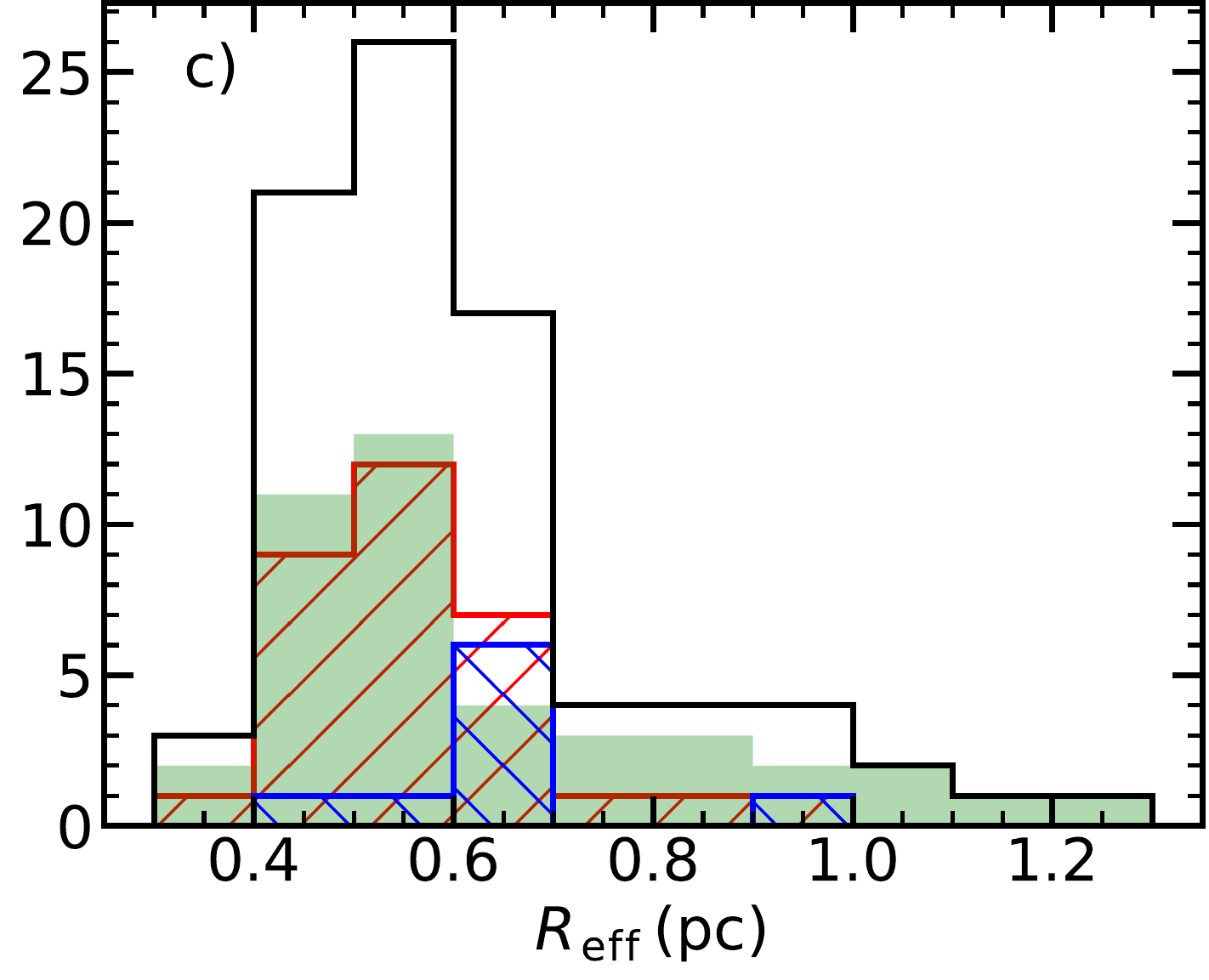}
\caption{Histograms of physical properties of \xxco clumps including ($a$) velocity dispersion, ($b$) excitation temperature, and ($c$) effective radius.}
\label{Fig:propertiesdistribution}
\end{figure}

The distributions of $dv$, $T_{\rm ex}$, and radius of the 83 \xxco clumps
are also plotted as histograms in Figure~\ref{Fig:propertiesdistribution}.
The velocity dispersions ($dv$) of the clumps are in the range of 0.17 to 0.72~\kms,
with the mean value of 0.33~\kms. Only a few clumps are similar to clumps seen in infrared dark clouds \citep[0.64--0.94~\kms][]{2011ApJS..193...10Z}. While the majority clumps are similar to those clumps
in low-mass protostars~\citep[0.13--0.89~\kms][]{2002A&A...389..908J}.
The clumps in sub-region A seem to have slightly larger values of $dv$ (mean value 0.35~\kms) than those clumps in
sub-regions B (mean 0.33~\kms) and C (mean 0.26~\kms).
The $T_{\rm ex}$ values of clumps in sub-region A
are typically $\sim$ 20~K, which are also much higher than those in sub-regions B and C.
Here, we should note that the conclusion for sub-region C is rather tentative, since the relatively small number of clumps in sub-region C (nine) likely restrict its statistical significance.
The values of $R_\mathrm{eff}$ for all sub-regions is between 0.3~pc and 1.3~pc, with the mean value of 0.59~pc.
We note that all of the detected \xxco clumps are above our detection resolution limit of 0.14~pc,
and match well with the definition of a clump, which have typical sizes of 0.3~pc~$\leqslant$~$R_\mathrm{eff}~\leqslant$~3~pc\citep[]
[]{2007ARA&A..45..339B}.
In contrast, the mean $R_\mathrm{eff}$ of ``\co clumps'',
identified by \cite{2013ApJ...772...45E}, is 4.79~pc, which actually matches typical cloud size.

To determine the dominant broadening mechanisms, we
calculate both thermal velocity dispersion ($\sigma_\mathrm{T}$) and non-thermal velocity dispersion ($\sigma_\mathrm{NT}$) using the following equations~\citep{1983ApJ...270..105M, 2015AJ....150...60L}:

\begin{gather}
  \sigma_\mathrm{T}\ =\ \frac{1}{1000}\sqrt{\frac{kT_\mathrm{kin}}{\mu m_\mathrm{H}}},\label{Equ:Thermal}\\
  \sigma_\mathrm{NT}\ =\ \sqrt{dv^2-\left(\frac{1}{1000} \sqrt{\frac{kT_\mathrm{kin}}{m_\mathrm{C^{18}O}}}\right)^2},\label{Equ:Non-Thermal}
\end{gather}

where $k$ is the Boltzmann constant, $\mu=2.4$, $T_\mathrm{kin}$ is kinetic temperature,
and $m_{\mathrm{C^{18}O}}$ is the mean molecular mass of C$^{18}$O.
We assumed $T_\mathrm{ex}$ as the lower limit of $T_\mathrm{kin}$ when estimating $\sigma_{\rm Non-thermal}$ and $\sigma_{\rm Thermal}$, which thus also both represent lower limits.

\begin{figure}[!ht]
\centering
\includegraphics[width=0.45\textwidth]{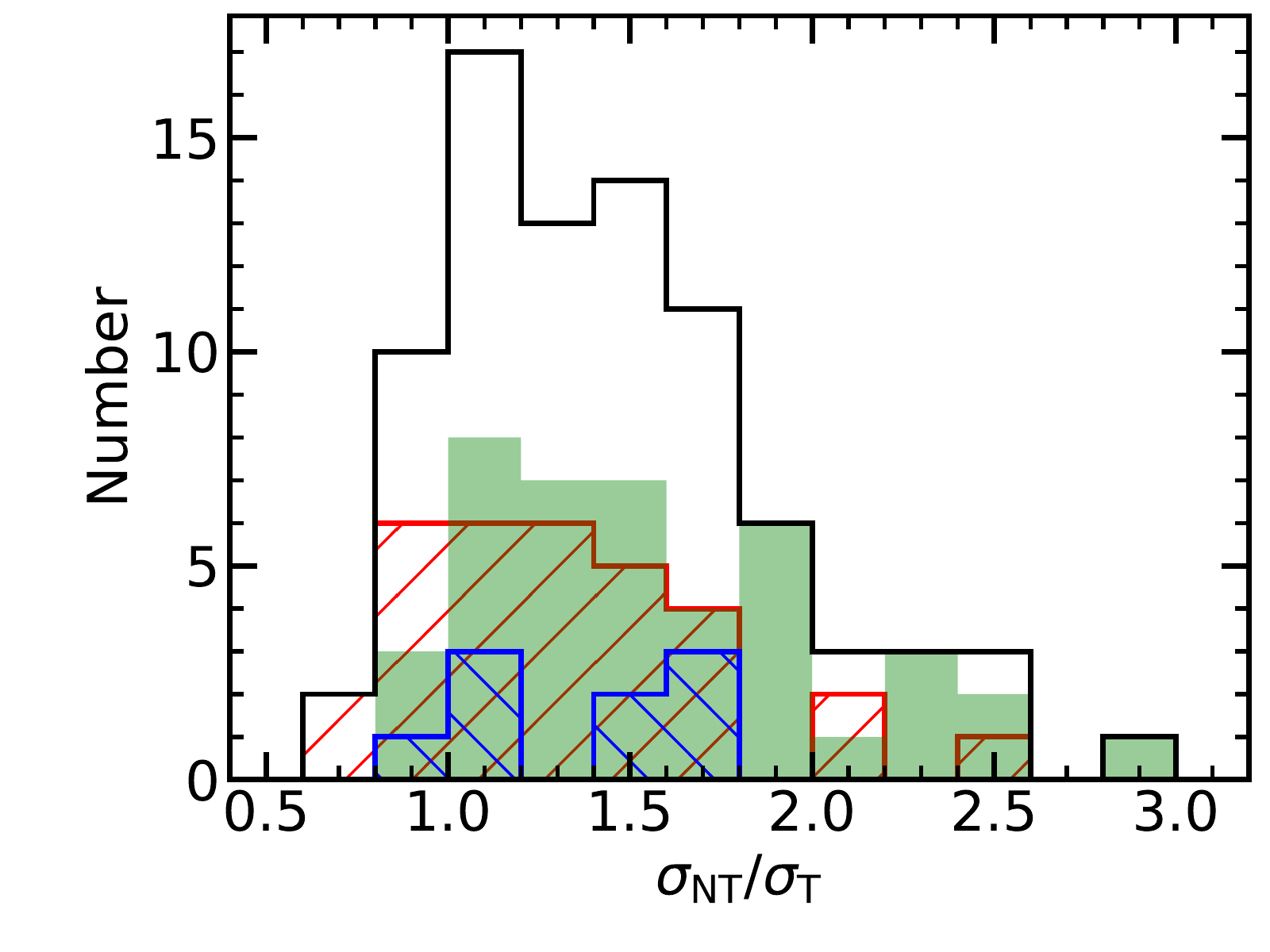}
\caption{Histograms of the ratio between non-thermal (blue) and thermal (red) velocity dispersions of the 83 \xxco clumps. The color codes are same as in Figure~\ref{Fig:propertiesdistribution}.}
\label{Fig:cpr3}
\end{figure}

Figure~\ref{Fig:cpr3} presents the ratios between $\sigma_{\rm NT}$ and $\sigma_{\rm T}$ for each sub-region.
Our statistical results reveal that 68 (82\%) clumps have $\sigma_\mathrm{NT}$ values larger than $\sigma_\mathrm{T}$.
Additionally, the ratios between $\sigma_\mathrm{NT}$ and $\sigma_\mathrm{T}$ of the \xxco clumps are in the
range of 0.64 to 2.84, with an average value of 1.44,
and a median value of 1.37, which implies that non-thermal broadening mechanism may play a dominant role in the CMa OB1 complex. In addition, we find that the clumps in sub-region B have the highest values of $\sigma_\mathrm{NT}$ to $\sigma_\mathrm{T}$ ratios (with an average value of
1.57, and median value of 1.46). Meanwhile, the $\sigma_\mathrm{NT}$ to $\sigma_\mathrm{T}$ ratios of the clumps in sub-region C (with an average value of 1.33, and median value of 1.47) have slightly larger values
than those of sub-region A (with an average value of 1.30, and median value of 1.25). However, as mentioned above, due to the limited number of clumps in sub-region C, the conclusion for sub-region C should be viewed with caution.

The virial parameter of each clump is illustrated in Figure~\ref{Fig:clumpdistribution}.
We find about 78 (94\%) clumps have $\alpha_{\rm vir}<$2, which are thought to be gravitationally bound.
This implies that these clumps may eventually collapse into stars.
In comparison, only 7\% of the ``\co clumps'' observed by \cite{2013ApJ...772...45E} were found to be gravitationally bound in the CMa OB1 complex. As mentioned above, the ``\co clumps'' with an average
radii of several parsecs actually represent typical MCs. Therefore, the discrepancy between the \xxco clumps
and ``\co clumps'' may lend support to the view that molecular clumps (i.e., the site in which star formation to take place) within MCs should be gravitationally bound; however, it is not necessary that the MCs as a whole be bound~\citep[\eg][]{2005MNRAS.361....2C,2011MNRAS.411...65B}.
Moreover, our results show the advantage of \xxco in tracing high density components.

\begin{figure}[!ht]
\centering
\includegraphics[width=0.45\textwidth]{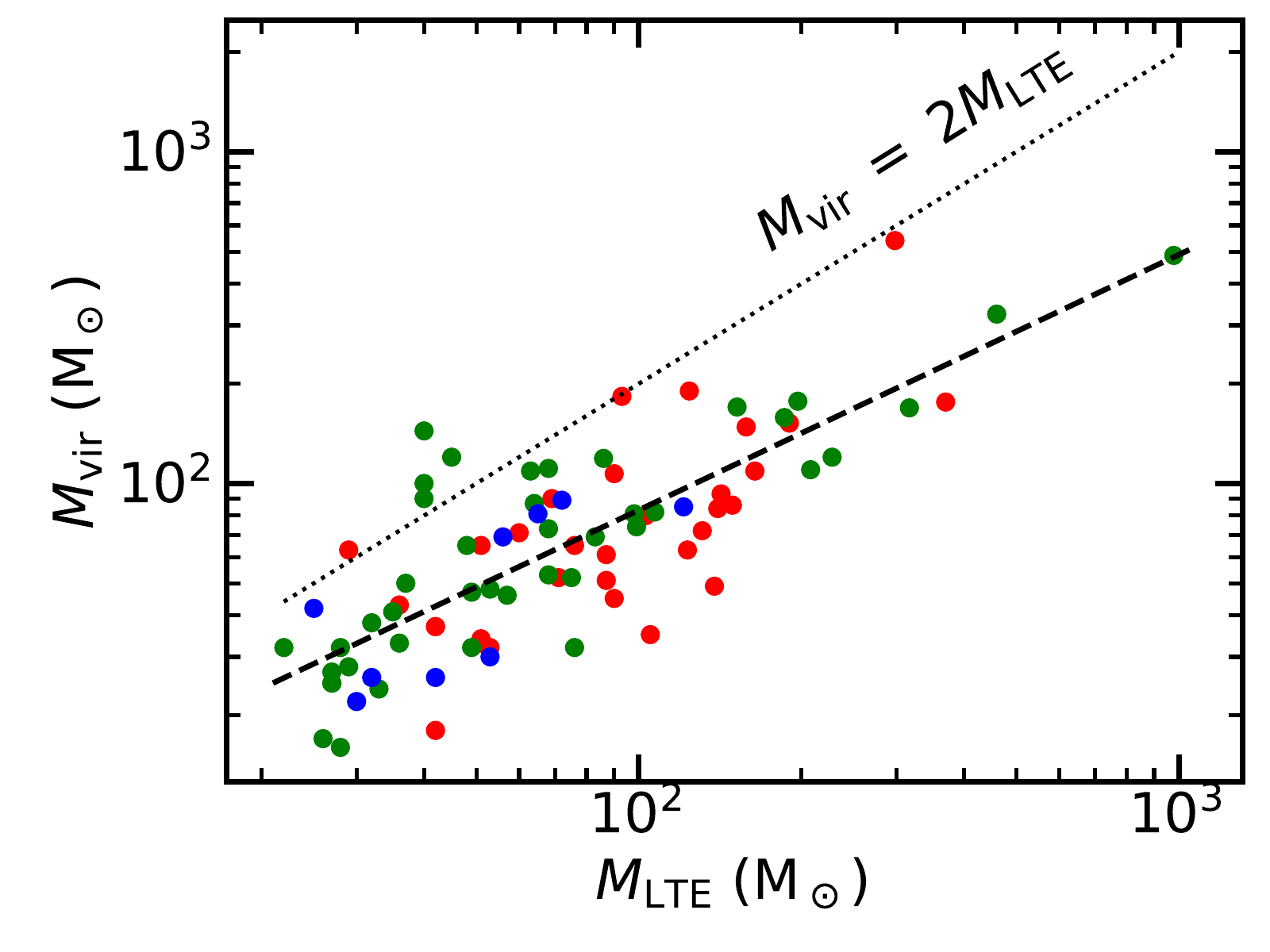}
\caption{Virial mass $M_\mathrm{vir}$ as a function of LTE mass $M_\mathrm{LTE}$ for \xxco clump.
The color codes are same as in Figure~\ref{Fig:propertiesdistribution}.}
\label{Fig:mm}
\end{figure}

The relationship between $M_\mathrm{LTE}$ and $M_\mathrm{vir}$ is shown in Figure~\ref{Fig:mm}.
The dashed line indicates the best-fitting power-law to all the 83 \xxco clumps, computed using a least-squares bisector.
The power index is 0.77$\pm$0.06, and the Pearson correlation coefficient of relation is 0.8.
We find that the virial parameters of \xxco clumps tend to decrease with the increase of LTE masses.
This indicates that \xxco clumps with larger LTE masses are more likely to be gravitationally bound.

\begin{figure}[!ht]
\centering
\includegraphics[width=0.45\textwidth]{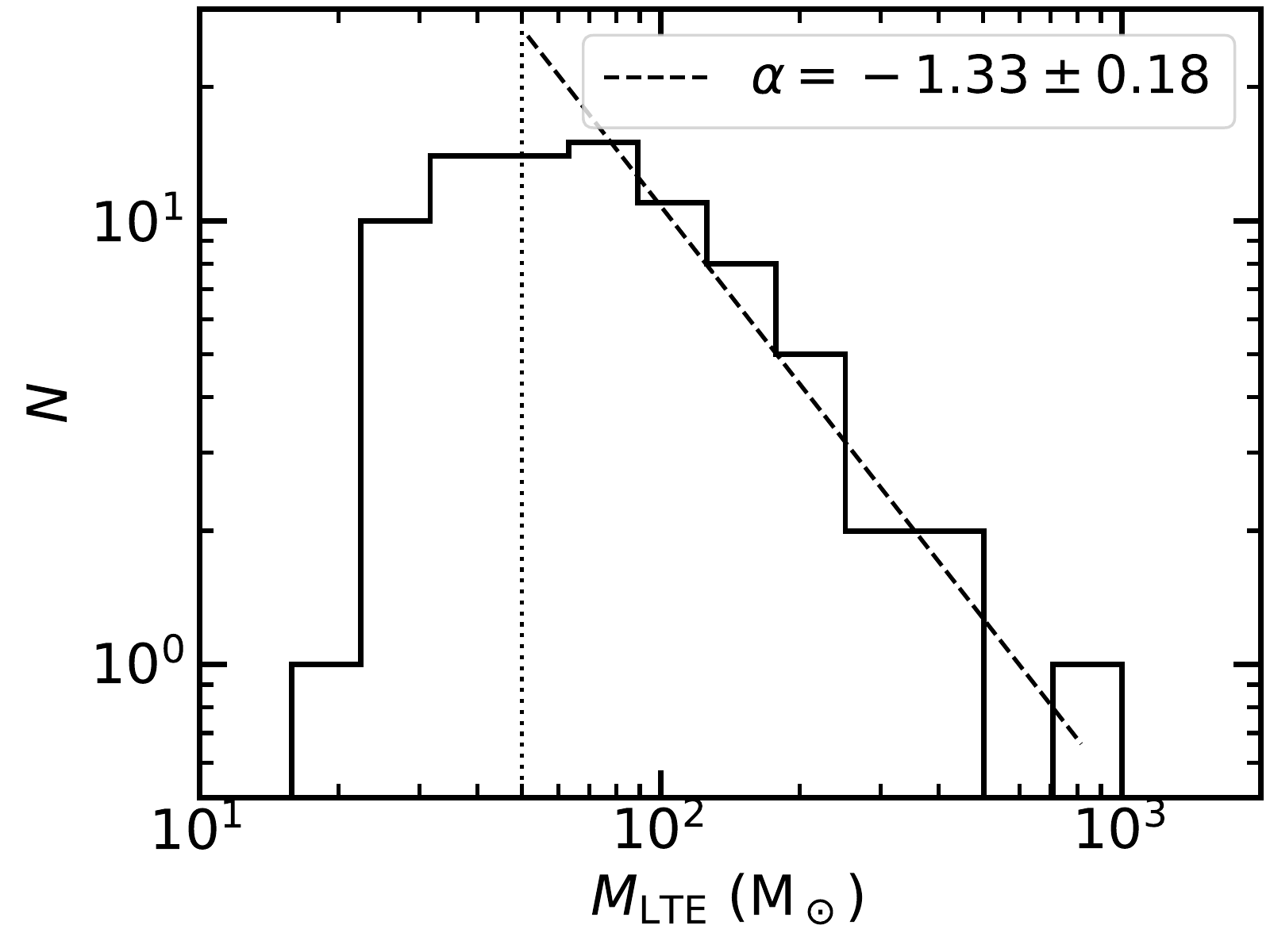}
\caption{Clump mass distribution in the CMa OB1 complex. The dotted line masks the minimum meaningful mass with a value of 50~M$_\sun$.}
\label{Fig:cpr2}
\end{figure}

The core/clump mass function (CMF) is an empirical function which describes the relative
frequency of cores/clumps with different masses. CMF resembles the stellar initial mass function (IMF)
which describes the initial distributions of masses of a stellar population. The CMF is defined as:

\begin{equation}\label{Equ:CMF}
  N(\log M) \propto M^{\alpha},
\end{equation}

where $N$ is the number of clumps per bin, and $M$ is the LTE mass here. For our case, we use the Markov Chain Monte Carlo (MCMC) algorithm to estimate the power-law index~\citep{2009MNRAS.397..495K,2020ApJ...898...80Y}.
We set the minimum meaningful mass as 50 M$_\odot$, and use 54 clumps to calculate the power-law index of the CMF.
We calculate two MCMC sampling chains with a total of 2000 sample numbers.
Consequently, the power-law index was found to be $\alpha = -1.33\pm$0.18 for the high-mass portion of the mass distribution of the \xxco clumps (see Figure~\ref{Fig:cpr2}).
Unlike the result of $\alpha= -0.7 \pm$ 0.3 based on ``\co clumps'' found by \citet{2013ApJ...772...45E},
the result derived from our \xxco clumps is more in agreement with that of the \textit{Hi-GAL} pre-stellar sources in the CMa OB1 complex~\citep[$\alpha=-1.0\pm$0.2,][]{2013ApJ...772...45E}.
In addition, the constrained CMF of the \xxco clumps in the CMa OB1 complex is similar to that of nearby star-forming regions, e.g., $\alpha = -1.3\pm$0.1 in Orion A \citep{2007ApJ...665.1194I}, $\alpha = -1.3\pm$0.2 in Perseus, Serpens, and Ophiuchus \citep{2008ApJ...684.1240E}.

Furthermore, we see that all those observed CMFs (from this study and literature) resemble
several typical IMF, e.g., $\alpha = -1.35$ \citep{1955ApJ...121..161S}, $\alpha = -1.3\pm$0.7 for
single stars~\citep{2001MNRAS.322..231K}, and $\alpha = -1.35\pm$0.3 for
multiple systems~\citep{2005ASSL..327...41C}. Such similarity between the CMF and IMF may imply that
the CMF can impact the origin of the IMF to some extent via turbulent fragmentation, which is consistent
with the prediction of turbulent fragmentation theories \citep[which have
been used to explain the CMF and cloud structure at various scales,][]{2014prpl.conf...53O}.

\begin{figure}[!ht]
\centering
\includegraphics[width=0.45\textwidth]{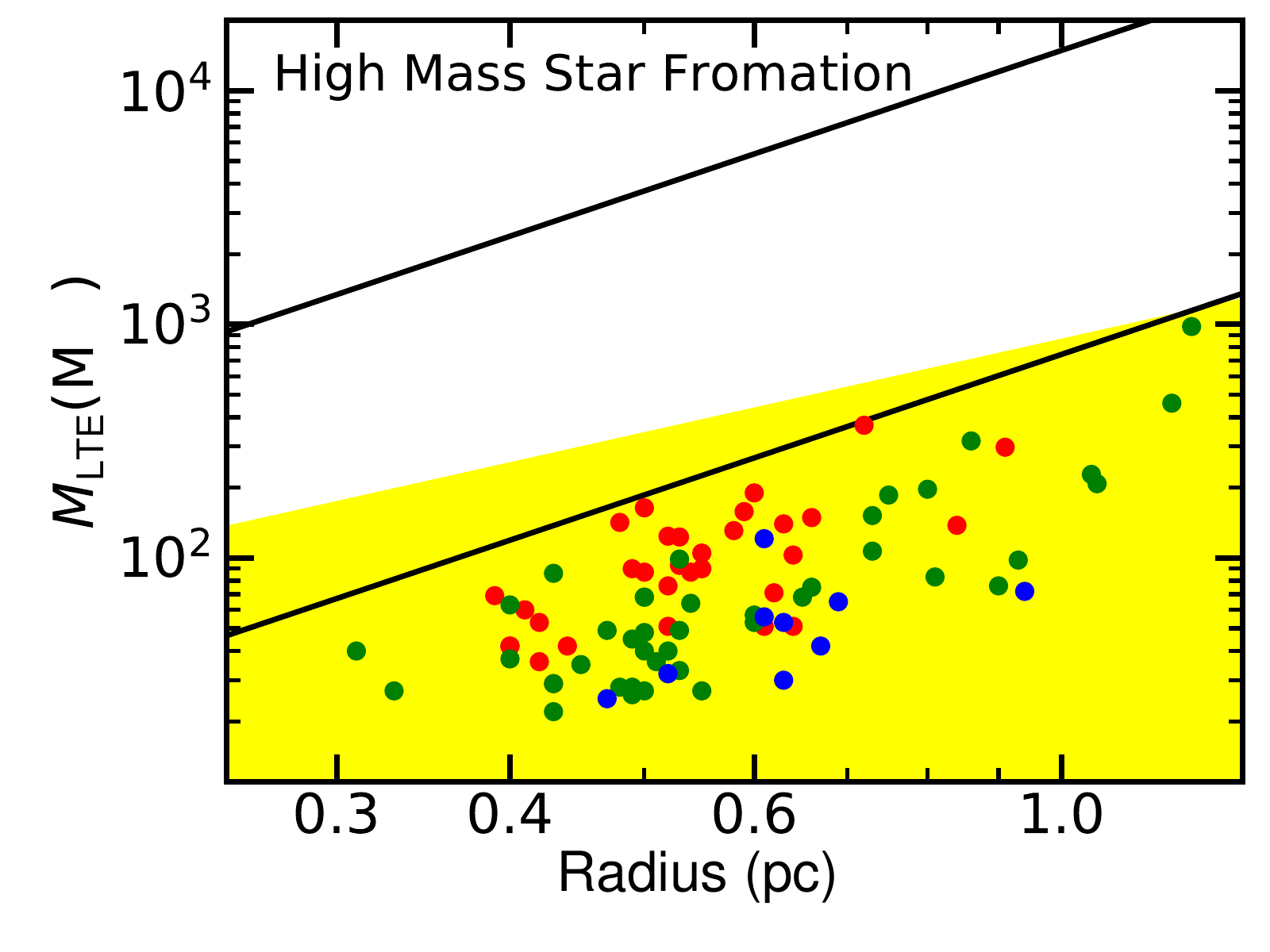}
\caption{Mass--size relationship of \xxco clumps in the CMa OB1 complex. The upper and lower black
lines represent the empirical surface densities required for high-mass star formation with values of
1 g\,cm$^{-2}$ \citep{2008Natur.451.1082K} and 0.05 g\,cm$^{-2}$ \citep{2013MNRAS.431.1752U}, respectively.
The yellow shaded region represents the region where low-mass star formation is more likely to take place~\citep{2010ApJ...723L...7K}. The color codes are same as in Figure~\ref{Fig:propertiesdistribution}.}
\label{Fig:cpr1}
\end{figure}

To examine how many clumps have the potential to form high-mass stars, we have presented the
mass--size relation of the \xxco clumps in Figure~\ref{Fig:cpr1}.
The upper and lower black solid lines indicate constant surface densities of 1~g\,cm$^{-2}$ \citep{2008Natur.451.1082K} and 0.05~g\,cm$^{-2}$ \citep{2013MNRAS.431.1752U}, respectively.
If these two lines provide the reliable empirical upper and lower bounds for the clump surface densities required for massive star formation, then none of the \xxco clump has enough surface densities to form high-mass stars.
Besides, when the empirical relationship suggested by \cite{2010ApJ...723L...7K} is considered, all of \xxco
clumps still lie in the region where low-mass star formation is more likely to take place.
These results would suggest that none of the \xxco clump has the potential to form high-mass stars in CMa OB1 complex.

\section{Outflow Candidates}\label{Section5}
\subsection{Identification of outflow candidates}\label{Section5.1}
Outflow is a universal phenomenon in star-forming regions, arising from deeply embedded protostellar objects
to optically visible young stars \citep{2001ARA&A..39..403R}. CO spectral lines can reveal outflow
activities in star-forming regions \citep[\eg][]{1987ARA&A..25...23S}, and compared to optical outflows,
CO outflows occur during the younger stages \citep{2011MNRAS.418.2121G}.
Large-scale outflows surveys can provide a large systematic sample for investigating the impact of outflow activities on their surrounding environments \citep[\eg][]{2010ApJ...715.1170A}.
Therefore, studying the large-scale properties of MCs and their associated star-forming
activities are necessary. However, to date there is no CO outflow survey in the CMa OB1 complex.

In this section, we identify \co molecular outflows by using a set of semi-automated IDL algorithm\footnote{\url{http://github.com/liyj09/outflow-survey-v1}} based on 3D $l-b-v$ data of \co and $^{13}$CO~\citep[][]{2018ApJS..235...15L, 2019ApJS..242...19L}. This algorithm suite allows the user to determine the central velocity and minimum velocity extents of each pixel by fitting optically-thin $^{13}$CO lines. Then, the CUPID clump-finding algorithm is used to automatically identify any extreme points (i.e., points that have maximum velocity extents compared with surrounding) in \co data~\citep[see details in][]{2018ApJS..235...15L, 2019ApJS..242...19L}.
Using such line and artificial diagnoses method, we search for and identify outflow candidates.
Then, we inspected the maps (e.g., intensity map, P--V diagram, and/or channel map) of each candidate by eye to exclude those with ambiguous outflow characteristics (\eg outflow extent velocities $<$ 1~\kms) or those seriously contaminated by other components.
In this step, about 2/3 raw outflow samples that automatically from the procedure are excluded.
Eventually, only 56 blue-lobe and 52 red-lobe candidates are included in this study.
If the separation between a blue-lobe and a red-lobe was smaller than 1.5 pc, then they were paired to constitute a bipolar outflow candidate. Finally, we identify 85 outflow candidates in the CMa OB1 complex,
including 23 bipolar outflow candidates, 33 blue and 29 red monopolar ones (see Appendix \ref{Appendix.B}).
All of these are new detections.

A search using the SIMBAD database revealed that only those three outflow candidates
associated with water masers (bipolar outflow candidates 43 and 47, and red monopolar candidates 83)
were single-pointing observed via SiO line by \cite{1998A&AS..132..211H} (in their table B1, indices of 19--21). Their observations resulted in SiO detections of each of these three candidates, which had SiO line widths up to 23.8~\kms. Moreover, only outflow candidate 47 was previously observed via NH$_3$ line \citep{2017ApJS..228...12T}.

\begin{figure}[!t]
\centering
\includegraphics[width=0.75\textwidth]{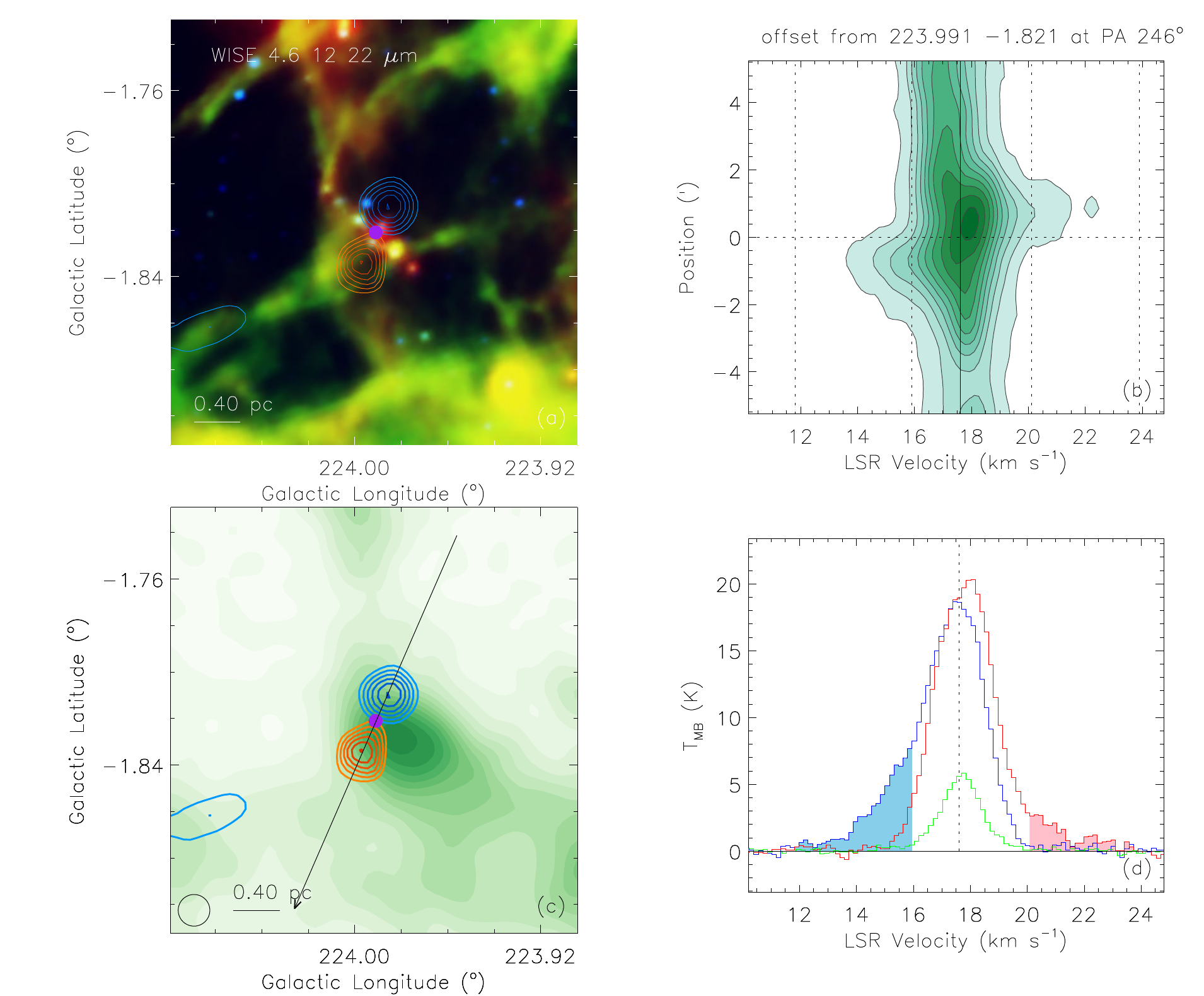}
\caption{An example outflow candidate (No. 25). ($a$) Integral intensity maps
of the red and blue lobes superposed on the WISE image (blue, green, and red for 4.6, 12, and 22 $\mu$m, respectively).
($b$) P--V diagram along position angle of the outflow
marked at the top the panel. ($c$) Integral intensity maps of the red and blue lobes superposed on the \xco intensity
map. The arrow indicates the position angle of the outflow. ($d$) \co spectra at the peak emission of the blue and
red lobes are coded in blue and red, respectively, while the green one is the averaged \xco spectrum over 3$\,\times\,$3 pixels centered
at the position of the outflow. The line wing velocity range of each lobe is specified with shading.}
\label{Fig:outfig1}
\end{figure}

\begin{figure}[!ht]
\centering
\includegraphics[width=0.75\textwidth]{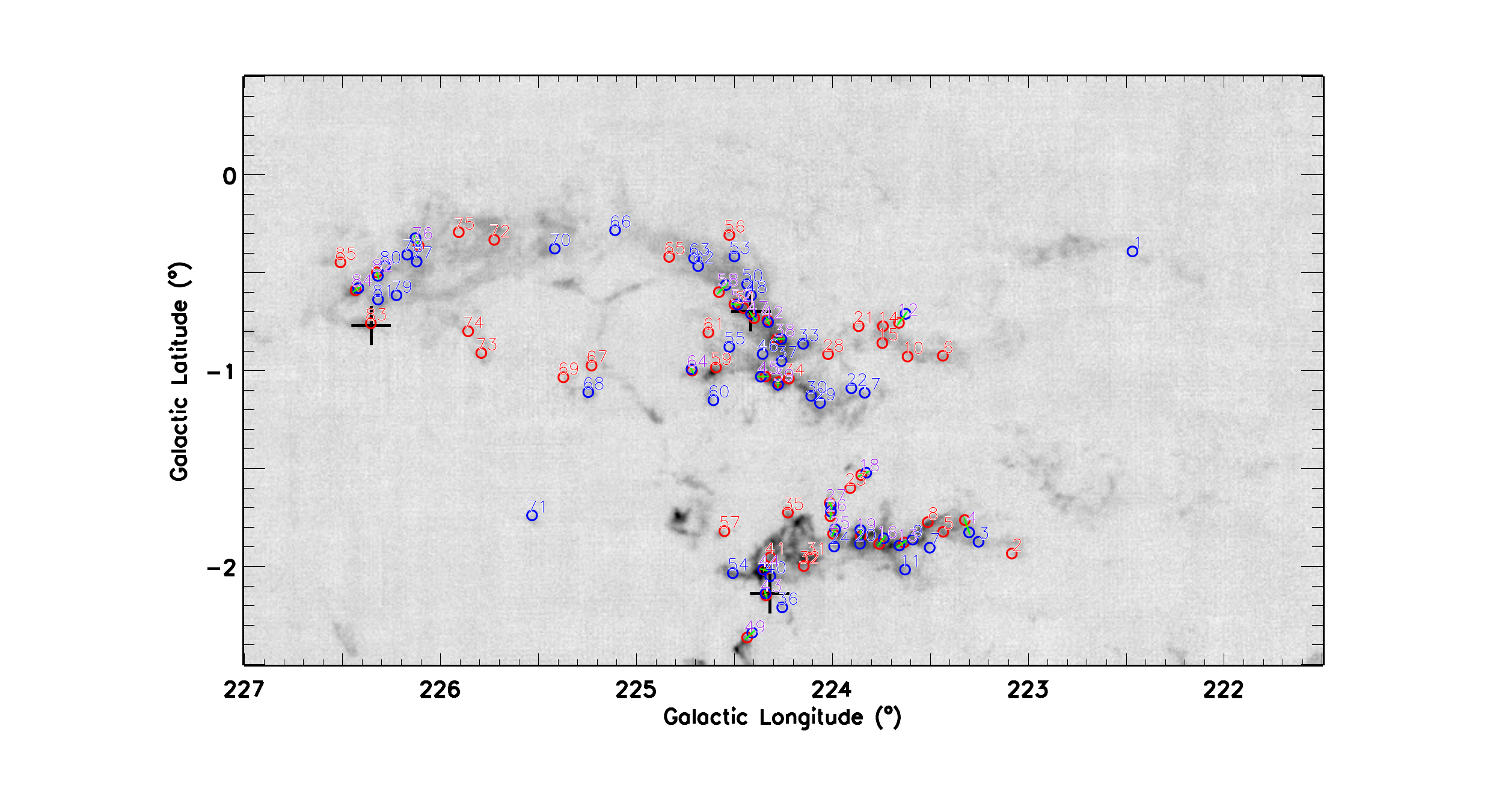}
\caption{Locations of outflow candidates. The grey-scale map presents the integrated intensity of \xco line,
integrated over (0, 30)~\kms. The blue and red circles represent the positions of the blue and red lobes, respectively. The green bars connect the blue and red lobes of the bipolar outflow candidates.
The blue/red and purple number present the index of blue/red monopolar and bipolar outflow candidates, respectively.
The black crosses represent of positions of the three known water masers \citep{1994A&AS..103..541B}.}
\label{Fig:outmap}
\end{figure}

The measured parameters for each outflow candidate are summarized in Table~\ref{table:outflow1}.
Figure~\ref{Fig:outfig1} presents the details of outflow candidate with the index of 25 as an example, and the spatial distributions of all detected \co outflow candidates are shown in Figure~\ref{Fig:outmap}.
Apparently, the outflow candidates are mainly distributed along filaments in the CMa OB1 complex.
Additionally, we find that the outflows are not uniformly distributed in different sub-regions,
and they are clearly concentrated in sub-regions A and B.
Using the equations listed in Appendix B of \cite{2018ApJS..235...15L}, the outflow candidates’ physical parameters were also calculated and tabulated in Table~\ref{table:outflow2}.
It is also worth noting that the calculated masses, momenta, and energies are lower limits, due to underestimated low-velocity outflows, as discussed in detail in section~4.2 of \cite{2018ApJS..235...15L}.

\begin{figure}[!ht]
\centering
\includegraphics[width=0.75\textwidth]{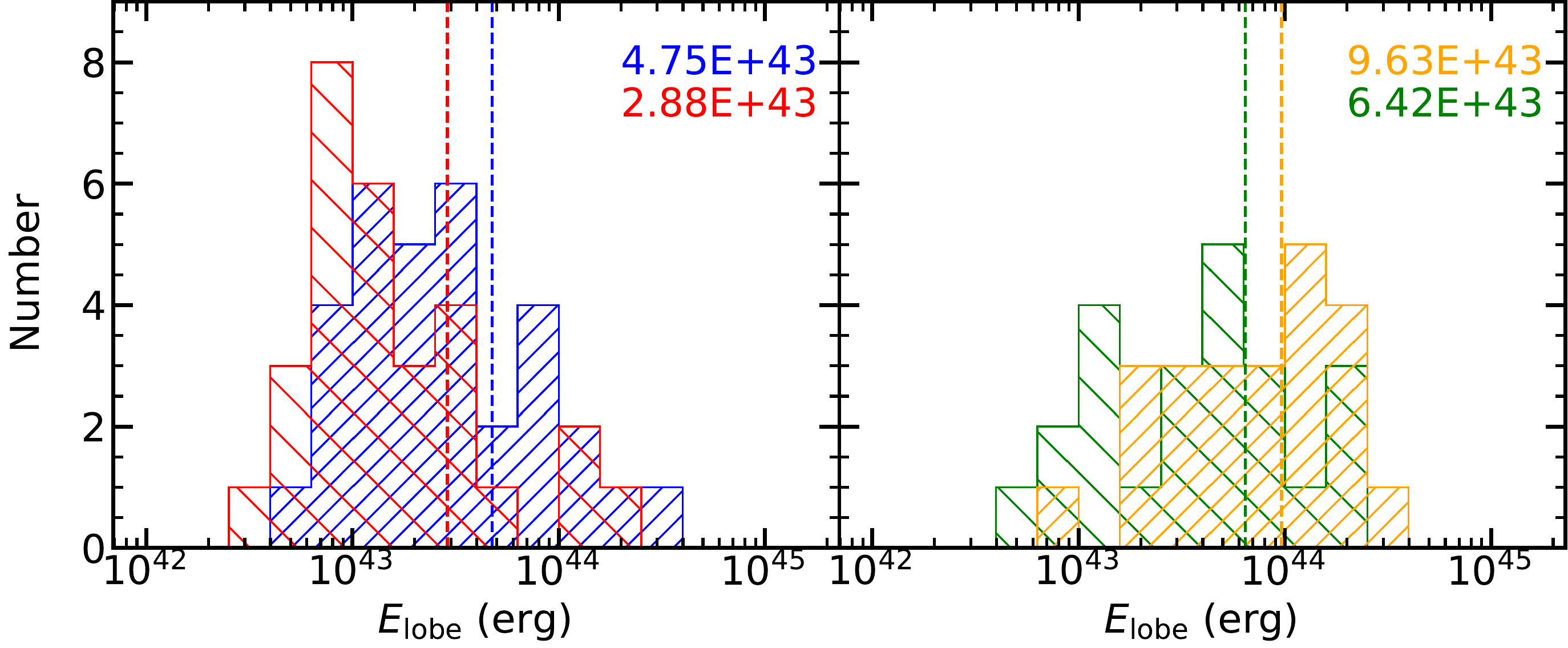}
\caption{Histograms of the kinetic energy of the outflow lobes.
The hatched regions in left panel show the distributions of the red-lobes (in red, sub-group 1)
and blue-lobes (in blue, sub-group 3) of the monopolar
outflow candidates, while the hatched regions in right panel show the red-lobes (in green, sub-group 2)
and blue-lobes (in yellow, sub-group 4) of the
bipolar outflow candidates. The mean values of each sub-group are also indicated in each panel.}
\label{Fig:outphy1}
\end{figure}

The 56+52 detected outflow lobes were split into
four sub-groups: (1) 29 red-lobes from monopolar outflows, (2) 23 red-lobes from bipolar outflows,
(3) 33 blue-lobes from monopolar outflows, and (4) 23 blue-lobes from bipolar outflows.
A comparison of the velocities, masses, kinetic energies, dynamical timescales and luminosities for the four sub-groups is shown in Table~\ref{table:outphy}.
We find that lobes from the bipolar outflow candidates (sub-groups 2 and 4) generally have larger velocities, masses, kinetic energies, and luminosities than those lobes of the monopolar outflow candidates (sub-groups 1 and 3).
Overall, the velocities of two lobes of bipolar outflows are symmetrical.
However, no significant difference between the lobes from bipolar and monopolar outflows was found in properties of dynamic timescale.
The kinetic-energy distributions, which exhibit significant differences between the 4 sub-groups,
are also presented in Figure~\ref{Fig:outphy1}, which are plotted in different colors.
The mean values of each sub-group are also indicated in each panel.
These findings are similar to the statistical results in Gemini OB1 \citep{2018ApJS..235...15L}
and W3/4/5 complex \citep{2019ApJS..242...19L}, implying that the components of bipolar outflows are
more massive, and thus more easily detectable.

\subsection{Star formation activity in the CMa OB1 complex}\label{Section5.2}

\begin{figure}[!ht]
\centering
\includegraphics[width=0.45\textwidth]{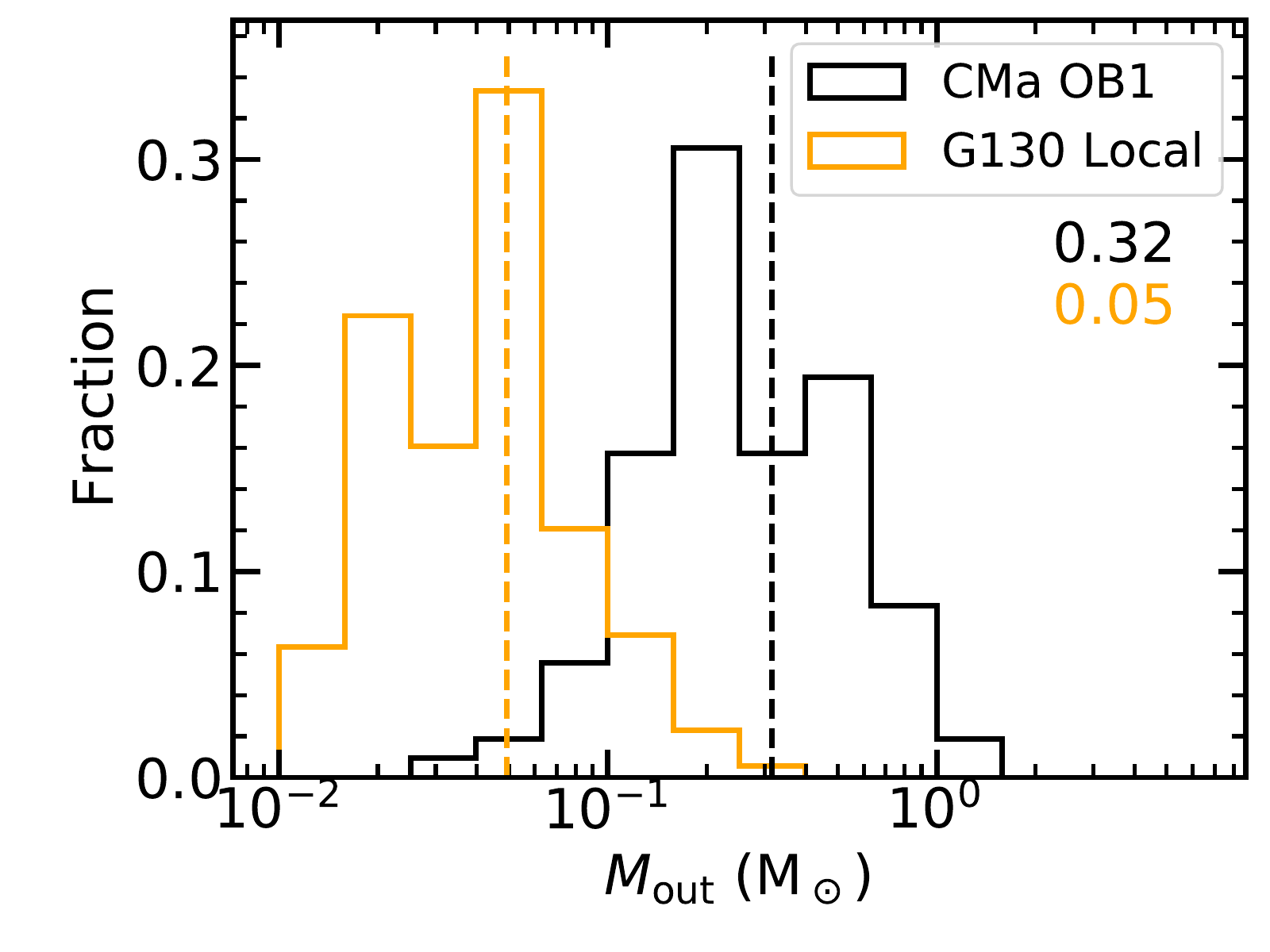}
\caption{Histograms of the mass of the outflow candidates in the CMa OB1 complex, and in G130 local MCs \citep[][]{2019ApJS..242...19L}}.
\label{Fig:outphy2}
\end{figure}

Histograms of the outflow masses for the CMa OB1 complex are presented in Figure~\ref{Fig:outphy2}.
For comparison, we have also overlaid a histogram of the local MCs in the G130 region (as mentioned in \S\ref{Section3.2}), which have a mean distance of $\sim$600~pc.~\citep[][]{2019ApJS..242...19L,2020ApJS..246....7S}.
The outflow masses range from 0.01 to 1.19 M$_\odot$, with an average value of 0.32~M$_\odot$ for the entire CMa OB1 complex,
which obviously have higher masses than G130 local MCs.
Since the two studies using data with uniform quality and the same identification method,
the outflow census may indicate that the CMa OB1 complex represents an active star-forming region in the Local arm.
Considering that the classification of an outflow is performed based on its mass~\citep[see][]{2018ApJS..235....3Y}, we suggest that the detected outflows in this study are likely to be dominated by intermediate- or low-mass outflows.

To quantify the level of star-formation activity in different sub-regions, and for the whole region, we
further carried out a statistical analysis. Herein, we have defined the number of outflows per unit area ($\eta_{\rm out}$) and per unit mass ($\zeta_{\rm out}$),
as $\eta_{\rm out}=N_{\rm out}$/$A_{\rm ^{12}CO}$, and $\zeta_{\rm out}=N_{\rm out}$/$M_{\rm ^{12}CO}$, respectively, where $N_{\rm out}$ is the number of outflow candidates, and $A_{\rm ^{12}CO}$ and $M_{\rm ^{12}CO}$ are the same parameters as defined in \S\ref{Section3.2}. Similarly, we define the outflow mass fraction, $\mathbb{M}_{\rm 12}^{\rm out}$ ($\mathbb{M}_{\rm 12}^{\rm out}$=$M_{\rm out}/M_{\rm ^{12}CO}$),
as the ratio of the mass of outflow and molecular gas derived from $^{12}$CO.
All these parameters ($\eta_{\rm out}$, $\zeta_{\rm out}$, and $M_{\rm out}^{\rm 12}$),
may offer valuable insights to the level of star-formation activity within MCs.
The derived parameters for the CMa OB1 complex and its sub-regions, and the G130 Local MCs from literature,
are listed in Table~\ref{table:odist}.

The results show that the value of$\eta_{\rm out}$ corresponding to the whole CMa OB1 complex is almost the same as that of the relatively
more nearby G130 local MCs \citep[0.037~pc$^{\rm -2}$;][]{2019ApJS..242...19L,2020ApJS..246....7S}.
The values of $\eta_{\rm out}$ of sub-regions A and B are similar and both larger than that of sub-region C.
Meanwhile, sub-region B possesses the largest number of outflows per unit mass.
As expected, sub-region C has the lowest value of $M_{\rm out}^{\rm 12}$, while the remaining two sub-regions
have similar values. All these strongly suggest that star-formation in sub-regions A and B are obviously more active relative
to that in sub-region C from the view of outflow activities.
Since star formation is thought to be closely related to the amount of dense gas present, we suggest
that significant variation of the mass ratio $M_\mathrm{C^{18}O}$/$M_\mathrm{^{12}CO}$ (refer to Table~\ref{tab:ratio})
may largely contribute to the outflow discrepancy across different sub-regions (refer to Table~\ref{table:odist}).

For our mapped region, sub-regions A and C are outside the \textit{spitzer/Herschel} fields. And in sub-region B, a total of 293 YSOs were
identified by \cite{2019ApJS..240...26S}, which resulted in the number YSOs per unit area
($\eta_{Y}$) of 0.31 pc$^{-2}$.
We further defined a conversion factor $C^{\rm Y}_{\rm O}$ here
(defined as the ratio $\eta_{\rm YSO}$/$\eta_{\rm out}$),
which is estimated to be $\sim$7.8 for sub-region B.
Since YSOs are always associated with outflows in nearby star-forming regions~\citep{2010MNRAS.408.1516C},
this conversion factor might reflect the relative completeness of the YSOs and/or outflows.
The disagreement between YSOs and outflows may indicate that we are likely to identify clustered
outflows as single or extended lobes, and/or fail to see the faint outflows.
Toward the more nearby MCs, e.g., the Perseus molecular cloud complex, the $C^{Y}_{O}$ with value of $\sim$
5.3$=$1.05/0.20~\citep[distance 250~pc;][]{2010ApJ...715.1170A}, and the Taurus molecular cloud with value of $\sim$
6.5$=$1.36/0.21~\citep[distance 140~pc;][]{2015ApJS..219...20L}, are both very
similar to that of the CMa OB1 complex. This may suggest that MWISP data with improved sensitivity are therefore able
to detect outflows at a distance of 1150~pc comparable to nearby Perseus and Taurus.

\section{Correlation between the compact \xxco clumps and the outflow candidates}\label{Section6}

If a \xxco clump overlapped with a monopolar outflow candidate (both spatially and in terms of velocity), we assume that this \xxco clump are correlated with the monopolar outflow candidate. And for bipolar outflow candidate, due to the central position of bipolar outflow candidate may easily confirm the parent MC, only the central position of bipolar outflow candidate located in \xxco clump, we assume that this \xxco clump is correlated with the bipolar outflow candidate. Consequently, we find 47 \xxco clumps (57\%) correlate with the outflow candidates. The correlation between the \xxco clumps and the outflow candidates are also presented in column 15 of Table~\ref{Appendix.A}.

Table~\ref{table:clump_out} presents the physical parameters of the dense \xxco clumps with and without associated outflow candidates. We find that the \xxco clumps with associated outflow candidates (i.e., the first sub-group) showed higher masses than the \xxco clumps without associated outflow candidates (i.e., the second sub-group). There are no significant differences in the other properties (i.e., $dv$, $T_\mathrm{ex}$, $\sigma_{\rm NT}$/$\sigma_{\rm T}$, $R_\mathrm{eff}$, $\alpha_\mathrm{vir}$) between the two categories.

There are 37 outflow candidates (44\%) correlated with the \xxco clumps, including 16 bipolar outflow candidates (70\%), 11 blue (33\%) and 10 red (34\%) monopolar ones. Similarly, the statistical outflow properties of the outflow candidates with and without associated \xxco clumps are presented in Table~\ref{table:out_clump}. The lobe candidates with associated \xxco clumps have slightly higher velocities, masses, kinetic energies, and luminosities than the lobes without associated \xxco clumps. However, the differences are still within their respective standard deviation (i.e., 1.31 \kms, 0.23 M$_\sun$, 6.66E+43 erg and 1.73E+32 erg s$^{-1}$, respectively).

\section{Summary}\label{Section7}
Using $^{\rm 12}$CO/$^{\rm 13}$CO/C$^{18}$O data from the ongoing MWISP project, we have reported a study of
the CMa OB1 complex (with sky coverage of 16.5 deg$^2$ in total), including a detail analysis of large-scale distributions
and properties of molecular gas, and diagnosis of the \xxco clumps and CO molecular outflows.
In this study, the CMa OB1 complex was roughly divided into three sub-regions according to the morphology of \xco emission.
The main results and conclusions of this study are summarized as follows:

(1) The total H$_{2}$ masses of the mapped area traced by the $^{12}$CO, $^{13}$CO, and \xxco emission were estimated to be 8.5 $\times\ 10^4$, 6.5 $\times\ 10^4$, and 6.6 $\times\ 10^3~\mathrm{M}_{\sun}$, respectively.
A large fraction of the observed molecular gas is emitted from regions with low excitation conditions.
A comparative study of the physical properties of the CMa OB1 complex reveals evidence for significant varying excitation temperature,
H$_2$ column density, surface gas density, $A_{\rm 18}^{\rm 12}$, and $M_{\rm 18}^{\rm 12}$ in different sub-regions.
The mean abundance ratio of $X_{\rm ^{13}CO}$/$X_{\rm C^{18}O}$ was estimated to be 14.9 for the entire mapped region,
which is much larger than that of the solar system value (5.5). Additionally, this ratio seemed to vary in different sub-regions, indicating different evolutionary stages of MCs in the CMa OB1 complex.

(2) A total of 83 \xxco clumps are identified using GaussClumps algorithm.
A non-thermal broadening mechanism was found to play an important role in the detected \xxco clumps.
We found about 94\% of the clumps are
gravitationally bound and may collapse to form stars. The size--mass relation of the clumps indicates that
the \xxco clumps in the CMa OB1 complex are unlikely to form high-mass stars. The power index of the \xxco clump mass
function, $N(\log M) \propto M^{\alpha}$, is found to be --1.33$\pm$0.18,
which is similar to that of the \textit{Hi-GAL} pre-stellar sources in the CMa OB1
complex~\citep[$\alpha$=--1.0$\pm$0.2,][]{2013ApJ...772...45E}, and also resembles the initial mass distribution
function in the Milk Way.

(3) A total of 85 outflow candidates were identified using a semi-automatic IDL algorithm, of which all were new detections, and 23 were bipolar outflow candidates. The total outflow mass of the mapped area was estimated to be 34.1~M$_{\sun}$.
Our outflow census indicated that the CMa OB1 complex represents an active star-forming region in the Local arm.
Our comparative study reveals that the level of star formation activity in the CMa OB1 complex seems to be closely related to the amount of relatively denser gas traced by C$^{18}$O.


\acknowledgments

We would like to thank all the staff members of Qinghai Radio Observing Station at Delingha for their help
during the observations.
We would like to thank the anonymous referee for the helpful comments and suggestions that helped to improve the paper.
This work is supported by the National Key Research \& Development Program of China (grant No. 2017YFA0402700),
Key Research Program of Frontier Sciences, CAS (grant No. QYZDJ-SSW-SLH047), the National Natural Science Foundation of
China (grant Nos. 11773077, 11933011, 11873019, 11673066, 11803091, and 11629302), the Key Laboratory for Radio Astronomy, CAS,
the Youth Innovation Promotion Association, CAS (2018355), and the ERC Advanced Investigator Grant GLOSTAR (247078).

\clearpage
\onecolumngrid

\vspace{-1cm}

\begin{deluxetable}{lccccccccccc}[!ht]
\tablecolumns{12}
\tabletypesize{\footnotesize}
\tablecaption{Observed properties of each region\label{table:cloud1}}
\tablehead{
\colhead{Observed properties} & \multicolumn{3}{c}{Sub-region A}& \colhead{} & \multicolumn{3}{c}{Sub-region B} & \colhead{} & \colhead{} & \colhead{Sub-region C} & \colhead{} \\ \cline{2-4} \cline{6-8} \cline{10-12}
\colhead{} & \colhead{\co} & \colhead{\xco} & \colhead{\xxco} & \colhead{} & \colhead{\co} & \colhead{\xco} & \colhead{\xxco} & \colhead{} & \colhead{\co} & \colhead{\xco} & \colhead{\xxco}
}
\startdata
$V_{\rm LSR}$ (km\,s$^{-1}$)      & 17.2 & 17.2 & 16.9 & & 15.3 & 15.5 & 14.6 & & 16.2 & 16.3 & 16.6 \\
$\Delta V$ (km\,s$^{-1}$)         &  3.1 &  2.6 &  2.5\tablenotemark{$\dagger$} & &  3.8 &  3.1 &  2.1 & &  3.6 &  2.8 &  3.2\tablenotemark{$\dagger$} \\
$A$ ($10^2$ arcmin$^{2}$)     &  54.7 & 22.5 & 1.2 & &  84.5 &  31.1 & 1.5 & &  52.0 & 23.5 & 0.3 \\
$I$ (K\,km\,s$^{-1}$\,arcmin$^{-2}$)& 8.0E4& 9.7E3& 1.2E2& & 7.9E4& 9.6E3& 2.1E2& & 4.7E4& 5.3E3& 0.2E2\\
\enddata
\tablecomments{Rows 1--2: Central velocity, and corresponding FWHM line width of the averaged $^{12}$CO/$^{13}$CO/\xxco spectra in each sub-region. Note that the averaged spectra of \xxco in sub-regions A and C, are not Gaussian-like, thus the FWHM line widths marked with ``$\dagger$" may be overestimated. Rows 3--4: Total area and total integrated intensity traced by $^{12}$CO/$^{13}$CO/\xxco in each sub-region.}
\end{deluxetable}


\begin{deluxetable}{lcccccccccc}[!ht]
\tablecolumns{11}
\tabletypesize{\normalsize}
\tablecaption{Statistics of excitation temperature, log[$N_\mathrm{H_2}$], mass, and surface density.\label{table:cloud2}}
\tablehead{
\colhead{Sub-region} & \colhead{Mask}  & \colhead{} & \colhead{$T_\mathrm{ex}$} & \colhead{} & \colhead{} & \colhead{} & \colhead{log[$N_\mathrm{H_2}$]}& \colhead{}  & \colhead{$M$} & \colhead{$\Sigma$}\\ \cline{3-5} \cline{7-9}
\colhead{} & \colhead{Region} & \colhead{min} & \colhead{max} & \colhead{mean} & \colhead{} & \colhead{min} & \colhead{max} & \colhead{mean} & \colhead{} & \colhead{} \\
\colhead{} & \colhead{} & \multicolumn{3}{c}{(K)} & \colhead{} & \multicolumn{3}{c}{(cm$^{-2}$)} & \colhead{($\times$10$^{3}$ M$_\odot$)} & \colhead{(M$_\odot$ pc$^{-2}$)}
}
\startdata
A & B   & 3.5 & 30.9 & 9.9 &  & 20.1 & 22.3 & 21.2 & 32.9 & 53.9\\
         & G  & 4.1 & 30.9 &14.7 &  & 20.2 & 22.7 & 21.3 & 28.3 &113.2\\
         & R & 9.6 & 29.1 &21.1 &  & 21.2 & 22.7 & 21.8 &  2.7 &192.9\\
\hline
B & B   & 3.4 & 18.8 & 7.4 &  & 19.7 & 22.1 & 21.0 & 32.5 & 34.6\\
         & G  & 4.1 & 18.8 & 9.6 &  & 20.2 & 22.4 & 21.1 & 23.7 & 67.7\\
         & R & 8.1 & 18.7 &11.1 &  & 21.1 & 22.5 & 21.7 &  3.6 &200.2\\
\hline
C & B   & 3.3 & 21.7 & 7.1 &  & 20.1 & 22.0 & 21.1 & 19.4 & 32.9\\
         & G  & 4.1 & 21.7 & 8.6 &  & 20.2 & 22.2 & 21.1 & 13.0 & 48.1\\
         & R & 6.4 & 14.7 & 9.7 &  & 21.2 & 22.0 & 21.4 &  0.3 &110.8\\
\enddata
\end{deluxetable}

\begin{deluxetable}{lllll}[!ht]
\tabletypesize{\small}
\setlength{\tabcolsep}{0.04in}
\tablewidth{0pt}
\tablecaption{Area and mass ratios.\label{tab:ratio}}
\tablehead{
Sub-region & $\mathbb{A}_{12}^{\rm 13}$ & $\mathbb{M}_{12}^{\rm 13}$ & $\mathbb{A}_{12}^{\rm 18}$
& $\mathbb{M}_{\rm 12}^{\rm 18}$  \\
& (\%) & (\%) & (\%) & (\%)
            }
\startdata
A & 41.2 & 86.0 & 2.2 & 8.2 \\
B & 36.8 & 72.9 &1.8 & 11.1\\
C & 45.3 & 67.0 & 0.5 & 1.5 \\
\enddata
\tablecomments{Refer to \S\ref{Section3.2} for the definations of the area and mass ratios.}
\end{deluxetable}

\begin{deluxetable}{lcccccccccccccccc}[!ht]
\tablecolumns{17}
\setlength{\tabcolsep}{0.03in}
\tabletypesize{\scriptsize}
\tablecaption{Statistics of outflow properties for the four sub-groups\label{table:outphy}}
\tablehead{
\colhead{Sub-group} & \colhead{Number}& \colhead{}  & \multicolumn{2}{c}{$\langle\Delta v_\mathrm{lobe}\rangle$} & \colhead{}  & \multicolumn{2}{c}{$M_\mathrm{lobe}$} & \colhead{}  & \multicolumn{2}{c}{$E_\mathrm{lobe}$} & \colhead{}  & \multicolumn{2}{c}{$t_\mathrm{lobe}$} & \colhead{}  & \multicolumn{2}{c}{$L_\mathrm{lobe}$}\\ \cline{4-5} \cline{7-8} \cline{10-11} \cline{13-14}  \cline{16-17}
\colhead{} & \colhead{} & \colhead{} & \colhead{mean} & \colhead{median} & \colhead{} & \colhead{mean} & \colhead{median} & \colhead{} & \colhead{mean} & \colhead{median} & \colhead{} & \colhead{mean} & \colhead{median} & \colhead{} & \colhead{mean} & \colhead{median}\\
\colhead{} & \colhead{}& \colhead{}  & \multicolumn{2}{c}{(km s$^{-1}$)} & \colhead{}  & \multicolumn{2}{c}{(M$_\odot$)} & \colhead{}  & \multicolumn{2}{c}{(erg)} & \colhead{}  & \multicolumn{2}{c}{(yr)} & \colhead{}  & \multicolumn{2}{c}{(erg s$^{-1}$)}
}
\startdata
1 & 29  &  &  3.24  &  2.93  &  & 0.20  & 0.16  &  & 2.88E+43 & 1.15E+43 &  & 2.20E+05 & 1.92E+05 &  & 7.04E+30 & 2.22E+30 \\
2 & 23  &  &  4.26  &  4.17  &  & 0.29  & 0.22  &  & 6.42E+43 & 4.36E+43 &  & 1.68E+05 & 1.25E+05 &  & 1.77E+31 & 9.27E+30 \\
3 & 33  &  & -3.39  & -3.26  &  & 0.31  & 0.26  &  & 4.75E+43 & 2.27E+43 &  & 2.72E+05 & 2.26E+05 &  & 9.06E+30 & 4.00E+30 \\
4 & 23  &  & -4.13  & -4.08  &  & 0.50  & 0.43  &  & 9.63E+43 & 8.75E+43 &  & 2.23E+05 & 1.78E+05 &  & 1.66E+31 & 1.08E+31 \\
\enddata
\tablecomments{Sub-groups 1--4 respectively represent: the (1) 29 red-lobes from monopolar outflows, (2) 23 red-lobes from bipolar outflows, (3) 33 blue-lobes from monopolar outflows, and (4) 23 blue-lobes from bipolar outflows.}
\end{deluxetable}

\begin{deluxetable}{lccccccc}[!ht]
\tablecolumns{8}
\setlength{\tabcolsep}{0.03in}
\tabletypesize{\scriptsize}
\tablecaption{Statistics of outflow candidates\label{table:odist}}
\tablehead{
\colhead{Region} & \colhead{Monopolar}  & \colhead{Bipolar} & \colhead{$N_{\rm out}$} &\colhead{$M_{\rm out}$}  & \colhead{$\eta_{\rm out}$} & \colhead{$\zeta_{\rm out}$} & \colhead{$\mathbb{M}_{\rm 12}^{\rm out}$}   \\
\colhead{} & \colhead{} & \colhead{} & \colhead{} & \colhead{} & \colhead{$\times10^{\rm -2}$} & \colhead{$\times$10$^{\rm -4}$} & \colhead{$\times$10$^{\rm -4}$}    \\
\colhead{} & \colhead{(B/R)} & \colhead{} & \colhead{} & \colhead{(M$_\sun$)}   &\colhead{(pc$^{\rm -2}$)} & \colhead{(M$_\sun^{-1}$)}   & \colhead{}
}
\startdata
sub-region A     & 9/9   & 11 & 29  &13.2 & 4.7 &  8.8  & 4.01 \\
sub-region B     & 16/12 & 9 & 37  &15.1 & 3.9 & 11.4  & 4.65 \\
sub-region C     & 8/8   &  3 & 19  &5.8  & 3.2 &  9.7  & 2.99 \\
whole region & 33/29 & 23 & 85  &34.1 & 3.7 & 10.0  & 4.02 \\
G130 Local   & 55/95 & 12 & 162 &17.8 & 3.7 & 17.1  & 1.80 \\
\enddata
\tablecomments{Refer to \S\ref{Section5.2} for the definitions of each column. Please refer to \cite{2019ApJS..242...19L}
and \cite{2020ApJS..246....7S} for the statistical results for the local MCs in the G130 region ($l$=[129\fdg75, 140\fdg25]
and $b$=[$-$5\fdg25, +5\fdg25]).}
\end{deluxetable}

\begin{deluxetable}{lccccccccccccccccccc}[!ht]
\tablecolumns{20}
\setlength{\tabcolsep}{0.03in}
\tabletypesize{\scriptsize}
\tablecaption{Statistical parameters of the \xxco clumps with and without associated outflow candidates. \label{table:clump_out}}
\tablehead{
\colhead{clump} & \colhead{number}& \colhead{}  & \multicolumn{2}{c}{$dv$} & \colhead{}  & \multicolumn{2}{c}{$T_\mathrm{ex}$} & \colhead{}  & \multicolumn{2}{c}{$\sigma_\mathrm{NT}$/$\sigma_\mathrm{T}$} & \colhead{}  & \multicolumn{2}{c}{$R_\mathrm{ref}$} & \colhead{}  & \multicolumn{2}{c}{$M_\mathrm{LTE}$} & \colhead{}  & \multicolumn{2}{c}{$\alpha_\mathrm{vir}$}\\ \cline{4-5} \cline{7-8} \cline{10-11} \cline{13-14}  \cline{16-17} \cline{19-20}
\colhead{} & \colhead{} & \colhead{} & \colhead{mean} & \colhead{median} & \colhead{} & \colhead{mean} & \colhead{median} & \colhead{} & \colhead{mean} & \colhead{median} & \colhead{} & \colhead{mean} & \colhead{median} & \colhead{} & \colhead{mean} & \colhead{median} & \colhead{} & \colhead{mean} & \colhead{median}\\
\colhead{} & \colhead{}& \colhead{}  & \multicolumn{2}{c}{(km s$^{-1}$)} & \colhead{}  & \multicolumn{2}{c}{(K)} & \colhead{}  & \multicolumn{2}{c}{} & \colhead{}  & \multicolumn{2}{c}{(pc)} & \colhead{}  & \multicolumn{2}{c}{(M$_\odot$)} & \colhead{}  & \multicolumn{2}{c}{}
}
\startdata
Correlated   & 47  &  & 0.34  & 0.32  &  & 15.26  & 13.38  &  & 1.46  & 1.35  &  & 0.64  & 0.60  &  & 126  & 86  &  & 0.96  & 0.80 \\
Uncorrelated & 36  &  & 0.31  & 0.28  &  & 14.45  & 13.15  &  & 1.32  & 1.17  &  & 0.54  & 0.50  &  &  70  & 49  &  & 1.08  & 0.92 \\
Total        & 83  &  & 0.33  & 0.31  &  & 14.90  & 13.38  &  & 1.40  & 1.30  &  & 0.59  & 0.53  &  & 102  & 68  &  & 1.01  & 0.85 \\
\enddata
\end{deluxetable}

\begin{deluxetable}{lccccccccccccccccccc}[!ht]
\tablecolumns{20}
\setlength{\tabcolsep}{0.03in}
\tabletypesize{\scriptsize}
\tablecaption{Statistical parameters of the lobe candidates with and without associated \xxco clumps.\label{table:out_clump}}
\tablehead{
\colhead{lobe} & \colhead{number}& \colhead{}  & \multicolumn{2}{c}{$\langle\Delta v_\mathrm{lobe}\rangle$} & \colhead{}  & \multicolumn{2}{c}{$M_\mathrm{lobe}$} & \colhead{}  & \multicolumn{2}{c}{$P_\mathrm{lobe}$} & \colhead{}  & \multicolumn{2}{c}{$E_\mathrm{lobe}$} & \colhead{}  & \multicolumn{2}{c}{$t_\mathrm{lobe}$} & \colhead{}  & \multicolumn{2}{c}{$L_\mathrm{lobe}$}\\ \cline{4-5} \cline{7-8} \cline{10-11} \cline{13-14}  \cline{16-17} \cline{19-20}
\colhead{} & \colhead{} & \colhead{} & \colhead{mean} & \colhead{median} & \colhead{} & \colhead{mean} & \colhead{median} & \colhead{} & \colhead{mean} & \colhead{median} & \colhead{} & \colhead{mean} & \colhead{median} & \colhead{} & \colhead{mean} & \colhead{median} & \colhead{} & \colhead{mean} & \colhead{median}\\
\colhead{(B/R)} & \colhead{}& \colhead{}  & \multicolumn{2}{c}{(km s$^{-1}$)} & \colhead{}  & \multicolumn{2}{c}{(M$_\odot$)} & \colhead{}  & \multicolumn{2}{c}{(M$_\odot$ km s$^{-1}$)} & \colhead{}  & \multicolumn{2}{c}{(erg)} & \colhead{}  & \multicolumn{2}{c}{(yr)} & \colhead{}  & \multicolumn{2}{c}{(erg s$^{-1}$)}
}
\startdata
Correlated B   & 27  &  & -3.87  & -3.86  &  & 0.44  & 0.37  &  & 1.82  & 1.71  &  & 7.96E+43 & 6.43E+43 &  & 2.28E+05 & 2.02E+05 &  & 1.41E+31 & 8.23E+30 \\
Uncorrelated B & 29  &  & -3.52  & -3.46  &  & 0.34  & 0.24  &  & 1.31  & 0.78  &  & 5.63E+43 & 2.68E+43 &  & 2.74E+05 & 2.47E+05 &  & 1.04E+31 & 3.76E+30 \\
Total B        & 56  &  & -3.69  & -3.69  &  & 0.39  & 0.32  &  & 1.56  & 1.00  &  & 6.75E+43 & 3.68E+43 &  & 2.51E+05 & 2.12E+05 &  & 1.22E+31 & 6.45E+30 \\ \cline{1-20}
Correlated R   & 26  &  &  4.44  &  4.22  &  & 0.32  & 0.25  &  & 1.50  & 1.01  &  & 7.60E+43 & 4.42E+43 &  & 1.69E+05 & 1.25E+05 &  & 2.11E+31 & 1.08E+31 \\
Uncorrelated R & 26  &  &  2.93  &  2.78  &  & 0.15  & 0.14  &  & 0.42  & 0.39  &  & 1.29E+43 & 1.05E+43 &  & 2.26E+05 & 1.97E+05 &  & 2.42E+30 & 1.55E+30 \\
Total R        & 52  &  &  3.69  &  3.32  &  & 0.24  & 0.18  &  & 0.96  & 0.58  &  & 4.45E+43 & 2.04E+43 &  & 1.97E+05 & 1.75E+05 &  & 1.17E+31 & 2.63E+30 \\
\enddata
\end{deluxetable}

\break
\clearpage
\onecolumngrid
\appendix
\section{The physical properties of \xxco clumps.}\label{Appendix.A}
\renewcommand{\thetable}{A}
\startlongtable
\begin{deluxetable}{lcccccccccccccl}
\setlength\tabcolsep{3pt}
\tabletypesize{\footnotesize}
\tablecolumns{15}
\tablecaption{The physical properties of \xxco clumps\label{table:clumps}}
\tablehead{
 \colhead{Clump} & \colhead{$V_\mathrm{LSR}$}& \colhead{$dv$}& \colhead{$T_{\mathrm{C}^{18}\mathrm{O}}$} & \colhead{$T_{^{12}\mathrm{CO}}$} & \colhead{$A$}  & \colhead{$T_\mathrm{ex}$} & \colhead{$\tau$} & \colhead{$\sigma_\mathrm{T}$} & \colhead{$\sigma_\mathrm{NT}$} & \colhead{$R_\mathrm{eff}$} & \colhead{$M_\mathrm{LTE}$} & \colhead{$M_\mathrm{vir}$} & \colhead{$\alpha_\mathrm{vir}$} & \colhead{Outflow Index}\\
 \colhead{Name} & \colhead{(km\,s$^{-1}$)}& \colhead{(km\,s$^{-1}$)} & \colhead{(K)} & \colhead{(K)} & \colhead{(arcmin$^2$)}  & \colhead{(K)} &  \colhead{} & \colhead{(km\,s$^{-1}$)} & \colhead{(km\,s$^{-1}$)} & \colhead{(pc)} & \colhead{(M$_\odot$)} & \colhead{(M$_\odot$)} & \colhead{} & \colhead{} \\
 \colhead{(1)} & \colhead{(2)} & \colhead{(3)} & \colhead{(4)} & \colhead{(5)} & \colhead{(6)} & \colhead{(7)} & \colhead{(8)} & \colhead{(9)} & \colhead{(10)} & \colhead{(11)} & \colhead{(12)} & \colhead{(13)} & \colhead{(14)} & \colhead{(15)}
}
\startdata
G223.483-01.767  & 18.78 & 0.33 & 2.00 & 14.73 & 10.11 & 18.17 & 0.15 & 0.25 & 0.32 & 0.58 & 131 &  72 & 0.55 & \\
G223.517-01.783  & 18.66 & 0.28 & 2.19 & 14.93 &  7.31 & 18.38 & 0.16 & 0.25 & 0.27 & 0.49 &  90 &  45 & 0.50 & 8,16 \\
G223.533-01.858  & 18.78 & 0.27 & 1.38 & 18.18 &  6.20 & 21.64 & 0.08 & 0.27 & 0.25 & 0.44 &  42 &  37 & 0.88 & 7 \\
G223.542-00.942  & 16.44 & 0.17 & 1.81 &  5.62 &  7.31 &  8.92 & 0.39 & 0.18 & 0.16 & 0.49 &  28 &  16 & 0.58 \\
G223.592-00.883  & 15.94 & 0.25 & 1.34 &  6.81 &  5.94 & 10.15 & 0.22 & 0.19 & 0.25 & 0.43 &  22 &  32 & 1.45 \\
G223.592-00.917  & 16.27 & 0.20 & 1.34 &  6.83 &  9.13 & 10.16 & 0.22 & 0.19 & 0.19 & 0.55 &  27 &  25 & 0.91 & 10 \\
G223.692-00.825  & 13.28 & 0.20 & 1.35 &  7.20 &  8.50 & 10.54 & 0.21 & 0.19 & 0.19 & 0.53 &  33 &  24 & 0.75 \\
G223.733-01.908  & 17.27 & 0.34 & 1.87 & 16.82 & 11.87 & 20.28 & 0.12 & 0.27 & 0.33 & 0.63 & 140 &  84 & 0.60 & 16\\
G223.758-01.875  & 17.27 & 0.33 & 1.73 & 16.42 & 12.99 & 19.87 & 0.11 & 0.26 & 0.33 & 0.66 & 149 &  86 & 0.58 & 16\\
G223.825-01.850  & 17.45 & 0.41 & 1.58 & 13.67 &  9.13 & 17.11 & 0.12 & 0.24 & 0.40 & 0.55 &  90 & 107 & 1.19 & 19,20\\
G223.875-01.858  & 18.11 & 0.72 & 1.27 & 18.72 & 23.77 & 22.19 & 0.07 & 0.28 & 0.71 & 0.91 & 298 & 540 & 1.82 & 19,20\\
G223.900-01.875  & 17.91 & 0.29 & 1.75 & 18.82 &  8.81 & 22.29 & 0.10 & 0.28 & 0.28 & 0.54 &  87 &  51 & 0.59 \\
G223.917-01.150  & 15.61 & 0.26 & 1.48 &  9.45 & 12.99 & 12.84 & 0.17 & 0.21 & 0.25 & 0.66 &  75 &  52 & 0.70 \\
G223.942-01.117  & 15.11 & 0.24 & 1.29 & 10.16 &  5.94 & 13.55 & 0.13 & 0.22 & 0.23 & 0.43 &  29 &  28 & 0.95 & 22\\
G223.975-01.833  & 17.94 & 0.43 & 2.07 & 19.82 &  7.60 & 23.30 & 0.11 & 0.28 & 0.42 & 0.50 & 164 & 109 & 0.66 & 25\\
G224.000-01.683  & 17.27 & 0.23 & 1.45 & 19.22 &  8.19 & 22.69 & 0.08 & 0.28 & 0.22 & 0.52 &  51 &  33 & 0.66 & 27\\
G224.008-01.708  & 16.79 & 0.33 & 1.98 & 14.96 &  7.60 & 18.41 & 0.14 & 0.25 & 0.32 & 0.50 &  87 &  61 & 0.70 & 26,27\\
G224.083-01.133  & 16.09 & 0.29 & 1.63 & 10.52 &  6.74 & 13.92 & 0.17 & 0.22 & 0.29 & 0.47 &  49 &  47 & 0.96 & 29,30\\
G224.117-01.100  & 15.31 & 0.35 & 2.13 &  9.51 &  8.50 & 12.90 & 0.25 & 0.21 & 0.34 & 0.53 &  99 &  74 & 0.75 & 30\\
G224.142-01.100  & 16.78 & 0.46 & 1.34 &  6.01 &  7.31 &  9.32 & 0.25 & 0.18 & 0.46 & 0.49 &  45 & 120 & 2.67 \\
G224.150-01.075  & 14.78 & 0.44 & 1.55 & 10.07 &  7.60 & 13.46 & 0.17 & 0.22 & 0.43 & 0.50 &  68 & 111 & 1.63 \\
G224.150-01.742  & 18.30 & 0.39 & 1.80 & 13.30 &  5.44 & 16.74 & 0.14 & 0.24 & 0.38 & 0.41 &  60 &  71 & 1.18 \\
G224.150-01.958  & 18.46 & 0.25 & 1.90 & 18.97 &  5.69 & 22.45 & 0.11 & 0.28 & 0.24 & 0.42 &  53 &  32 & 0.60 & 31,32\\
G224.167-01.758  & 19.61 & 0.55 & 1.50 & 16.54 &  8.50 & 20.00 & 0.10 & 0.26 & 0.54 & 0.53 &  93 & 183 & 1.97 \\
G224.183-01.050  & 14.60 & 0.35 & 1.70 & 10.37 &  7.60 & 13.77 & 0.18 & 0.22 & 0.35 & 0.50 &  68 &  73 & 1.07 \\
G224.192-00.933  & 13.11 & 0.21 & 1.41 &  8.12 &  7.60 & 11.48 & 0.19 & 0.20 & 0.20 & 0.50 &  27 &  25 & 0.90 \\
G224.217-01.733  & 20.62 & 0.35 & 1.35 &  9.05 &  5.94 & 12.43 & 0.16 & 0.21 & 0.35 & 0.43 &  29 &  63 & 2.13 & 35\\
G224.217-01.958  & 16.94 & 0.28 & 1.49 & 14.13 &  5.19 & 17.57 & 0.11 & 0.25 & 0.27 & 0.40 &  42 &  37 & 0.89 \\
G224.242-00.800  & 14.45 & 0.26 & 1.44 &  7.60 & 10.80 & 10.95 & 0.21 & 0.19 & 0.26 & 0.60 &  53 &  48 & 0.91 \\
G224.242-01.050  & 14.79 & 0.41 & 2.77 &  9.98 & 21.26 & 13.38 & 0.32 & 0.22 & 0.41 & 0.86 & 317 & 169 & 0.53 & 34,39\\
G224.250-00.933  & 13.61 & 0.31 & 1.47 &  8.56 & 15.80 & 11.94 & 0.19 & 0.20 & 0.30 & 0.73 & 107 &  82 & 0.77 & 37\\
G224.258-00.842  & 15.60 & 0.27 & 1.47 &  7.90 &  3.64 & 11.26 & 0.20 & 0.20 & 0.26 & 0.33 &  27 &  27 & 1.00 \\
G224.258-00.858  & 14.09 & 0.58 & 3.61 &  8.59 & 43.89 & 11.96 & 0.54 & 0.20 & 0.58 & 1.24 & 979 & 487 & 0.50 & 38\\
G224.267-01.900  & 17.28 & 0.33 & 1.27 & 15.84 & 12.24 & 19.29 & 0.08 & 0.26 & 0.32 & 0.64 & 103 &  80 & 0.78 \\
G224.275-00.742  & 14.12 & 0.25 & 1.39 &  7.57 &  8.19 & 10.92 & 0.20 & 0.19 & 0.24 & 0.52 &  32 &  38 & 1.18 \\
G224.275-00.867  & 15.12 & 0.33 & 1.42 &  8.52 &  7.60 & 11.89 & 0.18 & 0.20 & 0.33 & 0.50 &  48 &  65 & 1.34 & 38\\
G224.283-01.875  & 18.94 & 0.56 & 1.30 & 18.27 &  8.19 & 21.73 & 0.07 & 0.27 & 0.55 & 0.52 & 124 & 190 & 1.52 \\
G224.292-01.842  & 19.11 & 0.27 & 1.27 & 18.17 & 11.50 & 21.64 & 0.07 & 0.27 & 0.26 & 0.62 &  71 &  52 & 0.74 \\
G224.292-01.967  & 16.95 & 0.30 & 1.29 & 16.05 & 11.15 & 19.50 & 0.08 & 0.26 & 0.29 & 0.61 &  51 &  65 & 1.28 & 41\\
G224.300-00.817  & 14.26 & 0.48 & 2.41 &  8.13 & 41.12 & 11.50 & 0.35 & 0.20 & 0.48 & 1.20 & 460 & 324 & 0.70 \\
G224.300-00.892  & 13.95 & 0.49 & 1.43 &  8.33 &  3.44 & 11.70 & 0.19 & 0.20 & 0.49 & 0.31 &  40 &  90 & 2.24 & 38\\
G224.317-01.042  & 14.95 & 0.37 & 1.41 &  9.87 &  8.81 & 13.26 & 0.15 & 0.21 & 0.37 & 0.54 &  64 &  87 & 1.37 & 45\\
G224.317-01.942  & 16.95 & 0.21 & 1.43 & 12.35 & 12.24 & 15.78 & 0.12 & 0.23 & 0.20 & 0.64 &  51 &  34 & 0.67 & 41\\
G224.325-00.842  & 15.28 & 0.18 & 1.39 &  7.79 &  7.31 & 11.14 & 0.20 & 0.20 & 0.17 & 0.49 &  26 &  17 & 0.67 \\
G224.325-00.850  & 13.60 & 0.33 & 1.39 &  8.95 &  5.19 & 12.32 & 0.17 & 0.21 & 0.32 & 0.40 &  37 &  50 & 1.34 \\
G224.342-00.758  & 14.62 & 0.30 & 1.78 &  7.47 & 32.16 & 10.82 & 0.27 & 0.19 & 0.29 & 1.06 & 208 & 110 & 0.53 & 42\\
G224.342-01.025  & 15.94 & 0.44 & 2.20 & 10.06 & 18.43 & 13.45 & 0.25 & 0.22 & 0.43 & 0.80 & 197 & 177 & 0.90 & 45\\
G224.342-02.142  & 16.62 & 0.32 & 2.20 & 19.73 &  8.50 & 23.21 & 0.12 & 0.28 & 0.31 & 0.53 & 123 &  63 & 0.51 & 43\\
G224.350-02.008  & 16.91 & 0.46 & 2.85 & 17.10 & 15.38 & 20.56 & 0.18 & 0.27 & 0.45 & 0.72 & 370 & 176 & 0.48 & 44\\
G224.358-00.808  & 14.62 & 0.28 & 1.32 &  8.44 &  6.47 & 11.81 & 0.17 & 0.20 & 0.27 & 0.45 &  35 &  41 & 1.19 \\
G224.358-02.042  & 16.95 & 0.47 & 1.72 & 20.99 & 10.45 & 24.47 & 0.09 & 0.29 & 0.46 & 0.59 & 158 & 148 & 0.93 & 44\\
G224.375-00.475  & 14.12 & 0.24 & 1.39 &  4.94 &  7.02 &  8.22 & 0.33 & 0.17 & 0.24 & 0.48 &  28 &  32 & 1.15 \\
G224.392-02.033  & 16.64 & 0.22 & 1.79 & 16.37 & 20.29 & 19.83 & 0.12 & 0.26 & 0.21 & 0.84 & 138 &  49 & 0.36 \\
G224.400-00.692  & 13.44 & 0.49 & 1.27 &  9.06 &  8.19 & 12.44 & 0.15 & 0.21 & 0.49 & 0.52 &  40 & 144 & 3.61 & 47,51\\
G224.400-02.008  & 17.27 & 0.20 & 1.56 & 19.23 &  5.19 & 22.71 & 0.08 & 0.28 & 0.18 & 0.40 &  42 &  18 & 0.44 \\
G224.408-00.742  & 15.94 & 0.42 & 2.04 &  8.21 & 16.65 & 11.57 & 0.28 & 0.20 & 0.42 & 0.75 & 186 & 158 & 0.85 & 47\\
G224.425-00.800  & 15.28 & 0.26 & 1.44 & 10.23 & 10.80 & 13.63 & 0.15 & 0.22 & 0.25 & 0.60 &  57 &  46 & 0.81 \\
G224.433-00.700  & 15.77 & 0.23 & 1.73 &  9.59 &  8.50 & 12.98 & 0.20 & 0.21 & 0.22 & 0.53 &  49 &  32 & 0.65 & 47,51\\
G224.450-00.683  & 15.10 & 0.27 & 1.45 &  7.30 & 24.82 & 10.65 & 0.22 & 0.19 & 0.27 & 0.93 &  98 &  81 & 0.83 & 51,52,57\\
G224.458-00.592  & 15.77 & 0.26 & 1.50 &  7.64 & 12.61 & 10.99 & 0.22 & 0.20 & 0.26 & 0.65 &  68 &  53 & 0.79 & 50\\
G224.467-02.042  & 17.61 & 0.47 & 1.82 & 17.16 & 10.80 & 20.62 & 0.11 & 0.27 & 0.46 & 0.60 & 190 & 152 & 0.80 & 54\\
G224.467-02.400  &  9.94 & 0.33 & 1.50 & 21.91 &  8.19 & 25.40 & 0.07 & 0.30 & 0.32 & 0.52 &  76 &  65 & 0.86 & 49\\
G224.475-00.683  & 16.24 & 0.45 & 2.09 &  7.47 & 15.80 & 10.82 & 0.33 & 0.19 & 0.44 & 0.73 & 152 & 170 & 1.11 & 51,52,57\\
G224.483-00.642  & 15.60 & 0.31 & 1.66 &  7.32 & 31.56 & 10.66 & 0.26 & 0.19 & 0.31 & 1.05 & 228 & 120 & 0.53 & 51,52\\
G224.525-02.033  & 17.40 & 0.24 & 2.73 & 15.38 &  9.13 & 18.83 & 0.19 & 0.26 & 0.22 & 0.55 & 105 &  35 & 0.34 & 54\\
G224.533-00.575  & 15.44 & 0.27 & 1.47 &  7.06 & 18.89 & 10.40 & 0.23 & 0.19 & 0.27 & 0.81 &  83 &  69 & 0.83 & 58\\
G224.600-00.492  & 15.44 & 0.18 & 1.31 &  6.07 & 23.26 &  9.38 & 0.24 & 0.18 & 0.17 & 0.90 &  76 &  32 & 0.42 \\
G224.608-01.000  & 16.93 & 0.49 & 1.57 & 14.10 &  5.94 & 17.54 & 0.12 & 0.25 & 0.48 & 0.43 &  86 & 119 & 1.39 & 59\\
G224.633-01.033  & 16.94 & 0.48 & 1.44 & 13.38 &  5.19 & 16.81 & 0.11 & 0.24 & 0.48 & 0.40 &  63 & 109 & 1.72 \\
G224.767-01.750  & 15.60 & 0.41 & 2.40 & 17.18 &  7.02 & 20.65 & 0.15 & 0.27 & 0.40 & 0.48 & 142 &  93 & 0.65 \\
G224.783-01.817  & 15.79 & 0.30 & 1.44 & 12.24 &  5.69 & 15.66 & 0.13 & 0.23 & 0.29 & 0.42 &  36 &  43 & 1.21 \\
G224.792-01.733  & 15.27 & 0.44 & 1.64 & 19.49 &  4.95 & 22.97 & 0.09 & 0.28 & 0.43 & 0.39 &  69 &  90 & 1.30 \\
G224.908-01.250  & 12.61 & 0.24 & 1.30 &  9.28 &  7.89 & 12.66 & 0.15 & 0.21 & 0.23 & 0.51 &  36 &  33 & 0.92 \\
G224.917-01.233  & 13.27 & 0.42 & 1.40 & 10.13 &  7.60 & 13.52 & 0.15 & 0.22 & 0.41 & 0.50 &  40 & 100 & 2.53 \\
G225.225-01.100  & 13.94 & 0.18 & 1.30 &  7.82 & 13.37 & 11.18 & 0.18 & 0.20 & 0.17 & 0.67 &  42 &  26 & 0.61 & 68\\
G225.883-00.433  & 15.78 & 0.28 & 1.28 &  5.64 &  6.74 &  8.94 & 0.26 & 0.18 & 0.27 & 0.47 &  25 &  42 & 1.65 \\
G226.125-00.583  & 18.60 & 0.17 & 1.25 &  3.99 & 11.87 &  7.23 & 0.37 & 0.16 & 0.17 & 0.63 &  30 &  22 & 0.72 \\
G226.192-00.400  & 15.61 & 0.28 & 1.28 &  5.15 & 25.35 &  8.43 & 0.28 & 0.17 & 0.28 & 0.94 &  72 &  89 & 1.23 & 78\\
G226.292-00.608  & 16.45 & 0.21 & 1.36 &  5.70 &  8.19 &  9.00 & 0.27 & 0.18 & 0.20 & 0.52 &  32 &  26 & 0.80 & 81\\
G226.308-00.650  & 17.61 & 0.32 & 1.26 &  7.40 & 14.16 & 10.75 & 0.19 & 0.19 & 0.31 & 0.69 &  65 &  81 & 1.24 & 81\\
G226.333-00.683  & 17.93 & 0.31 & 1.45 &  5.73 & 11.15 &  9.04 & 0.29 & 0.18 & 0.31 & 0.61 &  56 &  69 & 1.24 & 81\\
G226.408-00.767  & 15.45 & 0.20 & 1.57 &  6.88 & 11.87 & 10.21 & 0.26 & 0.19 & 0.19 & 0.63 &  53 &  30 & 0.56 \\
G226.442-00.600  & 15.44 & 0.34 & 1.98 & 11.27 & 11.15 & 14.69 & 0.19 & 0.23 & 0.34 & 0.61 & 121 &  85 & 0.70 & 84
\enddata
\vspace{0.5cm}
\tabletypesize{\normalsize}
Note--
\begin{itemize}
  \item[--] Column (1): Name of each clump, as defined by its central Galactic Coordinates.
  \item[--] Columns (2)--(3): Central velocity and velocity dispersion of each clump.
  \item[--] Column (4): Angular area of each clump, estimated by GaussClumps.
  \item[--] Column (5): Peak main beam temperature of the \xxco emission peak fitted by GaussClumps.
  \item[--] Column (6): Main beam temperature of \co peak within the same area of clumps in \co data.
  \item[--] Columns (7)-(14): Excitation temperature, optical depth, thermal velocity dispersion, non-thermal velocity dispersion, effective radius, LTE mass, virial mass, and virial parameter of each clump (see details in \S\ref{Section4.2})
  \item[--] Column (15): Index of associated Outflow candidate (Table~\ref{Appendix.B},~\ref{Appendix.C} col. 1). 

\end{itemize}
\end{deluxetable}
\onecolumngrid
\section{The Outflow Candidates}\label{Appendix.B}
\renewcommand{\thetable}{B}
\startlongtable
\begin{deluxetable}{lccccccc}
\setlength\tabcolsep{3pt}
\tabletypesize{\footnotesize}
\tablecolumns{8}
\tablecaption{The outflow candidates.\label{table:outflow1}}
\tablehead{
\colhead{Index} & \colhead{$l$} & \colhead{$b$} & \colhead{bw0} & \colhead{bw1} & \colhead{rw0} & \colhead{rw1} & \colhead{tag}\\
\colhead{} & \colhead{(deg)} & \colhead{(deg)} & \colhead{(km\,s$^{-1}$)} & \colhead{(km\,s$^{-1}$)} & \colhead{(km\,s$^{-1}$)} & \colhead{(km\,s$^{-1}$)} & \colhead{}\\
\colhead{(1)} & \colhead{(2)} & \colhead{(3)} & \colhead{(4)} & \colhead{(5)} & \colhead{(6)} & \colhead{(7)} & \colhead{(8)}
}
\startdata
 1 & 222.468 & -0.391 & 14.1 & 15.2 & \nodata & \nodata & B \\
 2 & 223.084 & -1.934 & \nodata & \nodata & 18.6 & 20.2 & R \\
 3 & 223.254 & -1.874 & 15.1 & 16.6 & \nodata & \nodata & B \\
 4 & 223.313 & -1.794 & 13.8 & 16.3 & 20.1 & 21.7 & D \\
 5 & 223.434 & -1.823 & \nodata & \nodata & 20.6 & 24.0 & R \\
 6 & 223.436 & -0.924 & \nodata & \nodata & 18.4 & 21.1 & R \\
 7 & 223.503 & -1.904 & 12.8 & 14.6 & \nodata & \nodata & B \\
 8 & 223.513 & -1.774 & \nodata & \nodata & 20.3 & 21.6 & R \\
 9 & 223.590 & -1.863 & 13.4 & 15.5 & \nodata & \nodata & B \\
10 & 223.616 & -0.928 & \nodata & \nodata & 17.6 & 19.2 & R \\
11 & 223.629 & -2.016 & 11.6 & 15.5 & \nodata & \nodata & B \\
12 & 223.643 & -0.733 & 10.0 & 11.8 & 15.3 & 17.1 & D \\
13 & 223.646 & -1.885 & 12.6 & 14.9 & 19.8 & 22.4 & D \\
14 & 223.742 & -0.773 & \nodata & \nodata & 15.9 & 17.1 & R \\
15 & 223.745 & -0.859 & \nodata & \nodata & 14.7 & 17.4 & R \\
16 & 223.749 & -1.870 & 13.0 & 15.1 & 19.4 & 22.0 & D \\
17 & 223.835 & -1.113 & 12.1 & 14.4 & \nodata & \nodata & B \\
18 & 223.840 & -1.527 & 14.6 & 16.8 & 21.1 & 22.3 & D \\
19 & 223.856 & -1.829 & 12.3 & 14.8 & 20.3 & 21.7 & D \\
20 & 223.859 & -1.884 & 12.1 & 15.7 & \nodata & \nodata & B \\
21 & 223.866 & -0.773 & \nodata & \nodata & 14.2 & 17.3 & R \\
22 & 223.903 & -1.090 & 12.0 & 13.7 & \nodata & \nodata & B \\
23 & 223.909 & -1.600 & \nodata & \nodata & 20.4 & 21.7 & R \\
24 & 223.991 & -1.898 & 14.6 & 15.8 & \nodata & \nodata & B \\
25 & 223.991 & -1.821 & 11.8 & 15.9 & 20.1 & 23.9 & D \\
26 & 224.008 & -1.732 & 12.1 & 15.7 & 18.9 & 21.3 & D \\
27 & 224.010 & -1.681 & 11.7 & 15.2 & 19.4 & 23.5 & D \\
28 & 224.022 & -0.916 & \nodata & \nodata & 14.8 & 17.4 & R \\
29 & 224.063 & -1.165 & 10.8 & 13.6 & \nodata & \nodata & B \\
30 & 224.108 & -1.129 & 10.8 & 13.7 & \nodata & \nodata & B \\
31 & 224.109 & -1.954 & \nodata & \nodata & 20.6 & 23.3 & R \\
32 & 224.146 & -1.998 & \nodata & \nodata & 20.6 & 24.7 & R \\
33 & 224.149 & -0.863 & 8.0 & 10.4 & \nodata & \nodata & B \\
34 & 224.222 & -1.041 & \nodata & \nodata & 18.4 & 23.7 & R \\
35 & 224.227 & -1.725 & \nodata & \nodata & 22.1 & 24.8 & R \\
36 & 224.256 & -2.209 & 14.8 & 16.4 & \nodata & \nodata & B \\
37 & 224.259 & -0.951 & 8.0 & 12.1 & \nodata & \nodata & B \\
38 & 224.272 & -0.839 & 6.9 & 11.0 & 19.3 & 24.1 & D \\
39 & 224.276 & -1.066 & 7.3 & 11.0 & 19.4 & 25.1 & D \\
40 & 224.317 & -2.050 & 12.1 & 13.9 & \nodata & \nodata & B \\
41 & 224.317 & -1.953 & \nodata & \nodata & 20.2 & 22.7 & R \\
42 & 224.330 & -0.747 &  8.5 & 12.0 & 18.6 & 24.4 & D \\
43 & 224.340 & -2.143 & 10.5 & 15.4 & 19.4 & 24.9 & D \\
44 & 224.350 & -2.017 &  9.7 & 13.6 & 20.1 & 24.7 & D \\
45 & 224.354 & -1.030 &  8.6 & 13.9 & 18.4 & 21.7 & D \\
46 & 224.356 & -0.915 &  6.7 & 12.5 & \nodata & \nodata & B \\
47 & 224.407 & -0.721 &  7.1 & 11.8 & 19.4 & 22.4 & D \\
48 & 224.415 & -0.616 & 10.1 & 13.4 & \nodata & \nodata & B \\
49 & 224.422 & -2.351 &  5.8 & 7.6 & 11.4 & 14 & D \\
50 & 224.435 & -0.559 & 10.3 & 12.6 & \nodata & \nodata & B \\
51 & 224.469 & -0.671 & \nodata & \nodata & 19.5 & 22.4 & R \\
52 & 224.492 & -0.662 & 8.3 & 13.4 & 19.4 & 23.1 & D \\
53 & 224.500 & -0.417 & 10.1 & 13.3 & \nodata & \nodata & B \\
54 & 224.508 & -2.034 & 14.1 & 15.8 & \nodata & \nodata & B \\
55 & 224.526 & -0.879 & 12.3 & 13.5 & \nodata & \nodata & B \\
56 & 224.526 & -0.308 & \nodata & \nodata & 17.3 & 19.1 & R \\
57 & 224.551 & -1.820 & \nodata & \nodata & 20.9 & 23.4 & R \\
58 & 224.561 & -0.581 & 10.3 & 13.1 & 17.6 & 22.0 & D \\
59 & 224.594 & -0.984 & \nodata & \nodata & 20.3 & 26.7 & R \\
60 & 224.607 & -1.151 & 12.9 & 15.9 & \nodata & \nodata & B \\
61 & 224.633 & -0.805 & \nodata & \nodata & 17.8 & 20.9 & R \\
62 & 224.685 & -0.466 & 9.8 & 13.6 & \nodata & \nodata & B \\
63 & 224.706 & -0.426 & 9.3 & 13.4 & \nodata & \nodata & B \\
64 & 224.717 & -0.997 & 12.3 & 16.8 & 19.4 & 21.8 & D \\
65 & 224.833 & -0.419 & \nodata & \nodata & 19.4 & 20.7 & R \\
66 & 225.109 & -0.283 & 12.1 & 14.9 & \nodata & \nodata & B \\
67 & 225.229 & -0.974 & \nodata & \nodata & 15.8 & 17.2 & R \\
68 & 225.246 & -1.110 & 10.0 & 12.6 & \nodata & \nodata & B \\
69 & 225.373 & -1.034 & \nodata & \nodata & 16.2 & 17.7 & R \\
70 & 225.417 & -0.377 & 12.5 & 14.3 & \nodata & \nodata & B \\
71 & 225.533 & -1.739 & 13.4 & 14.6 & \nodata & \nodata & B \\
72 & 225.726 & -0.332 & \nodata & \nodata & 19.0 & 21.2 & R \\
73 & 225.792 & -0.910 & \nodata & \nodata & 16.3 & 17.7 & R \\
74 & 225.859 & -0.799 & \nodata & \nodata & 16.1 & 18.9 & R \\
75 & 225.907 & -0.293 & \nodata & \nodata & 18.1 & 19.7 & R \\
76 & 226.120 & -0.343 & 11.1 & 13.4 & 19.3 & 20.6 & D \\
77 & 226.122 & -0.443 & 9.6 & 13.1 & \nodata & \nodata & B \\
78 & 226.170 & -0.408 & 11.0 & 13.8 & \nodata & \nodata & B \\
79 & 226.225 & -0.615 & 12.6 & 15.8 & \nodata & \nodata & B \\
80 & 226.283 & -0.466 & 6.0 & 13.1 & \nodata & \nodata & B \\
81 & 226.319 & -0.637 & 13.8 & 15.3 & \nodata & \nodata & B \\
82 & 226.323 & -0.507 & 9.3 & 12.1 & 18.3 & 22.2 & D \\
83 & 226.356 & -0.758 & \nodata & \nodata & 18.7 & 22.7 & R \\
84 & 226.427 & -0.584 & 8.5 & 12.8 & 19.6 & 23.8 & D \\
85 & 226.511 & -0.447 & \nodata & \nodata & 18.6 & 21.6 & R
\enddata
\vspace{0.5cm}
\tabletypesize{\normalsize}
Note--
\begin{itemize}
  \item [--] Column (1): Index of outflow candidate.
  \item [--] Column (2)-(3): Central Galactic coordinates of outflow candidate.
  \item [--] Columns (4)-(5): Velocity range of the blue line wing.
  \item [--] Columns (6)-(7): Velocity range of the red line wing.
  \item [--] Column (8): Type of outflow candidate, where ``B", ``R", and ``D" denote blue-, red-monopolar, and bipolar outflow candidates, respectively.
\end{itemize}
\end{deluxetable}
\onecolumngrid
\section{Parameters of Outflow Candidates}\label{Appendix.C}
\renewcommand{\thetable}{C}
\startlongtable
\begin{deluxetable}{lccccccccccccc}
\setlength\tabcolsep{3pt}
\tabletypesize{\scriptsize}
\tablecolumns{14}
\tablecaption{Physical properties of the Outflow Candidates\label{table:outflow2}}
\tablehead{
\colhead{Index} & \colhead{Lobe} & \colhead{$l$} & \colhead{$b$} & \colhead{$v_\mathrm{peak}$}  & \colhead{$\psi$} & \colhead{$A_\mathrm{lobe}$} & \colhead{$\langle\Delta v_\mathrm{lobe}\rangle$} & \colhead{$l_\mathrm{lobe}$}& \colhead{$M_\mathrm{lobe}$} & \colhead{$P_\mathrm{lobe}$} & \colhead{$E_\mathrm{lobe}$} & \colhead{$t_\mathrm{lobe}$} & \colhead{$L_\mathrm{lobe}$} \\
\colhead{} & \colhead{} & \colhead{(deg)} & \colhead{(deg)} & \colhead{(km\,s$^{-1}$)} & \colhead{} & \colhead{(arcmin$^2$)} & \colhead{(km\,s$^{-1}$)} & \colhead{(pc)} & \colhead{(M$_\odot$)} & \colhead{(M$_\odot$ km\,s$^{-1}$)} & \colhead{(erg)} & \colhead{(yr)} & \colhead{(erg s$^{-1}$)} \\
 \colhead{(1)} & \colhead{(2)} & \colhead{(3)} & \colhead{(4)} & \colhead{(5)} & \colhead{(6)} & \colhead{(7)} & \colhead{(8)} & \colhead{(9)} & \colhead{(10)} & \colhead{(11)} & \colhead{(12)} & \colhead{(13)} & \colhead{(14)}
}
\startdata
1  & B & 222.468  & -0.391  & 16.16  & 0.75  & 1.47 & -1.62  & 0.72  & 3.66E-02 & 5.94E-02 & 9.54E+41 & 3.41E+05 & 8.87E+28 \\
2  & R & 223.084  & -1.934  & 17.60  & 0.75  & 2.17 &  1.64  & 0.87  & 3.45E-01 & 5.65E-01 & 9.18E+42 & 3.28E+05 & 8.87E+29 \\
3  & B & 223.254  & -1.874  & 17.82  & 0.60  & 7.33 & -2.12  & 1.60  & 4.43E-01 & 9.40E-01 & 1.98E+43 & 5.76E+05 & 1.09E+30 \\
4  & B & 223.302  & -1.826  & 17.80  & 0.50  & 4.39 & -3.00  & 1.24  & 2.06E-01 & 6.20E-01 & 1.84E+43 & 3.03E+05 & 1.93E+30 \\
   & R & 223.324  & -1.763  & 18.12  & 0.50  & 2.11 &  2.62  & 0.86  & 1.29E-01 & 3.37E-01 & 8.76E+42 & 2.35E+05 & 1.18E+30 \\
5  & R & 223.434  & -1.823  & 18.10  & 0.50  & 1.28 &  3.86  & 0.67  & 1.55E-01 & 5.97E-01 & 2.29E+43 & 1.11E+05 & 6.53E+30 \\
6  & R & 223.436  & -0.924  & 15.39  & 0.50  & 2.64 &  4.09  & 0.96  & 1.77E-01 & 7.24E-01 & 2.94E+43 & 1.65E+05 & 5.65E+30 \\
7  & B & 223.503  & -1.904  & 16.69  & 0.50  & 5.25 & -3.17  & 1.36  & 2.67E-01 & 8.47E-01 & 2.66E+43 & 3.41E+05 & 2.48E+30 \\
8  & R & 223.513  & -1.774  & 18.61  & 0.50  & 3.08 &  2.21  & 1.04  & 1.59E-01 & 3.51E-01 & 7.67E+42 & 3.41E+05 & 7.14E+29 \\
9  & B & 223.590  & -1.863  & 17.27  & 0.60  & 4.28 & -3.03  & 1.22  & 4.30E-01 & 1.30E+00 & 3.91E+43 & 3.10E+05 & 4.00E+30 \\
10 & R & 223.616  & -0.928  & 16.24  & 0.75  & 2.17 &  2.00  & 0.87  & 1.35E-01 & 2.70E-01 & 5.37E+42 & 2.88E+05 & 5.92E+29 \\
11 & B & 223.629  & -2.016  & 17.77  & 0.60  & 4.67 & -4.61  & 1.28  & 4.75E-01 & 2.19E+00 & 9.98E+43 & 2.03E+05 & 1.56E+31 \\
12 & B & 223.626  & -0.710  & 13.81  & 0.50  & 4.64 & -3.09  & 1.28  & 1.83E-01 & 5.66E-01 & 1.74E+43 & 3.27E+05 & 1.68E+30 \\
   & R & 223.660  & -0.756  & 13.82  & 0.75  & 1.64 &  2.20  & 0.76  & 8.61E-02 & 1.89E-01 & 4.13E+42 & 2.26E+05 & 5.80E+29 \\
13 & B & 223.658  & -1.893  & 17.60  & 0.50  & 5.69 & -4.08  & 1.41  & 5.30E-01 & 2.16E+00 & 8.75E+43 & 2.76E+05 & 1.00E+31 \\
   & R & 223.634  & -1.877  & 17.47  & 0.50  & 2.22 &  3.37  & 0.88  & 1.23E-01 & 4.16E-01 & 1.39E+43 & 1.75E+05 & 2.51E+30 \\
14 & R & 223.742  & -0.773  & 13.60  & 0.50  & 2.14 &  2.78  & 0.87  & 5.50E-02 & 1.53E-01 & 4.22E+42 & 2.42E+05 & 5.53E+29 \\
15 & R & 223.745  & -0.859  & 12.97  & 0.85  & 1.81 &  2.81  & 0.80  & 1.92E-01 & 5.40E-01 & 1.50E+43 & 1.76E+05 & 2.71E+30 \\
16 & B & 223.737  & -1.856  & 17.08  & 0.50  & 9.11 & -3.24  & 1.79  & 1.07E+00 & 3.46E+00 & 1.11E+44 & 4.28E+05 & 8.23E+30 \\
   & R & 223.761  & -1.885  & 16.99  & 0.60  & 2.61 &  3.45  & 0.96  & 1.82E-01 & 6.28E-01 & 2.15E+43 & 1.87E+05 & 3.64E+30 \\
17 & B & 223.835  & -1.113  & 16.01  & 0.90  & 1.44 & -2.99  & 0.71  & 1.02E-01 & 3.04E-01 & 9.03E+42 & 1.78E+05 & 1.61E+30 \\
18 & B & 223.827  & -1.521  & 18.28  & 0.50  & 4.47 & -2.80  & 1.25  & 4.23E-01 & 1.18E+00 & 3.29E+43 & 3.32E+05 & 3.14E+30 \\
   & R & 223.853  & -1.533  & 19.22  & 0.50  & 1.89 &  2.36  & 0.81  & 1.80E-01 & 4.23E-01 & 9.89E+42 & 2.59E+05 & 1.21E+30 \\
19 & B & 223.857  & -1.814  & 17.25  & 0.50  & 3.28 & -3.95  & 1.07  & 2.95E-01 & 1.17E+00 & 4.56E+43 & 2.12E+05 & 6.83E+30 \\
   & R & 223.855  & -1.845  & 18.30  & 0.50  & 5.06 &  2.56  & 1.33  & 3.96E-01 & 1.01E+00 & 2.57E+43 & 3.83E+05 & 2.12E+30 \\
20 & B & 223.859  & -1.884  & 17.34  & 0.75  & 4.17 & -3.80  & 1.21  & 4.49E-01 & 1.71E+00 & 6.43E+43 & 2.25E+05 & 9.04E+30 \\
21 & R & 223.866  & -0.773  & 12.29  & 0.80  & 2.08 &  3.15  & 0.85  & 1.35E-01 & 4.26E-01 & 1.33E+43 & 1.67E+05 & 2.53E+30 \\
22 & B & 223.903  & -1.090  & 14.72  & 0.60  & 3.36 & -2.04  & 1.09  & 2.55E-01 & 5.20E-01 & 1.05E+43 & 3.90E+05 & 8.55E+29 \\
23 & R & 223.909  & -1.600  & 19.37  & 0.50  & 3.14 &  1.55  & 1.05  & 1.18E-01 & 1.82E-01 & 2.80E+42 & 4.41E+05 & 2.01E+29 \\
24 & B & 223.991  & -1.898  & 16.90  & 0.50  & 4.83 & -1.82  & 1.30  & 3.90E-01 & 7.09E-01 & 1.28E+43 & 5.54E+05 & 7.32E+29 \\
25 & B & 223.986  & -1.810  & 17.58  & 0.50  & 1.86 & -4.14  & 0.81  & 4.88E-01 & 2.02E+00 & 8.29E+43 & 1.37E+05 & 1.92E+31 \\
   & R & 223.996  & -1.833  & 17.56  & 0.50  & 1.75 &  4.06  & 0.78  & 2.67E-01 & 1.08E+00 & 4.36E+43 & 1.21E+05 & 1.14E+31 \\
26 & B & 224.007  & -1.720  & 17.03  & 0.55  & 1.78 & -3.49  & 0.79  & 3.02E-01 & 1.05E+00 & 3.64E+43 & 1.57E+05 & 7.37E+30 \\
   & R & 224.010  & -1.743  & 17.27  & 0.75  & 2.00 &  2.59  & 0.84  & 1.86E-01 & 4.81E-01 & 1.24E+43 & 2.03E+05 & 1.93E+30 \\
27 & B & 224.008  & -1.687  & 16.94  & 0.50  & 1.58 & -3.84  & 0.75  & 1.65E-01 & 6.33E-01 & 2.41E+43 & 1.39E+05 & 5.48E+30 \\
   & R & 224.013  & -1.676  & 16.77  & 0.50  & 1.83 &  4.27  & 0.80  & 1.89E-01 & 8.06E-01 & 3.41E+43 & 1.17E+05 & 9.27E+30 \\
28 & R & 224.022  & -0.916  & 12.58  & 0.95  & 1.39 &  3.26  & 0.70  & 1.06E-01 & 3.46E-01 & 1.12E+43 & 1.42E+05 & 2.50E+30 \\
29 & B & 224.063  & -1.165  & 14.78  & 0.75  & 5.42 & -2.86  & 1.38  & 8.02E-01 & 2.29E+00 & 6.49E+43 & 3.39E+05 & 6.07E+30 \\
30 & B & 224.108  & -1.129  & 15.35  & 0.70  & 1.75 & -3.39  & 0.78  & 1.99E-01 & 6.73E-01 & 2.26E+43 & 1.68E+05 & 4.25E+30 \\
31 & R & 224.109  & -1.954  & 18.08  & 0.50  & 2.14 &  3.60  & 0.87  & 1.50E-01 & 5.41E-01 & 1.93E+43 & 1.62E+05 & 3.77E+30 \\
32 & R & 224.146  & -1.998  & 18.38  & 0.50  & 3.53 &  3.86  & 1.11  & 2.25E-01 & 8.71E-01 & 3.34E+43 & 1.72E+05 & 6.14E+30 \\
33 & B & 224.149  & -0.863  & 13.45  & 0.60  & 3.06 & -4.49  & 1.04  & 1.72E-01 & 7.72E-01 & 3.43E+43 & 1.86E+05 & 5.86E+30 \\
34 & R & 224.222  & -1.041  & 14.70  & 0.70  & 1.50 &  5.82  & 0.73  & 4.16E-01 & 2.42E+00 & 1.40E+44 & 7.89E+04 & 5.61E+31 \\
35 & R & 224.227  & -1.725  & 20.81  & 0.50  & 5.39 &  2.37  & 1.37  & 4.25E-01 & 1.01E+00 & 2.36E+43 & 3.37E+05 & 2.22E+30 \\
36 & B & 224.256  & -2.209  & 17.18  & 0.75  & 3.86 & -1.74  & 1.16  & 3.57E-01 & 6.20E-01 & 1.07E+43 & 4.79E+05 & 7.06E+29 \\
37 & B & 224.259  & -0.951  & 13.50  & 0.90  & 1.39 & -3.86  & 0.70  & 2.16E-01 & 8.33E-01 & 3.19E+43 & 1.24E+05 & 8.15E+30 \\
38 & B & 224.260  & -0.838  & 14.32  & 0.50  & 6.72 & -5.78  & 1.54  & 9.02E-01 & 5.21E+00 & 2.98E+44 & 2.02E+05 & 4.67E+31 \\
   & R & 224.284  & -0.840  & 13.97  & 0.50  & 3.39 &  7.25  & 1.09  & 3.94E-01 & 2.86E+00 & 2.06E+44 & 1.05E+05 & 6.19E+31 \\
39 & B & 224.277  & -1.072  & 14.57  & 0.50  & 3.83 & -5.79  & 1.16  & 4.79E-01 & 2.77E+00 & 1.59E+44 & 1.56E+05 & 3.24E+31 \\
   & R & 224.275  & -1.059  & 14.80  & 0.50  & 2.14 &  6.88  & 0.87  & 4.52E-01 & 3.11E+00 & 2.12E+44 & 8.22E+04 & 8.17E+31 \\
40 & B & 224.317  & -2.050  & 16.47  & 0.75  & 1.47 & -3.65  & 0.72  & 1.09E-01 & 3.98E-01 & 1.44E+43 & 1.61E+05 & 2.85E+30 \\
41 & R & 224.317  & -1.953  & 17.07  & 0.50  & 3.19 &  4.13  & 1.06  & 2.37E-01 & 9.78E-01 & 4.01E+43 & 1.84E+05 & 6.91E+30 \\
42 & B & 224.328  & -0.752  & 14.78  & 0.75  & 1.56 & -4.88  & 0.74  & 2.28E-01 & 1.11E+00 & 5.37E+43 & 1.15E+05 & 1.48E+31 \\
   & R & 224.332  & -0.742  & 14.81  & 0.50  & 4.22 &  6.11  & 1.22  & 6.79E-01 & 4.15E+00 & 2.51E+44 & 1.24E+05 & 6.40E+31 \\
43 & B & 224.341  & -2.138  & 16.62  & 0.55  & 1.94 & -4.16  & 0.83  & 6.23E-01 & 2.59E+00 & 1.07E+44 & 1.32E+05 & 2.57E+31 \\
   & R & 224.338  & -2.148  & 16.68  & 0.50  & 2.25 &  4.92  & 0.89  & 5.57E-01 & 2.74E+00 & 1.34E+44 & 1.06E+05 & 4.01E+31 \\
44 & B & 224.354  & -2.016  & 16.79  & 0.50  & 2.25 & -5.53  & 0.89  & 3.68E-01 & 2.04E+00 & 1.12E+44 & 1.23E+05 & 2.89E+31 \\
   & R & 224.347  & -2.018  & 16.79  & 0.55  & 2.39 &  5.15  & 0.92  & 3.16E-01 & 1.63E+00 & 8.32E+43 & 1.13E+05 & 2.33E+31 \\
45 & B & 224.365  & -1.031  & 15.57  & 0.65  & 3.31 & -4.85  & 1.08  & 7.15E-01 & 3.47E+00 & 1.67E+44 & 1.51E+05 & 3.50E+31 \\
   & R & 224.342  & -1.030  & 15.55  & 0.50  & 4.50 &  4.17  & 1.26  & 2.75E-01 & 1.15E+00 & 4.74E+43 & 2.00E+05 & 7.51E+30 \\
46 & B & 224.356  & -0.915  & 16.11  & 0.80  & 3.61 & -7.09  & 1.13  & 6.32E-01 & 4.48E+00 & 3.14E+44 & 1.17E+05 & 8.51E+31 \\
47 & B & 224.415  & -0.710  & 14.78  & 0.50  & 5.56 & -5.80  & 1.40  & 6.93E-01 & 4.02E+00 & 2.31E+44 & 1.78E+05 & 4.12E+31 \\
   & R & 224.399  & -0.732  & 15.26  & 0.60  & 2.39 &  5.34  & 0.92  & 1.60E-01 & 8.56E-01 & 4.53E+43 & 1.25E+05 & 1.15E+31 \\
48 & B & 224.415  & -0.616  & 15.15  & 0.85  & 2.22 & -3.73  & 0.88  & 4.13E-01 & 1.54E+00 & 5.68E+43 & 1.71E+05 & 1.05E+31 \\
49 & B & 224.409  & -2.338  &  8.30  & 0.50  & 2.75 & -1.78  & 0.98  & 2.40E-01 & 4.28E-01 & 7.56E+42 & 3.84E+05 & 6.24E+29 \\
   & R & 224.435  & -2.363  &  9.56  & 0.50  & 4.11 &  2.88  & 1.20  & 9.03E-01 & 2.60E+00 & 7.44E+43 & 2.64E+05 & 8.92E+30 \\
50 & B & 224.435  & -0.559  & 15.13  & 0.65  & 6.22 & -3.91  & 1.48  & 5.32E-01 & 2.08E+00 & 8.05E+43 & 2.99E+05 & 8.53E+30 \\
51 & R & 224.454  & -0.680  & 15.02  & 0.50  & 7.53 &  5.64  & 1.62  & 5.29E-01 & 2.99E+00 & 1.67E+44 & 2.15E+05 & 2.46E+31 \\
52 & B & 224.483  & -0.663  & 15.17  & 0.55  & 2.11 & -4.83  & 0.86  & 4.28E-01 & 2.06E+00 & 9.87E+43 & 1.23E+05 & 2.55E+31 \\
   & R & 224.500  & -0.661  & 15.35  & 0.50  & 1.64 &  5.53  & 0.76  & 1.48E-01 & 8.17E-01 & 4.48E+43 & 9.57E+04 & 1.48E+31 \\
53 & B & 224.500  & -0.417  & 16.42  & 0.90  & 1.47 & -5.04  & 0.72  & 1.61E-01 & 8.10E-01 & 4.05E+43 & 1.11E+05 & 1.15E+31 \\
54 & B & 224.508  & -2.034  & 17.23  & 0.85  & 1.50 & -2.45  & 0.73  & 2.21E-01 & 5.42E-01 & 1.32E+43 & 2.26E+05 & 1.84E+30 \\
55 & B & 224.526  & -0.879  & 15.22  & 0.65  & 2.94 & -2.44  & 1.02  & 1.09E-01 & 2.67E-01 & 6.46E+42 & 3.40E+05 & 6.02E+29 \\
56 & R & 224.526  & -0.308  & 15.05  & 0.80  & 1.81 &  2.97  & 0.80  & 8.23E-02 & 2.44E-01 & 7.18E+42 & 1.92E+05 & 1.18E+30 \\
57 & R & 224.551  & -1.820  & 18.97  & 0.50  & 1.67 &  2.93  & 0.76  & 1.35E-01 & 3.96E-01 & 1.15E+43 & 1.69E+05 & 2.16E+30 \\
58 & B & 224.544  & -0.563  & 15.28  & 0.50  & 8.47 & -3.86  & 1.72  & 6.96E-01 & 2.68E+00 & 1.03E+44 & 3.39E+05 & 9.59E+30 \\
   & R & 224.579  & -0.599  & 15.28  & 0.85  & 1.75 &  4.08  & 0.78  & 2.23E-01 & 9.09E-01 & 3.68E+43 & 1.14E+05 & 1.02E+31 \\
59 & R & 224.594  & -0.984  & 17.12  & 0.50  & 1.89 &  5.74  & 0.81  & 4.20E-01 & 2.41E+00 & 1.37E+44 & 8.31E+04 & 5.23E+31 \\
60 & B & 224.607  & -1.151  & 17.36  & 0.65  & 2.14 & -3.26  & 0.87  & 1.31E-01 & 4.27E-01 & 1.38E+43 & 1.90E+05 & 2.30E+30 \\
61 & R & 224.633  & -0.805  & 16.27  & 0.50  & 6.19 &  2.77  & 1.47  & 3.56E-01 & 9.86E-01 & 2.71E+43 & 3.12E+05 & 2.75E+30 \\
62 & B & 224.685  & -0.466  & 15.61  & 0.65  & 1.69 & -4.29  & 0.77  & 2.04E-01 & 8.74E-01 & 3.71E+43 & 1.30E+05 & 9.07E+30 \\
63 & B & 224.706  & -0.426  & 15.77  & 0.50  & 7.64 & -4.83  & 1.64  & 8.35E-01 & 4.04E+00 & 1.93E+44 & 2.47E+05 & 2.48E+31 \\
64 & B & 224.719  & -0.993  & 17.82  & 0.55  & 2.50 & -3.72  & 0.94  & 4.13E-01 & 1.54E+00 & 5.66E+43 & 1.66E+05 & 1.08E+31 \\
   & R & 224.716  & -1.000  & 17.82  & 0.50  & 3.78 &  2.54  & 1.15  & 2.21E-01 & 5.61E-01 & 1.41E+43 & 2.83E+05 & 1.58E+30 \\
65 & R & 224.833  & -0.419  & 15.81  & 0.55  & 2.92 &  4.11  & 1.01  & 7.10E-02 & 2.92E-01 & 1.19E+43 & 2.02E+05 & 1.86E+30 \\
66 & B & 225.109  & -0.283  & 15.96  & 0.55  & 5.92 & -2.74  & 1.44  & 2.43E-01 & 6.68E-01 & 1.82E+43 & 3.65E+05 & 1.58E+30 \\
67 & R & 225.229  & -0.974  & 14.52  & 0.55  & 6.42 &  1.84  & 1.50  & 2.19E-01 & 4.03E-01 & 7.34E+42 & 5.48E+05 & 4.25E+29 \\
68 & B & 225.246  & -1.110  & 13.67  & 0.50  & 6.44 & -2.63  & 1.50  & 3.11E-01 & 8.17E-01 & 2.13E+43 & 4.00E+05 & 1.69E+30 \\
69 & R & 225.373  & -1.034  & 14.14  & 0.75  & 1.39 &  2.66  & 0.70  & 5.74E-02 & 1.53E-01 & 4.02E+42 & 1.92E+05 & 6.64E+29 \\
70 & B & 225.417  & -0.377  & 15.94  & 0.50  & 2.94 & -2.72  & 1.02  & 8.73E-02 & 2.37E-01 & 6.40E+42 & 2.89E+05 & 7.02E+29 \\
71 & B & 225.533  & -1.739  & 15.28  & 0.50  & 3.42 & -1.40  & 1.09  & 2.38E-01 & 3.32E-01 & 4.59E+42 & 5.71E+05 & 2.55E+29 \\
72 & R & 225.726  & -0.332  & 16.79  & 0.60  & 1.64 &  3.09  & 0.76  & 6.97E-02 & 2.15E-01 & 6.60E+42 & 1.68E+05 & 1.24E+30 \\
73 & R & 225.792  & -0.910  & 14.30  & 0.90  & 2.56 &  2.56  & 0.95  & 1.46E-01 & 3.74E-01 & 9.51E+42 & 2.72E+05 & 1.11E+30 \\
74 & R & 225.859  & -0.799  & 14.50  & 0.90  & 1.39 &  2.72  & 0.70  & 1.49E-01 & 4.06E-01 & 1.09E+43 & 1.55E+05 & 2.23E+30 \\
75 & R & 225.907  & -0.293  & 16.09  & 0.80  & 1.69 &  2.65  & 0.77  & 1.42E-01 & 3.77E-01 & 9.92E+42 & 2.09E+05 & 1.51E+30 \\
76 & B & 226.128  & -0.323  & 15.48  & 0.65  & 2.92 & -3.46  & 1.01  & 2.26E-01 & 7.82E-01 & 2.68E+43 & 2.26E+05 & 3.76E+30 \\
   & R & 226.112  & -0.362  & 15.59  & 0.95  & 1.69 &  4.23  & 0.77  & 6.83E-02 & 2.89E-01 & 1.21E+43 & 1.51E+05 & 2.55E+30 \\
77 & B & 226.122  & -0.443  & 15.53  & 0.50  & 4.69 & -4.53  & 1.28  & 5.11E-01 & 2.32E+00 & 1.04E+44 & 2.11E+05 & 1.56E+31 \\
78 & B & 226.170  & -0.408  & 15.70  & 0.85  & 2.78 & -3.58  & 0.99  & 2.65E-01 & 9.48E-01 & 3.36E+43 & 2.05E+05 & 5.19E+30 \\
79 & B & 226.225  & -0.615  & 17.34  & 0.85  & 1.67 & -3.46  & 0.76  & 1.91E-01 & 6.62E-01 & 2.27E+43 & 1.58E+05 & 4.57E+30 \\
80 & B & 226.283  & -0.466  & 15.07  & 0.65  & 1.61 & -6.23  & 0.75  & 3.37E-01 & 2.10E+00 & 1.30E+44 & 8.10E+04 & 5.08E+31 \\
81 & B & 226.319  & -0.637  & 16.70  & 0.75  & 1.92 & -2.30  & 0.82  & 1.62E-01 & 3.72E-01 & 8.46E+42 & 2.77E+05 & 9.69E+29 \\
82 & B & 226.320  & -0.516  & 14.59  & 0.50  & 9.44 & -4.17  & 1.82  & 1.19E+00 & 4.97E+00 & 2.05E+44 & 3.36E+05 & 1.94E+31 \\
   & R & 226.325  & -0.498  & 14.78  & 0.75  & 1.75 &  5.08  & 0.78  & 1.95E-01 & 9.92E-01 & 4.99E+43 & 1.03E+05 & 1.53E+31 \\
83 & R & 226.356  & -0.758  & 15.61  & 0.90  & 1.39 &  4.69  & 0.70  & 1.79E-01 & 8.38E-01 & 3.90E+43 & 9.63E+04 & 1.28E+31 \\
84 & B & 226.420  & -0.579  & 14.93  & 0.50  & 3.81 & -4.71  & 1.16  & 6.10E-01 & 2.87E+00 & 1.34E+44 & 1.76E+05 & 2.41E+31 \\
   & R & 226.434  & -0.590  & 15.05  & 0.50  & 2.22 &  6.23  & 0.88  & 2.39E-01 & 1.49E+00 & 9.19E+43 & 9.88E+04 & 2.95E+31 \\
85 & R & 226.511  & -0.447  & 17.38  & 0.50  & 3.19 &  2.42  & 1.06  & 1.59E-01 & 3.84E-01 & 9.21E+42 & 2.45E+05 & 1.19E+30
\enddata
\vspace{0.5cm}
Note--
\begin{itemize}
  \item [--] Column (1): Index of outflow candidate.
  \item [--] Column (2): Type of lobe candidate, where ``B", and ``R" denote blue and red lobe candidate, respectively.
  \item [--] Columns (3)--(4): Central Galactic coordinates of lobe candidate.
  \item [--] Column (5): Velocity of \co peak emission.
  \item [--] Column (6): Lowest contour boundary of each lobe, ranges from 50\% to 95\% of the peak intensity. For the isolated lobes, we use $\psi=0.5$. If a lobe candidate suffered from confusion due to adjacent components, then a higher value of $\psi$ was tested to ensure that the lobe could be distinguished from the cloud.
  \item [--] Column (7): Angular area of each lobe, defined by the lowest contour level of each lobe.
  \item [--] Columns (8)-(14): Velocity, length, mass, momentum, kinetic energy, dynamical timescale, and luminosity of each outflow candidate. As suggested by many outflow studies~\citep[e.g.,][]{1984ApJ...284..176S,2018ApJS..235...15L,2019ApJS..242...19L}, we assume a constant $T_\mathrm{ex}$ of 30~K for all detected outflow candidates.
\end{itemize}

\end{deluxetable}
\end{CJK*}
\end{document}